\documentclass[twocolumn,prb,showpacs,superscriptaddress]{revtex4}
\usepackage{graphicx}
\usepackage{amsmath}
\usepackage{amsfonts}
\usepackage{amssymb}
\usepackage{enumerate}
\usepackage{longtable}
\usepackage{dcolumn}
\usepackage{bm}
\usepackage[colorlinks,bookmarks=false,citecolor=blue,linkcolor=blue,urlcolor=blue]{hyperref}
\usepackage{color}
\usepackage{subfigure}
\begin{document}
\title{Microscopic mechanism for the 1/8 magnetization plateau in SrCu$_2$(BO$_3$)$_2$}

\author{M.~Nemec}
\author{G.~R.~Foltin}
\author{K.~P.~Schmidt}
\email{schmidt@fkt.physik.uni-dortmund.de}
\affiliation{Lehrstuhl f\"ur Theoretische Physik 1, TU Dortmund, Germany}

\date{\rm\today}

\begin{abstract}
The frustrated quantum magnet SrCu$_2$(BO$_3$)$_2$ shows a remarkably rich phase diagram
 in an external magnetic field including a sequence of magnetization plateaux. The by far experimentally 
most studied and most prominent magnetization plateau is the 1/8 plateau. Theoretically, one expects that this material
 is well described by the Shastry-Sutherland model. But recent microscopic calculations indicate that the 1/8 plateau is
 energetically not favored. Here we report on a very simple microscopic mechanism which naturally leads to a 
1/8 plateau for realistic values of the magnetic exchange constants. 
We show that the 1/8 plateau with a diamond unit cell benefits most compared to other plateau structures from
 quantum fluctuations which to a large part are induced by Dzyaloshinskii-Moriya interactions. Physically, 
such couplings result in kinetic terms in an effective hardcore boson description leading to a 
renormalization of the energy of the different plateaux structures which we treat in this work 
on the mean-field level. The stability of the resulting plateaux are discussed. Furthermore, our results 
indicate a series of stripe structures above 1/8 and a stable magnetization plateau at 1/6. Most qualitative
 aspects of our microscopic theory agree well with a recently formulated phenomenological 
theory for the experimental data of SrCu$_2$(BO$_3$)$_2$. Interestingly, our calculations point to a rather large ratio of the magnetic couplings in the Shastry-Sutherland model such that non-perturbative effects become essential for the understanding of the frustrated quantum magnet SrCu$_2$(BO$_3$)$_2$. 
\end{abstract}

\pacs{05.30.Jp, 03.75.Kk, 03.75.Lm, 03.75.Hh}

\maketitle

\section{Introduction}
\label{Sect:Intro}
Strongly frustrated quantum magnets in an external field are fascinating systems because the 
interplay between interactions and kinetics can lead to very rich phase diagrams. The magnetization can
be described as a gas of bosonic particles whose density is controlled by the external magnetic field,
and because frustration typically reduces the kinetic energy, Mott insulating phases (corresponding to 
magnetization plateaux)\cite{miyahara99,momoi00a,honecker}, superfluid or even
supersolid phases have been predicted to occur\cite{momoi00b,schmidt08}. 
The experimental observation of these phases is an on-going challenge. A major player in the field is the layered copper
oxide SrCu$_2$(BO$_3$)$_2$, in which several magnetization plateaux 
have been observed\cite{onizuka00,kodama02,sebastian07,takigawa08,levy08,jaime12}.
However, the definitive sequence of plateaux and the presence of supersolid phases remain 
open issues that call for further experimental and theoretical investigation.

\begin{figure}
   \begin{center}
   \includegraphics[width=0.95\columnwidth]{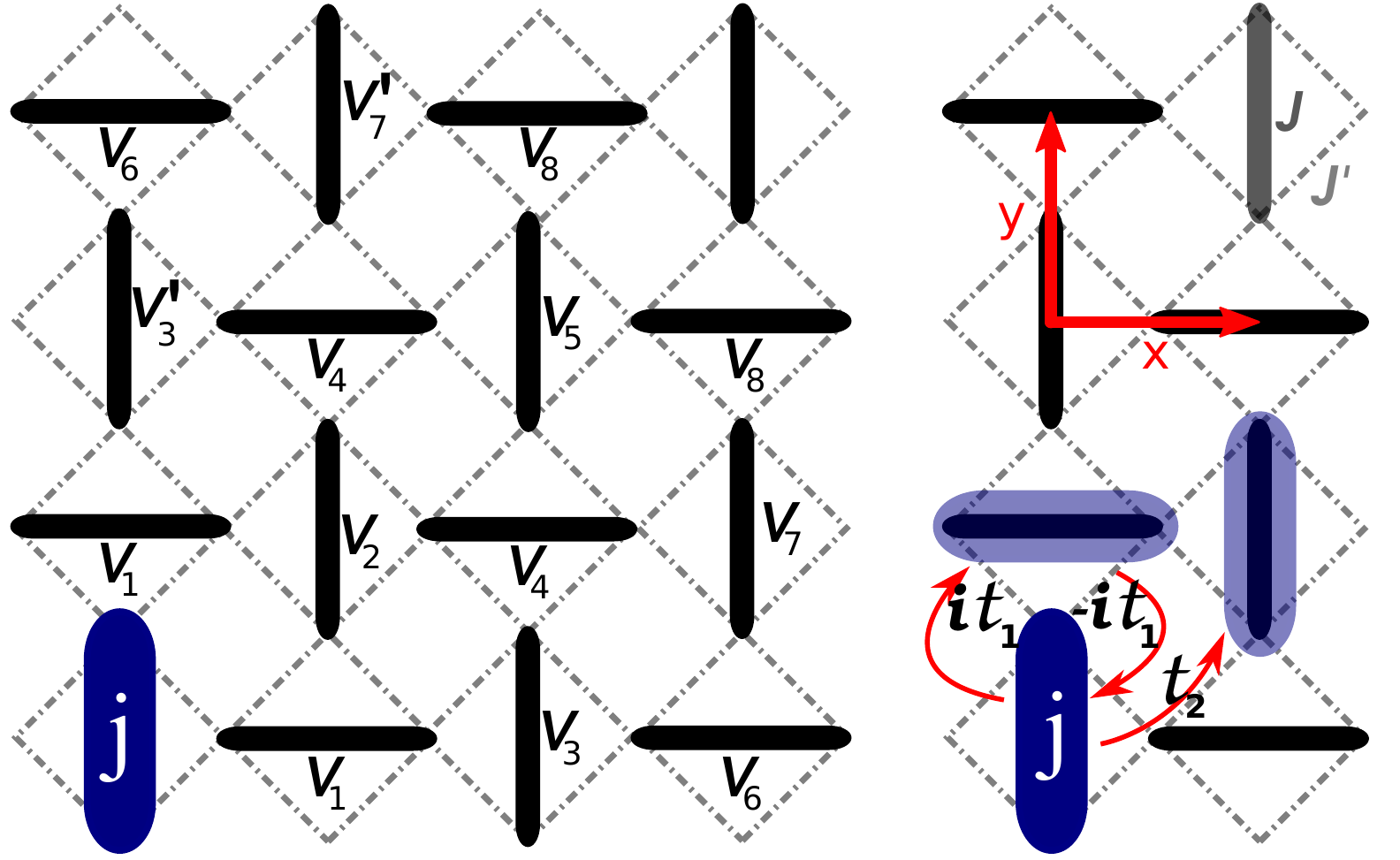}
   \end{center}
    \caption{(Left) Illustration of the Shastry-Sutherland lattice and of the two-body interactions $V_\delta \hat{n}_j \hat{n}_{j+\delta}$. Thick solid lines (dot-dashed lines) correspond to the magnetic exchange coupling $J$ ($J'$).  The two-body interaction labelled by $V_\delta$ are defined as the density-density interaction between the thick dimer labeled as $j$ and the dimer labeled  $V_\delta$. (Right) The hopping amplitudes $t_1$ and $t_2$ are illustrated.}
    \label{fig:Interactions_V+hoppings+axis}
\end{figure}

The magnetization of SrCu$_2$(BO$_3$)$_2$ is expected to be described by 
 the two-dimensional spin-1/2 antiferromagnetic Heisenberg model known as the Shastry-Sutherland model\cite{shastry82} 
in a magnetic field \[H=J'\sum_{<i,j>}\bm S_{i}\cdot \bm S_{j}+J\sum_{\ll i,j\gg}\bm
S_{i}\cdot \bm S_{j}-B\sum_{i}S_i^z\quad,\] with $J'/J\simeq 0.65$, where the $\ll$i,j$\gg$ bonds 
build an array of orthogonal dimers while the $<$i,j$>$ bonds are best seen as inter-dimer couplings 
(see Fig.~\ref{fig:Interactions_V+hoppings+axis}). For $J'/J$ smaller than the phase transition point at $\sim 0.7$ \cite{zheng99,miyahara00,koga00,lauchli02}, the ground state of the model is exactly given by the product of dimer singlets, and the magnetization
process can be described in terms of hardcore bosons which represent 
polarized triplons $|t^1$$\rangle=|\uparrow\uparrow\rangle$
on the dimers interacting and moving on an effective square lattice\cite{momoi00b,miyahara03R,dorier08}. 

All theoretical approaches agree on the presence of magnetization
plateaux at 1/3 and 1/2\cite{momoi00b,misguich01,miyahara00,miyahara03R,miyahara03,isacsson06}, 
in agreement with experiments\cite{onizuka00,sebastian07,jaime12}. Additionally, a plateau at 2/5 has been recently proposed 
 in Ref.~\onlinecite{jaime12}. However, the structure below 1/3 is rather controversial. On the experimental
side, the original pulsed field data have only detected two anomalies interpreted as
plateaux at 1/8 and 1/4\cite{onizuka00}, but the presence of additional phase transitions and of
a broken translational symmetry above the 1/8 plateau has been established by recent
torque and NMR measurements up to 31 T\cite{takigawa08,levy08}. The possibility of additional plateaux has been pointed out by Sebastian et al\cite{sebastian07}, who have interpreted their high-field torque measurements as evidence for plateaux at $1/q$ with $2\le q\le 9$ and at $2/9$.

On the theoretical side, the situation is not settled either. The finite clusters 
available to exact diagonalizations prevent reliable predictions for high-commensurability
plateaux, and the accuracy of the Chern-Simons mean-field approach
initiated by Misguich {\it et al.}\cite{misguich01} and recently used by Sebastian {\it et al.}\cite{sebastian07} to explain additional plateaux is hard to access. The essential difficulty lies in the fact that, since plateaux come from repulsive interactions between triplons, an accurate determination of the low-density, high-commensurability plateaux requires a precise knowledge of the long-range part of the interaction. 

A quantitative step in calculating this long-range part of the interaction has been achieved by high-order series expansions \cite{dorier08} as well as by contractor renormalization \cite{capponi08}. In Ref.~\onlinecite{dorier08} an effective low-energy hardcore boson model for the Shastry-Sutherland model in a magnetic field is derived. The classical solution of the effective model (which is expected to work very well at low densities $n \leq 1/6$ and not too large values of the perturbation $J'/J\approx 0.5$) gives a sequence of magnetization plateaux at 1/9, 2/15, and 1/6. Interestingly, the experimentally most prominent and most studied 1/8 plateau is not favored energetically \cite{footnote}.

Recently, a phenomenological theory based on interpreting boron and copper NMR data of SrCu$_2$(BO$_3$)$_2$ in high magnetic fields reveals an even more complex magnetization process \cite{takigawa12}. Evidences for a sequence of plateaux at densities 1/8, 2/15, and 1/6 is found. Furthermore, the regime between 2/15 and 1/6 is interpreted as an infinite hierarchy of stripe structures. Interestingly, the structure of the different plateaux is in disagreement with the proposed plateau structures in the classical limit \cite{dorier08}.   

In this work, we are aiming at a microscopic theory which we want to compare with the phenomenological findings. To this end, we treat the quantum fluctuations of the pure Shastry-Sutherland model present at low densities and we extend the effective low-energy model derived by perturbative continuous unitary transformations (pCUTs)\cite{dorier08} by the dominant effect of additional magnetic couplings. 
Most importantly, we find that the inclusion of Dzyaloshinskii-Moriya (DM) interactions results in a natural microscopic mechanism for the stability of a 1/8 plateau with a diamond unit cell. Physically, such unfrustrated terms give rise to kinetic processes in the effective description and therefore they introduce quanum fluctuations in the classical plateau structures. Additionally, it is shown that our theory predicts a stable 1/6 plateau whose structure is only consistent with the phenomenological theory if the coupling ratio $J'/J$ is rather large. This is not a consequence of the induced quantum fluctuations. It results from the fact that the two-body density-density interactions do not respect anymore the perturbative hierarchy close to the phase transition point and therefore novel structures are stabilized. Finally, our theory predicts a series of stripe structures between 1/8 and 1/6. Altogether, our theory shares many similarities with the recent findings for SrCu$_2$(BO$_3$)$_2$ in a magnetic field. Nevertheless, several discrepancies between the microscopic and the phenomenological theory still remain. This is either due to the limited access to the most complicated non-perturbative regime of the studied microscopic model or due to the fact that the physics of SrCu$_2$(BO$_3$)$_2$ in a magnetic field depends on further subtle details of the material.

The paper is organized as follows. In Sect.~\ref{Sect:ClassicalPlateaux}, we explain the origin of the classical plateaux in the effective model and we introduce the relevant plateau structures.  In Sect.~\ref{Sect:QF}, the additional couplings incorporated in this work are introduced in detail and their most important physical effect is illustrated. Additionally, the details of the mean-field approach are given Sect.~\ref{Sect:MF}. We finally present our results in Sect.~\ref{Sect:Results} and we summarize the major findings in Sect.~\ref{Sect:Summary}.

\section{Classical plateaux} 
\label{Sect:ClassicalPlateaux}

\subsection{pCUT+CA} 
\label{SSect:pCUT}
The pCUT transforms the Shastry-Sutherland model into an effective model conserving the number of triplons\cite{dorier08}. 
The relevant processes for the physics in a finite magnetic field have maximum total spin 
and total $S_z$. This is true as long as bound states are not essential at low energies \cite{manmana11} which we assume in the following. The effective Hamiltonian obtained by the pCUT takes then the form of an interacting hardcore boson 
model where the amplitudes of the hardcore boson model are given as a high-order series expansion in $J'/J$. 

Typically, any kind of kinetic and interaction processes are present in such an effective model. The effective Hamiltonian $H_{\rm eff}$ is by no means simpler than the original one in general, but it is in the limit of small density. In Ref.~\onlinecite{dorier08} all terms with up to three creation and annihilation operators and all four-body interactions that first appear up to order $\leq 8$ have been kept. But in the small density limit $n \leq 1/6$, the magnetization is not affected by the three-particle and four-particle interactions \cite{dorier08}. Furthermore, in that limit the kinetic terms are very small, and they can be considered as a perturbation of the interaction part. It is thus appropriate to use a Hartree approximation  in which the variational ground state is a product of local boson wave-functions since this approximation becomes exact in the limit of vanishing kinetic energy.

This Hartree approximation is most simply implemented by mapping the effective model onto a spin 1/2 model 
using the Matsubara-Matsuda representation\cite{matsubara56,dorier08} of hardcore bosons. In the spin language it 
then translates into the classical approximation (CA) where the spins are treated as classical vectors of length 1/2.  

It has been found in Ref.~\onlinecite{dorier08} that only magnetization plateaux are realized in the classical limit (except a tiny superfluid in the dilute limit) although superfluid or supersolid phases can be described within the CA.
 The magnetization plateaux correspond to Wigner crystals where triplons are frozen in a periodic fashion in the ground state breaking 
the discrete translational symmetry of the problem. The classical energy of all plateaux at low densities $n \leq 1/6$ can be quantitatively captured by the following 
effective Hamiltonian
\begin{equation}
 \label{eq:ham}
 \frac{\hat{H}_{\rm eff}^{\rm cl}}{J} = - (\mu-\mu_0 ) \sum_{j} \hat{n}_j+\frac{1}{2} \sum_{\delta} V_\delta \sum_{j} \hat{n}_j \hat{n}_{j+\delta} \quad ,
\end{equation}
where the sum over $j$ runs over all the sites of the effective square lattice formed by the dimers. The first term represents the chemical potential $(\mu -\mu_0 )$ of hard-core bosons where $\mu$ corresponds to the external magnetic field $B$ and where $\mu_0$ originates from the Shastry-Sutherland model. We have calculated $\mu_0$ up to order 17 in $J'/J$ (see appendix \ref{sect:Appendix}). The second term denotes the two-triplon density-density interactions $V_{\delta}$. Note that the sum over $\delta$ contains all interaction terms illustrated in Fig.~\ref{fig:Interactions_V+hoppings+axis} plus their three symmetric counterparts. The classical energy of a given magnetization plateau depends therefore only on the arrangement of the local densities $\hat{n}=b^\dagger b^{\phantom{\dagger}}$. 

\begin{figure}
   \begin{center}
     \includegraphics[width=\columnwidth]{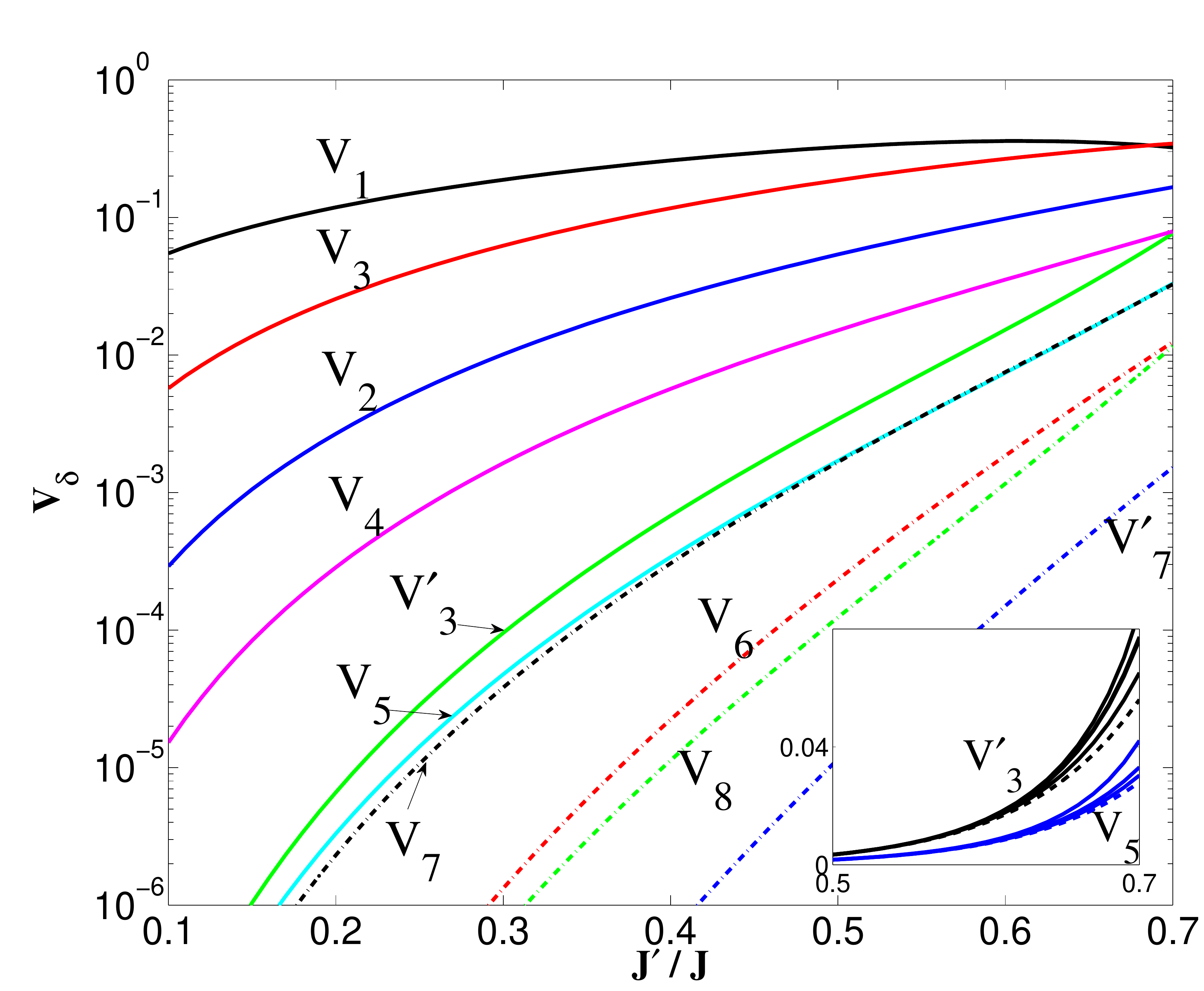}
   \end{center}
   \caption{Amplitude $V_\delta$ of the extrapolated two-body interactions as a function of
     $J'/J$. Inset: Different dlogPad\'{e} extrapolants (solid lines) as well as the bare series (dashed lines) for $V'_3$
     and $V_5$. The displayed curves correspond to the data given in Ref.~\onlinecite{dorier08}.}
    \label{fig:Interactions_V_PLOT}
\end{figure}

All two-body interactions $V_\delta$ which first appear at order less or equal to 10 have been calculated up to order 15 (except $V_1$ which has been determined up to order 14). The evolution with $J'/J$ of the two-body interactions defined in Fig.~\ref{fig:Interactions_V+hoppings+axis} is depicted in 
Fig.~\ref{fig:Interactions_V_PLOT}. At small $J'/J$, interactions beyond $V_4$ are small and may be neglected, but for larger $J'/J$  the higher order terms $V'_3$, $V_5$, and $V_7$ (appearing at order 6) become important and contribute to the formation of low-density plateaux. For these terms, the bare series and the dlogPad\'{e} extrapolations are basically indistinguishable below $J'/J=0.6$. Beyond that value, various extrapolations still give consistent results for the two-body interactions. Interestingly, we find that the perturbative hierarchy becomes invalid for $J'/J \geq 0.65$, i.e.~one observes $V_3\approx V_1$ and $V'_3 \approx V_4$ although these interactions do not originate from the same perturbative order. In the following we deduce several indications that the frustrated quantum magnet SrCu$_2$(BO$_3)_2$ is most likely situated in this most challenging but also most interesting non-perturbative regime. 

\subsection{Plateaux structures and energies} 
\label{SSect:CA}

\begin{figure}
   \begin{center}
\subfigure[$\,J'/J=0.5$]{
    \includegraphics[width=\columnwidth]{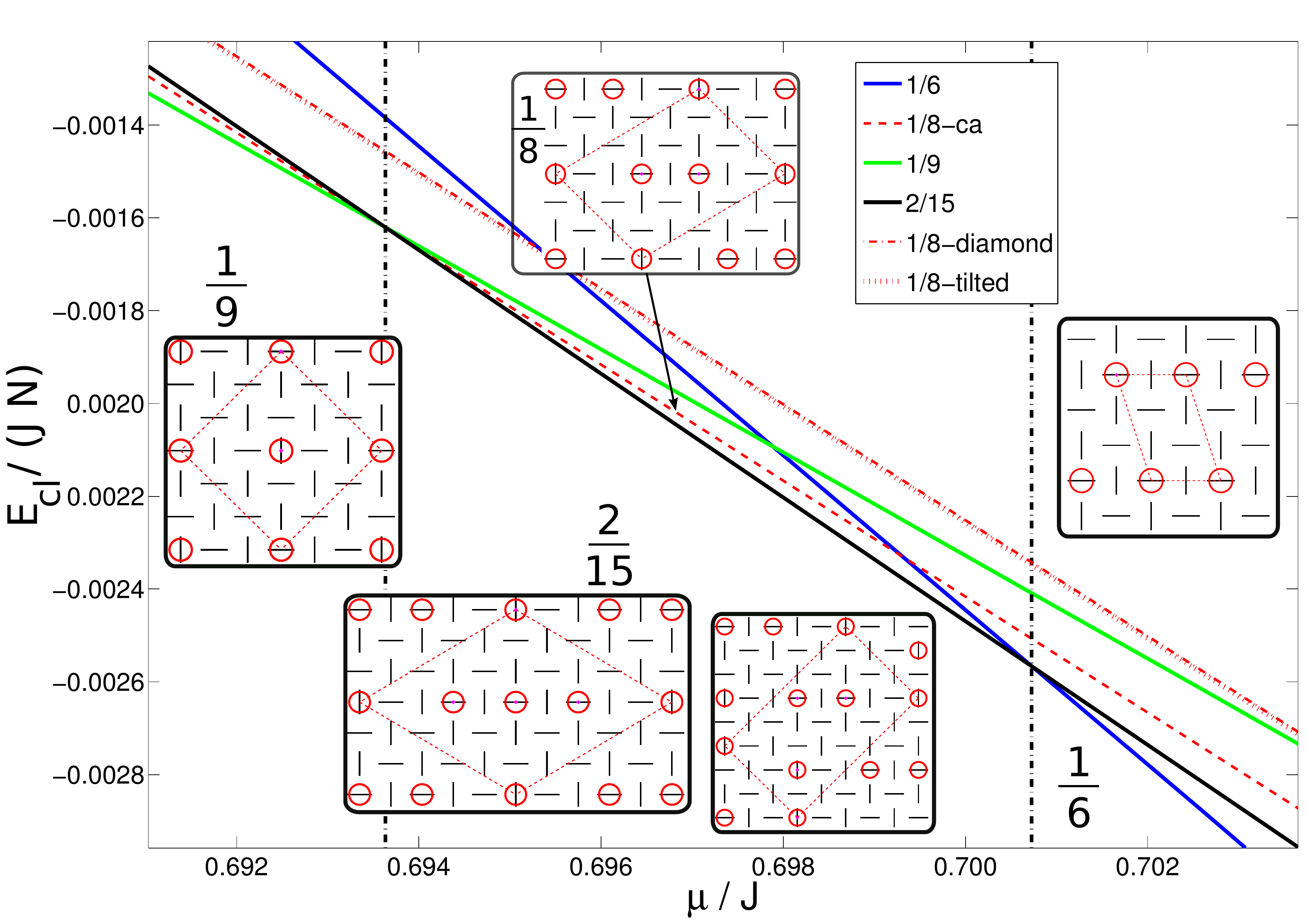}
\label{fig:ClassR2_J50}
}
\subfigure[$\,J'/J=0.68$]{
   \includegraphics[width=\columnwidth]{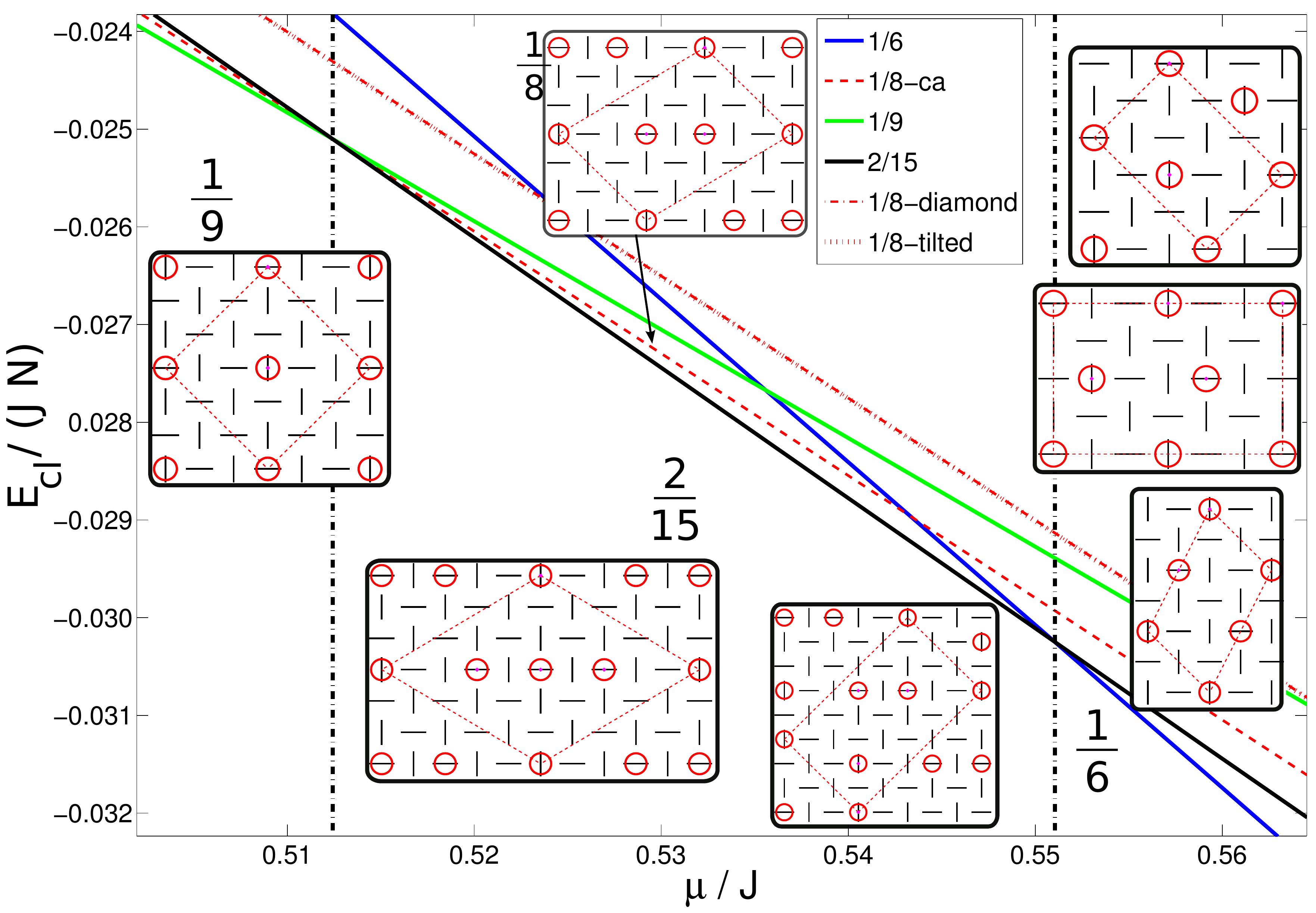}
\label{fig:ClassR2_J68}
}
   \end{center}
   \caption{Lowest classical energies $E_{\rm cl}/(JN)$ per dimer of all plateaux in the density regime $1/9 \leq n \leq 1/6$ as a function of $\mu/J$ for (a) $J'/J = 0.5$ and (b) $J'/J = 0.68$ using Eq.~\eqref{eq:H_MF_final} setting all kinetic terms to zero. The energies of the different $2/15$ respectively $1/6$ structures shown are exactly degenerate. The circle on a dimer denotes the presence of a particle.}
    \label{fig:ClassR2}
\end{figure}

\begin{figure}
   \begin{center}
   \includegraphics[width=\columnwidth]{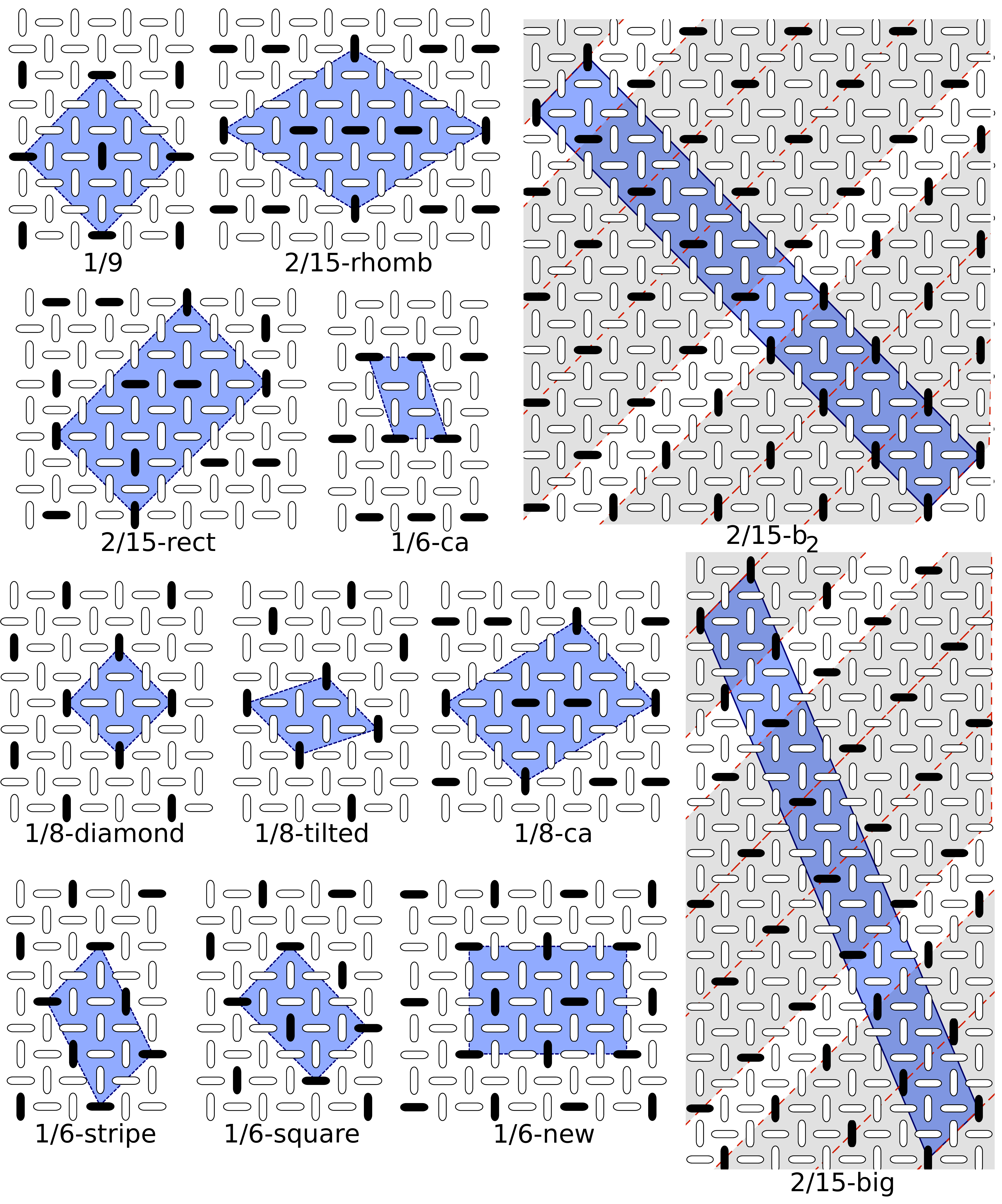}
   \end{center}
   \caption{Classical plateaux which are relevant at low densities. The unit cells of the different structures are shown in dark gray (blue). Note that not all plotted structures are realized within the CA. This includes all plateaux at density $n=1/8$. Additionally, for the structures $2/15$-b2 ($2/15$-big) is built alternatingly by three $1/8$-diamond (three $1/8$-tilted) and one $1/6$. Each $1/8$-stripe is shaded in light gray.}
    \label{fig:classical_plateaux}
\end{figure}

Representative results for the classical limit at $J'/J=0.5$ and $J'/J = 0.68$ are shown in Fig.~\ref{fig:ClassR2}. Physically, the realization of all relevant plateaux in the classical limit at low magnetizations follow the same guiding principle for $J'/J<0.67$. Triplons are placed such that only two-body density-density interactions $V_\delta$ which start in order 6 or higher have to be paid. Interactions starting in order 6 are $V^\prime_3$, $V_5$, and $V_7$ (see Figs.~\ref{fig:Interactions_V+hoppings+axis} and \ref{fig:Interactions_V_PLOT}). Note that only the classical plateau at 1/9 does not pay any order-6 interaction. The explicit expressions for the classical energies of all the plateaux are given in the appendix \ref{sect:Appendix}. Additionally, we display all the relevant classical structures in Fig.~\ref{fig:classical_plateaux}.

We obtain a sequence of plateaux at densities $1/9$, $2/15$, and $1/6$ in the range $0.5 \leq J'/J \leq 0.68$ using Eq.~\eqref{eq:ham} in full agreement with the results given in Ref.~\onlinecite{dorier08} for $J'/J=0.5$. The details of the calculation will be discussed below. It is remarkable that there are only two transitions with two multi-intersectional points (see Fig.~\ref{fig:ClassR2}). Note that at these two multi-intersectional points many other magnetization plateaux are degenerate which we do not display in Fig.~\ref{fig:ClassR2} for clarity reasons. The energy of the $1/8$-ca structure shown in Fig.~\ref{fig:ClassR2} just intersects the other curves but the $1/8$ plateau is not favored. The two plateaux $1/8$-diamond and $1/8$-tilted are almost degenerate and slightly above the $1/8$-ca plateau. The existence of the 1/8 plateau in the frustrated quantum magnet SrCu$_2$(BO$_3$)$_2$ can therefore not be explained solely by the effective interactions \cite{dorier08}. The origin must be a consequence of quantum fluctuations which either originate from the Shastry-Sutherland model itself or from additional magnetic terms like the DM-interaction. Below, we will indeed show that the DM-interaction plays a central role for the appearance of the 1/8 plateau with a diamond unit cell.  

Although the sequence of plateaux is unchanged when $J'/J$ is increased, the structure of the classical $1/6$ plateau becomes different for $J'/J \geq 0.67$. This is a direct consequence of the fact that the two-body interaction $V^\prime_{3}$ strongly increases for large $J'/J$ compared to the competing terms $V_4$, $V_5$, and $V_7$ (see Figs.~\ref{fig:Interactions_V+hoppings+axis} and \ref{fig:Interactions_V_PLOT}). It therefore becomes attractive to realize structures containing the latter interactions and to avoid $V^\prime_{3}$. This is the reason why the structure of the $1/6$ plateau changes for $J'/J \geq 0.67$ already in the CA as can be seen in the difference between Fig.~\ref{fig:ClassR2_J50} and Fig.~\ref{fig:ClassR2_J68}. 

Most interestingly, the phenomenological theory of the experimental data \cite{takigawa12} proposes the $1/6$-stripe plateau which does not contain any $V^\prime_{3}$ interaction but is built by $V_4$ and $V_5$ interactions similarly to the $1/6$-square and the $1/6$-new structure shown in Fig.~\ref{fig:ClassR2_J68}. All the three plateaux have exactly the same classical energy. Therefore, we have found strong evidences that the ratio $J'/J$ must be rather large being close to the phase transition $J'/J \gtrsim 0.67$. Let us remark that the accuracy of the value $J'/J=0.67$, where the 1/6 structure does change in the classical limit, of course depends on the accuracy of the two-body interactions $V_\delta$ which might lead to a little shift of this value.    

\section{Quantum fluctuations} 
\label{Sect:QF}
In the following we discuss the quantum fluctuations which are relevant for the magnetization
 plateaux at low densities. We study first the kinetic terms of the pure Shastry-Sutherland model. 
Afterwards we discuss the effect of additional magnetic couplings like DM interactions 
or Heisenberg exchanges to more distant neighbors.

\subsection{Kinetic terms in the Shastry-Sutherland model} 
\label{SSect:KineticTerms}
\begin{figure}
   \begin{center}
   \includegraphics[width=\columnwidth]{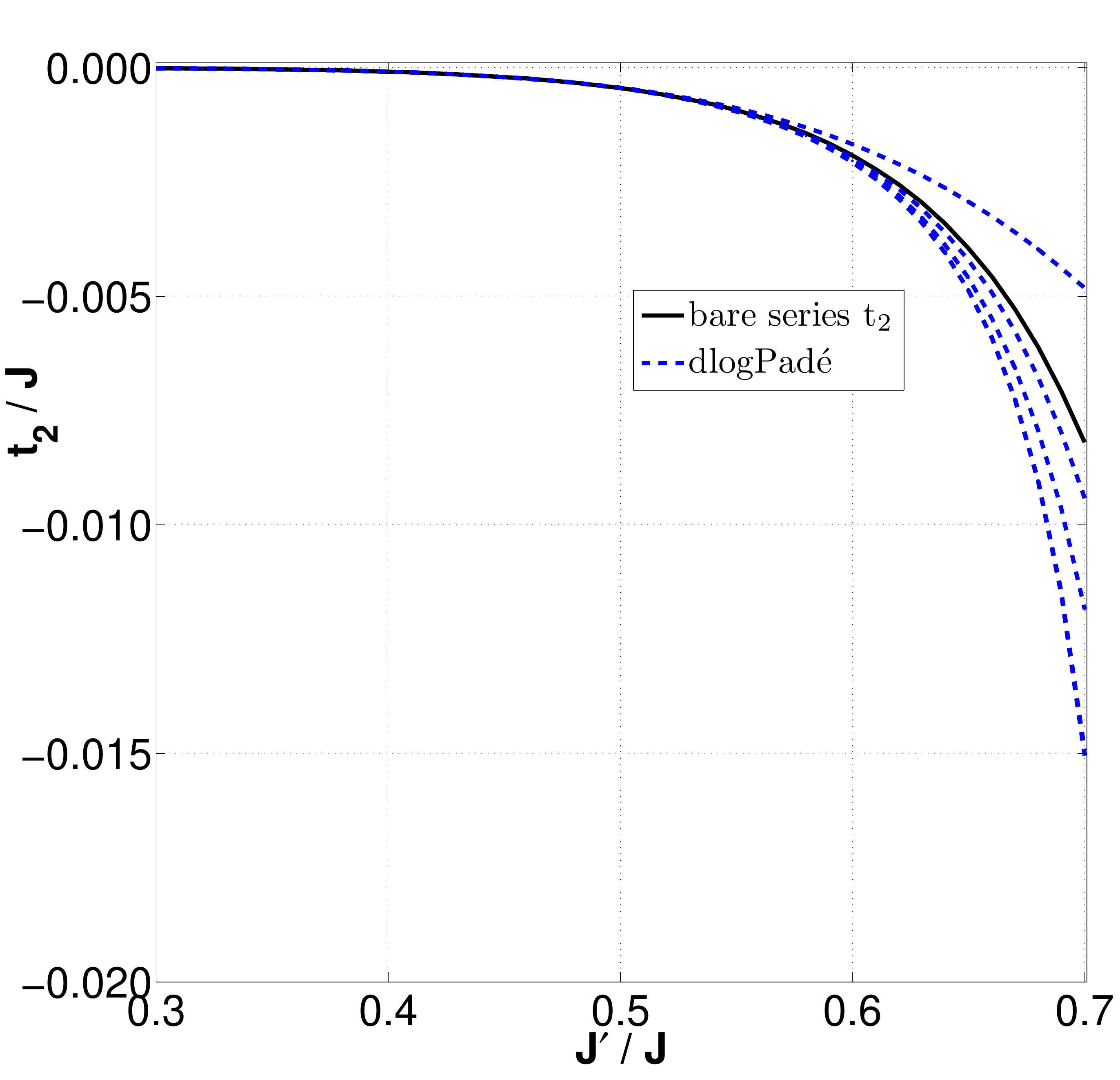}
   \end{center}
   \caption{One-triplon hopping amplitude $t_2/J$ as a function of $J'/J$. Solid line represents the bare series of order 17 while the dashed curves correspond to various dlogPad\'{e} extrapolations.}
    \label{fig:t2_extrapolated}
\end{figure}

It is known that there exists only one very weak standard hopping process $t_2$ in the Shastry-Sutherland model
 due to the strong frustration \cite{miyahara99,knetter00_2}. This kinetic process represents a hopping over the diagonal (see Fig.~\ref{fig:Interactions_V+hoppings+axis})  which starts only
 in order six perturbation theory with a very small prefactor
\begin{align}
\frac{\hat{H}_{\rm eff, t_2}}{J} &= t_2 \sum_{\alpha = \pm x \pm y} \sum_{j } \hat{b}^{\dagger}_{j + \alpha} \hat{b}^{\phantom{\dagger}}_j\, .
\end{align}
We have calculated this hopping amplitude up to order 17 in $J'/J$ using Takahashi's degenerate perturbation theory \cite{takahashi77,klagges12}. Note that one would get exactly the same series using pCUTs. For the given problem, Takahashi's expansion is more efficient for one-particle properties and one is able to determine a higher perturbative order. The extrapolated series are shown in Fig.~\ref{fig:t2_extrapolated}. The amplitude is very small in a broad range of couplings $J'/J \leq 0.6$. Nevertheless, in the most important range $J'/J > 0.6$ we observe a rapid increase of the hopping amplitude $t_2$. The coupling is found to be $-0.015 \leq t_2/J \leq -0.005$ for $J'/J = 0.65-0.70$. 

Let us remark that the dominant kinetic terms in the effective model are actually so-called correlated hopping terms where one triplon is able to hop if another particle is present but remains static \cite{momoi00b,knetter00_2,dorier08}. Indeed, correlated hopping processes arise already in order two perturbation theory. But such terms are not important in the limit of low densities. Here triplons in the Wigner crystals are rather far apart and the effects of correlated hopping is suppressed. As a consequence, the importance of correlated hopping increases with increasing density. Below, we show that correlated hopping is unimportant for $n\leq 1/8$ but it might be of relevance for $n=1/6$. Clearly, at larger densities like $n=1/4$ correlated hopping is expected to be essential \cite{foltin12}.

\subsection{Additional magnetic couplings} 
\label{SSect:AddCouplings}

The pure Shastry-Sutherland model is a good but not perfect microscopic model for the description of SrCu$_2$(BO$_3$)$_2$. It is known that a finite DM interaction is present which is expected to be a few percent of the dominant nearest-neighbor Heisenberg exchange $J$ \cite{cepas01,cheng07,mazurenko08,takigawa08,romhanyi11}. Additionally, also Heisenberg exchange interactions between next-nearest neighbor dimers might be of a similar order \cite{mazurenko08}. 

It is important to realize that all of these interactions are typically unfrustrated and, as a consequence, will pump
 kinetic energy into the system. Here we aim at treating the first-order effects on the effective low-energy model. This is expected to be a reasonable approximation since all the additional couplings are small perturbations of the order $J/100$. Let us stress that these perturbations can nevertheless be important at low densities because perturbatively the two-body interactions stabilizing the low-density plateaux appear in a rather high order in $J'/J$ as discussed in detail above.

\begin{figure}
   \begin{center}
   \includegraphics[width=0.9\columnwidth]{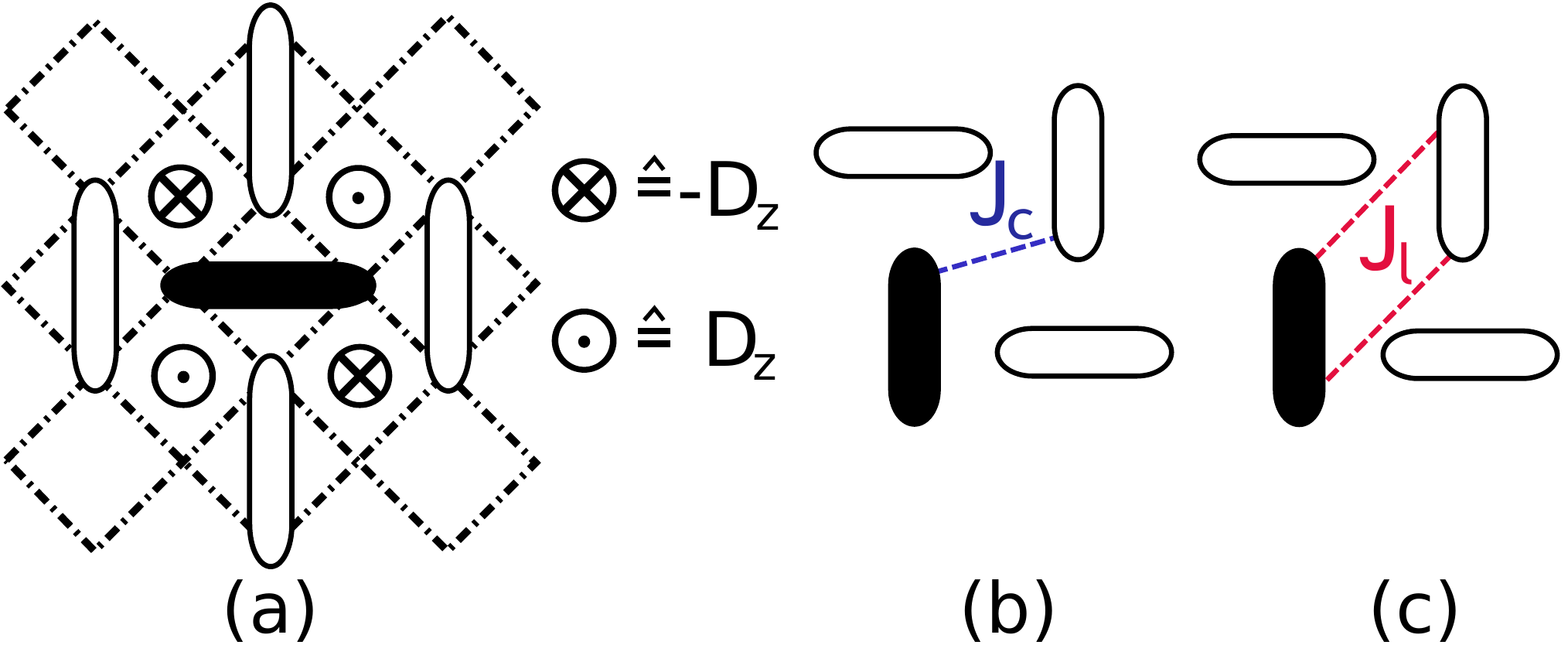}
   \end{center}
   \caption{Illustration of the additional magnetic couplings beyond the ones included in the Shastry-Sutherland model: (a) inter-dimer DM interaction $D_{\rm z}$ in $z$-direction. (b-c) Heisenberg exchanges to next-nearest neighbor dimers are denoted by $J_{\rm c}$ (chain-like coupling) and $J_{\rm l}$ (ladder-like coupling).}
    \label{fig:illustrations_couplings}
\end{figure} 

We start by discussing the effects of DM interactions $\vec{D}\cdot \vec{S}_i\times \vec{S}_j$ 
which turn out to
 be the most relevant correction for SrCu$_2$(BO$_3$)$_2$. One has to distinguish between
 intra- and inter-dimer DM interactions.

Taking the DM interaction as a perturbation of the bare Shastry-Sutherland model, it can be easily checked that the intra-dimer DM interactions leads to a mixing of singlet and triplet states on the single dimers. The effective low-energy model is therefore only affected in second order perturbation theory for this perturbation. 

This is different for the inter-dimer interaction between two dimers. We assume that the interaction is largest in
 $z$-direction and that it has a different sign for the two interactions between nearest-neighbor
 dimers as illustrated in Fig.~\ref{fig:illustrations_couplings} \cite{cepas01,cheng07,mazurenko08,takigawa10}. In the effective model the inter-dimer DM interaction results in a nearest-neighbor hopping $t_1=D_z/2$ which reads
\begin{align}
 \frac{\hat{H}_{\rm eff, t_1}}{J} &= + i t_1 \sum_{\substack{\alpha = \pm x; \\ \alpha = \pm y;}} \sum_{j} (-1)^j \hat{b}^{\dagger}_{j + \alpha} \hat{b}^{\phantom{\dagger}}_j \quad ,
\end{align}
where $(-1)^j$ positive (negative) corresponds to effective sites $j$ representing a vertical (horizontal) dimer as shown in Fig.~\ref{fig:Interactions_V+hoppings+axis}. Additionally, we note that the 
inter-dimer DM interactions $D_{\rm x}$ and $D_{\rm y}$ do not give any boson-conserving terms in the effective model in leading order. These couplings result in processes of the form $(b^\dagger b^\dagger b^{\phantom{\dagger}} + {\rm h.c.})$. Since the $D_{\rm x,y}$ components are expected to be smaller than $D_{\rm z}$ and since they change the terms in the effective model only in order $D_{\rm x,y} (J'/J)^n$ with $n>1$, effects of the transverse components of the inter-dimer DM interactions are expected to be small.

Another aspect for a microscopic description of the material SrCu$_2$(BO$_3$)$_2$ are additional Heisenberg exchanges to more distant dimers. The couplings to next nearest-neighbor dimers are denoted by $J_c$ and by $J_l$. They are shown in Fig.~\ref{fig:illustrations_couplings}. One can easily see that both couplings contribute in first order to the already discussed diagonal hopping element $t_2$ present in the pure Shastry-Sutherland model
\begin{align}
 \frac{t_2^{\rm total}}{J} &= \frac{1}{J} \left(t_2 - \frac{J_c}{4} + \frac{J_l}{2} \right) \quad .
\end{align}
Recently, ab-initio calculations estimated $J_c/J \approx 0.023$ and $J_l/J \approx 0.008$ for $J'/J \approx 0.56$ \cite{mazurenko08}. The latter ratio for $J'/J$ is likely too small. We nevertheless trust the order of magnitude for the additional Heisenberg couplings. Then the total diagonal hopping $t_2^{\rm total}$ is expected to be similar to the amplitude $t_2$ of the pure Shastry-Sutherland model.

Altogether, the most important kinetic terms at low densities correspond to a nearest-neighbor hopping $t_1$ originating from the inter-dimer DM interaction and  to a next-nearest neigbor hopping $t_2$ already present in the pure Shastry-Sutherland model. Both couplings are expected to be of the order $J/100$ for a realistic coupling ratio $J'/J\approx 0.65$. 

Let us finally stress that the effective Hamiltonian to first order in the additional magnetic interactions has still the same symmetries as the pure Shastry-Sutherland model in a field. This is especially true for the U(1) symmetry which is only broken once order 2 processes in the DM interactions are taken into account. As a consequence, in our effective low-energy description it is still possible that the model displays true superfluid or supersolid phases where the U(1) symmetry is spontaneously broken at zero temperature.

In the CA, the energy of the magnetization plateaux is independent of the kinetic terms. Thus, the quantum fluctuations induced by the just discussed kinetic processes are not at all captured by the CA for the Mott insulating phases. In the following we describe our mean-field approach aiming at treating the quantum fluctuations on the different magnetization plateaux.

\section{Approach} 
\label{Sect:MF}

We are interested in studying the effects of the kinetic terms $t_1$ and $t_2$ on the classical magnetization plateaux. Let us stress that we do not want to describe the melting of the Wigner crystals clearly present for large kinetic terms. We focus on the competition between different low-density plateaux because experimentally a sequence of plateaux is found at densities 1/8, 2/15, and 1/6. 

The basic idea is that all symmetrically equivalent triplons of a Wigner crystal benefit from the kinetic processes in an identical fashion. Additionally, the kinetic processes $t_1$ and $t_2$ are small compared to the largest two-body interactions $V_1$ and $V_3$ and therefore quantum fluctuations of triplons inside a Wigner crystal are expected to be rather local (but not necessarily small) about their classical positions. Consequently, we use a mean-field approach. 

\begin{figure}
   \begin{center}
   \includegraphics[width=0.95\columnwidth]{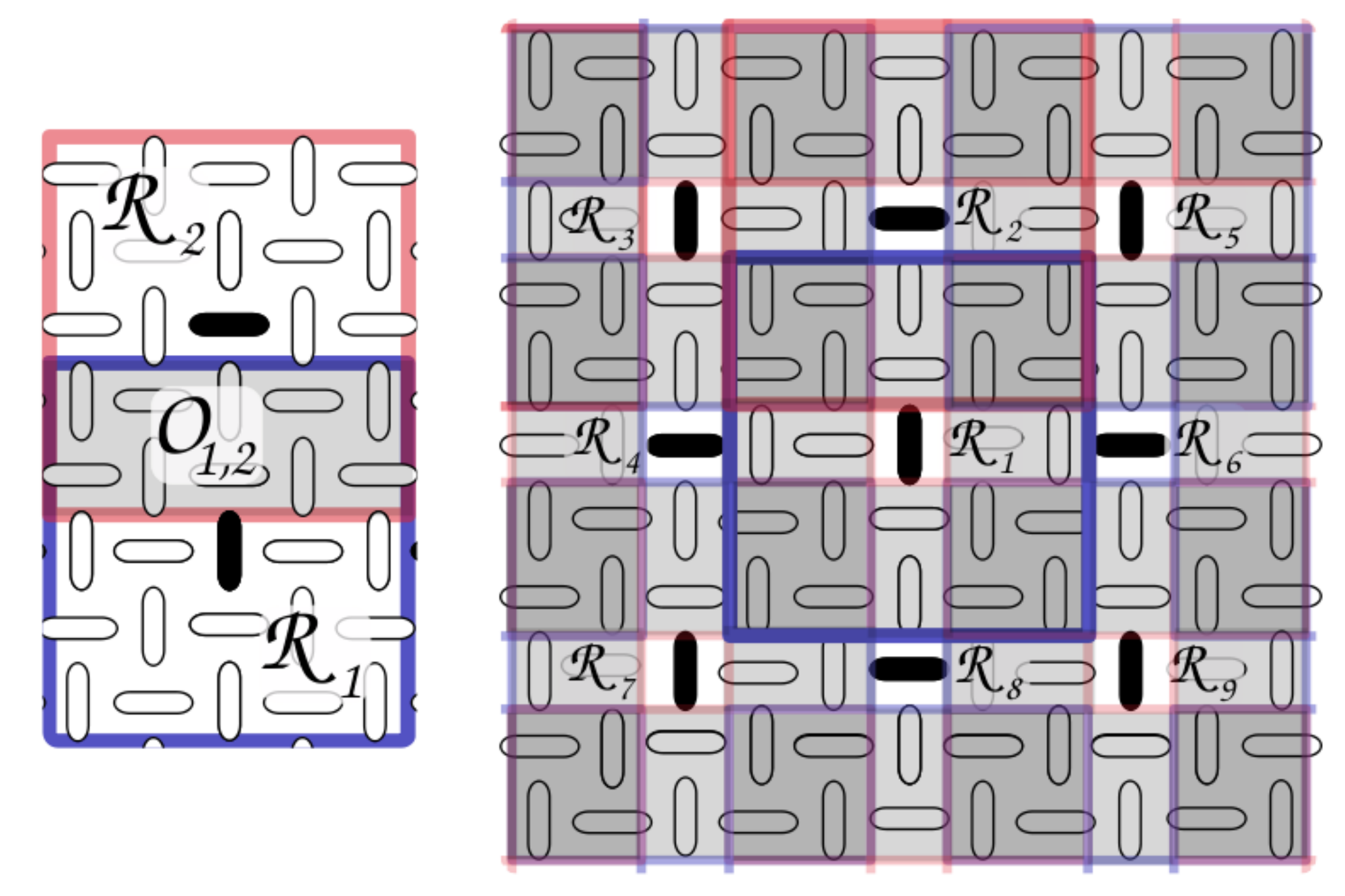}
   \end{center}
   \caption{Illustrations of the regions $\mathcal{R}_r$ having the dimension dim($\mathcal{R}_r) = 25$. Left: Figure displays two regions $\mathcal{R}_1$ and $\mathcal{R}_2$ of two triplets (black dimers) having an overlap $\mathcal{O}_{12}$ (shaded area). Right: Illustration of the regions $\mathcal{R}_r$ for the classical plateau at density $n=1/9$. The regions with an odd respectively even number are periodically equivalent. Note that regions $\mathcal{R}_r$ with $r$ odd and $r$ even display exactly the same behaviour due to rotational symmetry of the lattice.}
    \label{fig:regions}
\end{figure} 

We split the Hilbert space into periodically equivalent finite regions $\mathcal{R}_r$ (see Fig.~\ref{fig:regions}) and we investigate the dynamics (quantum fluctuations) of a single triplon inside these regions assuming that all other triplons of the Wigner crystal remain static. Physically, this corresponds to an effective one-body problem in a static external potential which is given by the two-body interactions $V_\delta$ of the surrounding triplons. The kinetic hopping amplitudes $t_1$ and $t_2$ give rise to hoppings of triplons around their classical position leading to an overall energy reduction of the Wigner crystal. This procedure is iterated selfconsistently until convergence is reached.

To be specific, the full effective Hamiltonian
\begin{align} \label{eq:Heff}
 \nonumber \frac{\hat{H}_{\rm eff}}{J} &= \frac{1}{2} \sum_{\delta} V_\delta \sum_{j } \hat{n}_j \hat{n}_{j+\delta} + i t_1 \sum_{\substack{\alpha = \pm x; \\ \alpha = \pm y;}} \sum_{j } (-1)^j \hat{b}^{\dagger}_{j + \alpha} \hat{b}^{\phantom{\dagger}}_j \\
 &\quad + t_2 \sum_{\alpha = \pm x \pm y} \sum_{j } \hat{b}^{\dagger}_{j + \alpha} \hat{b}^{\phantom{\dagger}}_j - (\mu -\mu_0) \sum_{j } \hat{n}_j \\
  &= \sum_r \left( \hat{H}^{(r)} + \hat{H}^{(\mathcal{O}_{r,r'})}+ \hat{H}_{\rm kin}^{(r,r' )}  \right) - (\mu -\mu_0) \sum_{j } \hat{n}_j \nonumber
\end{align}
is split into three different parts. We stress that each lattice site is at least part of one region $\mathcal{R}_r$, i.e. different regions are allowed to overlap. The overlap of one region $\mathcal{R}_r$ with all other regions $\mathcal{R}_{r'}$ is denoted by $\mathcal{O}_{r,r'}$. Formally, if a dimer $j$ is contained in the overlap $j\in \mathcal{O}_{r,r'}$, it follows  $j \in \mathcal{R}_r$ and $j \in \mathcal{R}_{r'}$ with $r'\neq r$. 

The explicit expressions of the three terms in the Hamiltonian are given by  
\begin{align}\label{Eq:Eff_Ham_Term1}
\nonumber \hat{H}^{(r)} &= \frac{1}{2} \sum_{\delta} V_\delta \sum_{\substack{j \in \mathcal{R}_r \\ j \notin \mathcal{O}_{r,r'}}} \hat{n}_j \hat{n}_{j+\delta} \\ 
\nonumber &\quad+ i t_1 \sum_{\substack{\alpha = \pm x; \\ \alpha = \pm y;}} \sum_{\substack{j \in \mathcal{R}_r \\ j, j+\alpha \notin \mathcal{O}_{r,r'}}} (-1)^j \hat{b}^{\dagger}_{j + \alpha} \hat{b}^{\phantom{\dagger}}_j \\
 &\quad + t_2 \sum_{\alpha = \pm x \pm y} \sum_{\substack{j \in \mathcal{R}_r \\ j, j+\alpha \notin \mathcal{O}_{r,r'}}} \hat{b}^{\dagger}_{j + \alpha} \hat{b}^{\phantom{\dagger}}_j \, ;
\end{align}
\begin{align}\label{Eq:Eff_Ham_Term2}
\nonumber \hat{H}^{(\mathcal{O}_{r,r'})} &= \frac{1}{2} \sum_{\delta} V_\delta \sum_{\substack{j \in \mathcal{O}_{r,r'}}} \frac{1}{N_j}\, \hat{n}_j \hat{n}_{j+\delta} \\
\nonumber &\quad + i t_1 \sum_{\substack{\alpha = \pm x; \\ \alpha = \pm y;}} \sum_{\substack{j \notin \mathcal{R}_{r'} \\ j+\alpha \in \mathcal{O}_{r,r'}}} (-1)^j \hat{b}^{\dagger}_{j + \alpha} \hat{b}^{\phantom{\dagger}}_j \\
\nonumber &\quad + i t_1 \sum_{\substack{\alpha = \pm x; \\ \alpha = \pm y;}} \sum_{\substack{j \in \mathcal{O}_{r,r'}}}   \frac{(-1)^j }{N_j}\, \hat{b}^{\dagger}_{j + \alpha} \hat{b}^{\phantom{\dagger}}_j \\
\nonumber &\quad + t_2 \sum_{\alpha = \pm x \pm y} \sum_{\substack{j \notin \mathcal{R}_{r'} \\ j+\alpha \in \mathcal{O}_{r,r'}}} \hat{b}^{\dagger}_{j + \alpha} \hat{b}^{\phantom{\dagger}}_j \\
&\quad + t_2 \sum_{\substack{\alpha = \pm x \pm y}} \sum_{\substack{j \in \mathcal{O}_{r,r'}}} \frac{1}{N_j}\, \hat{b}^{\dagger}_{j + \alpha} \hat{b}^{\phantom{\dagger}}_j \, ;
\end{align}
\begin{align}\label{Eq:Eff_Ham_Term3}
\nonumber \hat{H}_{\rm kin}^{(r,r' )} &= i t_1 \sum_{\substack{\alpha = \pm x; \\ \alpha = \pm y;}} \sum_{\substack{j \notin \mathcal{R}_{r'} \\ j+\alpha \notin \mathcal{R}_r}} (-1)^j \hat{b}^{\dagger}_{j + \alpha} \hat{b}^{\phantom{\dagger}}_j \\
&\quad + t_2 \sum_{\substack{\alpha = \pm x \pm y}} \sum_{\substack{j \notin \mathcal{R}_{r'} \\ j+\alpha \notin \mathcal{R}_r}} \hat{b}^{\dagger}_{j + \alpha} \hat{b}^{\phantom{\dagger}}_j \, .
\end{align}

The first term $\hat{H}^{(r)}$ of the effective Hamiltonian given by Eq.~\eqref{Eq:Eff_Ham_Term1} contains all hopping processes where the initial and the final dimer is only part of region $\mathcal{R}_r$. 
Additionally, the sum over $j$ of the interaction part is restricted to the dimers of region $\mathcal{R}_r$ which do not belong to any other region. The second part $\hat{H}^{(\mathcal{O}_{r,r'})}$ includes all kinetic processes taking place in the overlap $\mathcal{O}_{r,r'}$ of regions $\mathcal{R}_r$ and $\mathcal{R}_{r'}$. Similarly, this part contains the interactions $V_\delta \hat{n}_j \hat{n}_{j+\delta}$ where the index $j$ is a dimer belonging to the overlap of regions $\mathcal{R}_r$ and $\mathcal{R}_{r'}$. Let us remark that $N_j$ denotes the number of regions $\mathcal{R}_r$ containing dimer $j$ which ensures no double counting is done. Finally, the third part of the Hamiltonian $\hat{H}_{\rm kin}^{( r , r' ) }$ takes into account all kinetic hopping terms from region $\mathcal{R}_r$ to region $\mathcal{R}_{r'}$ such that initial and final dimers are not part of the overlap $\mathcal{O}_{r,r'}$ of different regions. 

\begin{figure}
   \begin{center}
\subfigure[$\,$Start: hop from $|0\rangle$ to $|1\rangle$]{
    \includegraphics[width=0.45\columnwidth]{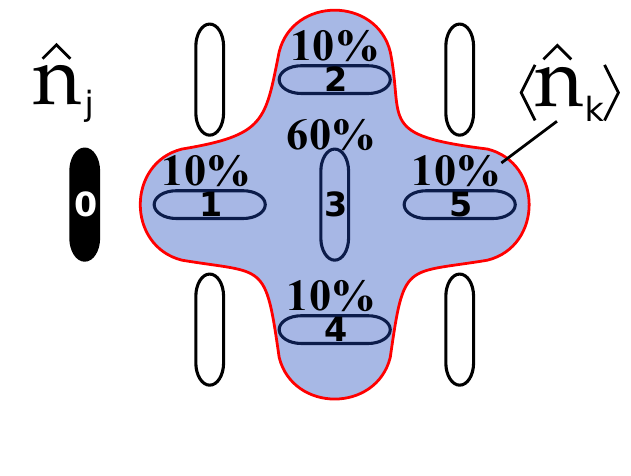}
\label{fig:renorm_A}
}
\subfigure[$\,$Hartree decoupling without renormalization]{
   \includegraphics[width=0.45\columnwidth]{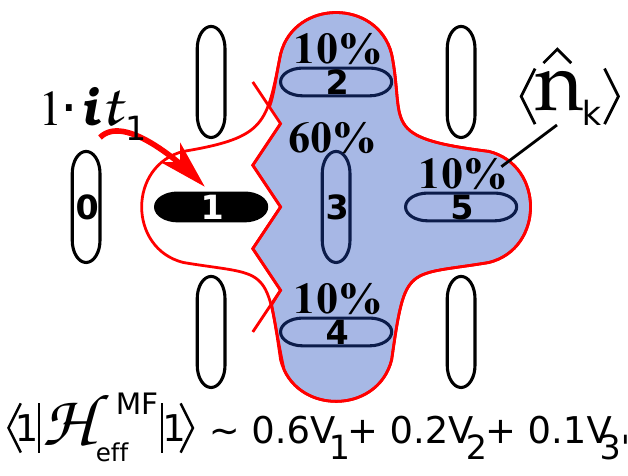}
\label{fig:renorm_B}
}\\
\subfigure[$\,$Representative renormalization scheme]{
    \includegraphics[width=0.5\columnwidth]{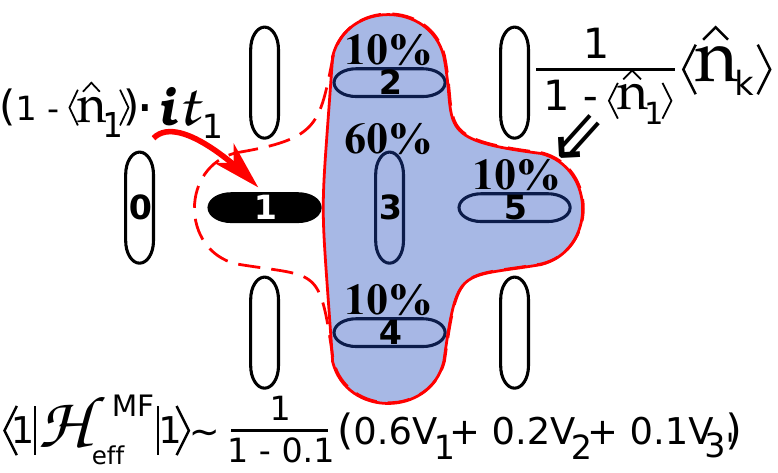}
\label{fig:renorm_C}
}
\subfigure[$\,$Explicit change due to renormalization]{
   \includegraphics[width=0.4\columnwidth]{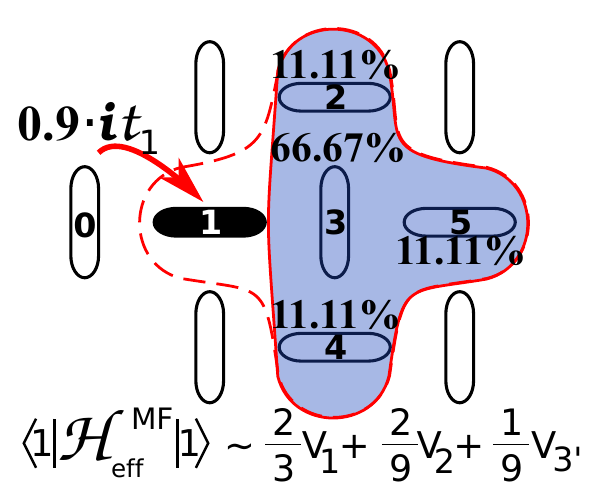}
\label{fig:renorm_D}
}
   \end{center}
   \caption{Qualitative illustration of the renormalization factors due to the overlap of different regions. 
(a): We consider the initial situation that a particle on dimer $0$ hops to the dimer $1$ which is part of the overlap with a second region. All dimers contained in this second region are shaded in dark grey (blue) and the number attached on top of these dimers correspond to the probablity (density) that the particle in this second region is on a specific dimer. 
(b): If all renormalization factors are absent, the particle which has moved to dimer $1$ is unaffected by the $10 \%$ probability that another particle of the second region is already present on dimer $1$. As a consequence, the particle does not pay the potential energy resulting from this $10 \%$ of the density distribution on dimer $1$. Because the hopping amplitude is $i t_1$ it would be attractive to hop on dimers with a large density of another particle in order to avoid high potential barriers. This is clearly unphysical.
(c)-(d): These two figures illustrate the same situtation but with the two renormalization factors for the kinetic part and for the interaction part of the mean-field Hamiltonian. The hopping amplitude $i t_1$ for the hopping from dimer $0$ to dimer $1$ is renormalized to $(1-\langle\hat{n}_1 \rangle )i t_1 $. At the same time, this leads to a change of the density distribution of the particle inside the second region. The density expectation value $\langle \hat{n}_k \rangle$ of the particle in the second region is effectively increased by $\langle \hat{n}_k \rangle/ (1-\langle\hat{n}_1 \rangle) $.
}
    \label{fig:renorm}
\end{figure} 

In the following we neglect the term $\hat{H}_{\rm kin}^{( r , r' )}$, because we want to study an effective one-body problem, i.e. we study the one-triplon quantum fluctuations of one region assuming that all other regions remain frozen. Physically, this is reasonable because the triplons are expected to only fluctuate strongly in the close vicinity of the classical positions of the Wigner crystals due to the presence of strong repulsive density-density interactions in the effective Hamiltonian. 

The next step is to decouple the two-body interaction in the Hartree channel
\begin{align} \label{eq:decouple}
 \hat{n}_j \hat{n}_k \approx 2 \hat{n}_j \langle \hat{n}_k \rangle - \langle \hat{n}_j \rangle \langle \hat{n}_k \rangle \, .
\end{align}
It is important to study how the hardcore constraint is violated in the effective mean-field description and how one possibly can correct such violations. Clearly, there is no conflict in applying Eq.~\eqref{eq:decouple} to the first term $\hat{H}^{(r)}$ because no overlap is involved. This is different for the second term $\hat{H}^{(\mathcal{O}_{r,r'})}$. Here it is in principle possible to violate the hardcore constraint by placing the particle of the region under consideration to a dimer where the probability distribution of another particle from a different region is close to one. As a consequence, two particles would be on top of each other which is energetically favored because the potential term $V_\delta$ vanishes for $\delta = 0$. This is clearly unphysical. Furthermore, it is also problematic that the probability of such a hopping process would be equal to one (see Figs.~\ref{fig:renorm}(a-b)).

We are aiming at partially repairing this violation of the hardcore constraint via two renormalization factors, one for the kinetic part and one for the interaction part. The hopping of a particle should depend on whether there already is a finite probability for the presence of another particle on the involved dimers. Consequently, we put the renormalization factor ($1-\langle \hat{n}_j \rangle) (1-\langle \hat{n}_{j+\alpha})$ for the hopping term $\hat{b}^{\dagger}_{j + \alpha} \hat{b}^{\phantom{\dagger}}_j$. The second renormalization factor $1/(1 - \langle \hat{n}_j \rangle)$ corrects the potential terms $V_\delta \hat{n}_j \langle \hat{n}_{j+\delta}\rangle $ where the dimer $j$ belongs to the overlap $\mathcal{O}_{r,r'}$. There are three reasons to add a factor $1/(1 - \langle \hat{n}_j \rangle)$ for the interaction part: i) If two particles are placed on the same dimer, the potential should be infinite due to the hardcore constraint. This is exactly the case for $\langle\hat{n}_j\rangle=1$. ii)  If two particles do not overlap, then the renormalization factor should be $1$ which is true for $\langle\hat{n}_j\rangle =0$. iii)  If the particle under consideration in region $\mathcal{R}_r$ is placed on a dimer $j$ being part of the overlap $\mathcal{O}_{r,r'}$ and if, simultaneously, the density of the particle in region $\mathcal{R}_{r'}$ is finite on this dimer, the effective potential in the mean-field description should increase on this dimer. This is again guaranteed by the renormalization factor (see Fig.~\ref{fig:renorm_C}-\ref{fig:renorm_D}). In the following we use $C_j=1-\langle\hat{n}_j\rangle$ in order to lighten the equations.\\
 
\begin{figure}
   \begin{center}
\subfigure[$\,$Density]{
    \includegraphics[width=0.45\columnwidth]{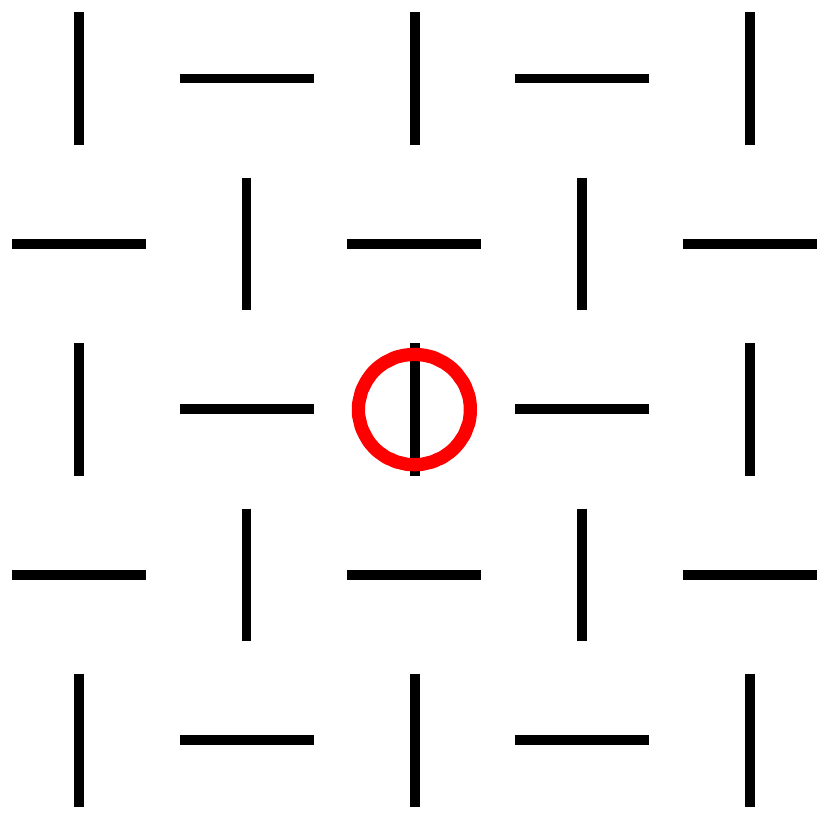}
}
\subfigure[$\,$Magnetization]{
    \includegraphics[width=0.45\columnwidth]{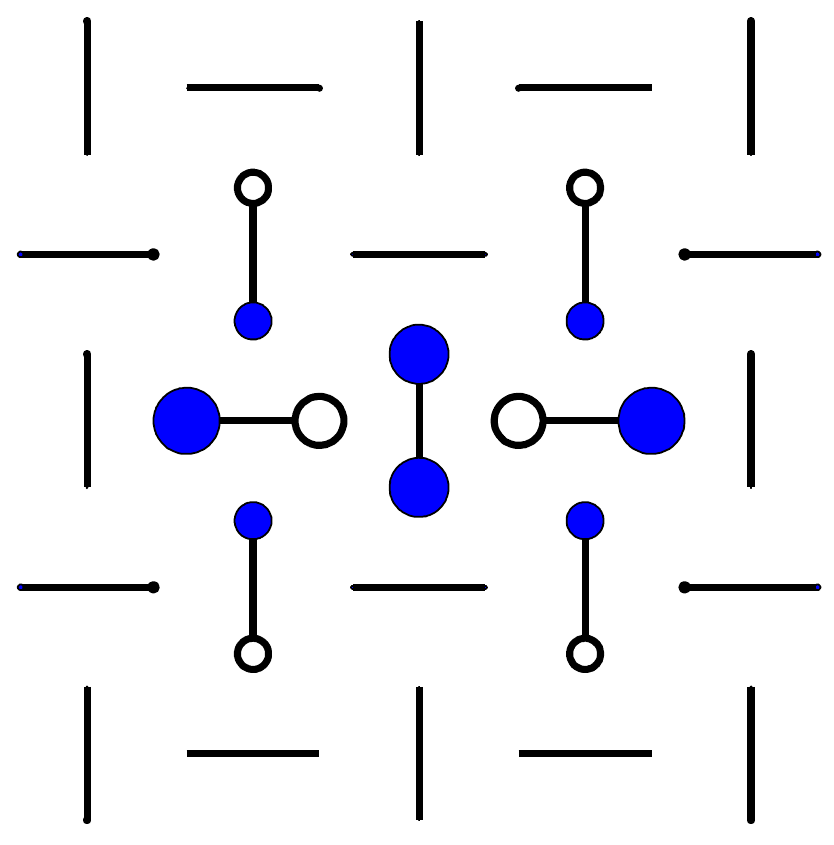}
}
   \end{center}
   \caption{Schematic illustration of the effective observables $\hat{S}_{1,j}^{{\rm eff},z}$ and $\hat{S}_{2,j}^{{\rm eff},z}$. Left figure represents a local density $\hat{n}$ from a (mean-field) state of the low-energy description. The right figure corresponds to the distribution of the local magnetization in the effective model originating from this finite density. The radius of the circles is proportional to the square root of the density (left) / local magnetization (right) on the dimer (left) / spin sites (right). In the right figure, filled (empty) circles denote a local magnetization pointing outside (inside) the plane.}
    \label{fig:magnetScheme}
\end{figure} 

As mentionned above, we study the quantum dynamics of one particle in a region $\mathcal{R}_{r_0}$ assuming that all other particles remain frozen, 
i.e.~these particles are not allowed to hop and we replace the density operators $\hat{n}$ by expectation values $\langle\hat{n}\rangle$ for these particles. Consequently, the many-body problem is replaced by an effective one-body problem in the finite Hilbert space $\mathcal{R}_{r_0}$. We therefore obtain the following mean-field Hamiltonian
\begin{align} \label{eq:H_MF_final}
 \frac{\hat{H}_{\rm eff}^{\rm (mf)}}{JN_r} &=  \hat{H}^{(r_0)}_{\rm mf} + \hat{H}^{(\mathcal{O}_{r_0,r'})}_{\rm mf} - (\mu -\mu_0) \sum_{j } \hat{n}_j 
\end{align}
where $N_r$ corresponds to the total number of regions and the sum in the last term is over all dimers $j$ contained in region $\mathcal{R}_{r_0}$. The explicit expressions for the first two terms are given by
\begin{align}
\nonumber \hat{H}^{(r_0)}_{\rm mf} &= \sum_{\delta} V_\delta \sum_{\substack{j \in \mathcal{R}_{r_0} \\ j \notin \mathcal{O}_{r_0,r'}}}   \left(-\frac{1}{2} \langle \hat{n}_j \rangle \langle \hat{n}_{j+\delta} \rangle + \hat{n}_j \langle \hat{n}_{j+\delta} \rangle \right)    \\ 
\nonumber &\quad+ i t_1 \sum_{\substack{\alpha = \pm x; \\ \alpha = \pm y;}} \sum_{\substack{j \in \mathcal{R}_{r_0} \\ j, j+\alpha \notin \mathcal{O}_{r_0,r'}}} (-1)^j \hat{b}^{\dagger}_{j + \alpha} \hat{b}^{\phantom{\dagger}}_j \\
 &\quad + t_2 \sum_{\alpha = \pm x \pm y} \sum_{\substack{j \in \mathcal{R}_{r_0} \\ j, j+\alpha \notin \mathcal{O}_{r_0,r'}}} \hat{b}^{\dagger}_{j + \alpha} \hat{b}^{\phantom{\dagger}}_j 
\end{align}
and
\begin{align}
 & \hat{H}^{(\mathcal{O}_{r_0,r'})}_{\rm mf} =\nonumber\\
          &\quad   \sum_{\delta} V_\delta \sum_{\substack{j \in \mathcal{O}_{r_0,r'}}} \frac{1}{C_j} \left(-\frac{1}{2} \langle \hat{n}_j \rangle \langle \hat{n}_{j+\delta} \rangle + \hat{n}_j \langle \hat{n}_{j+\delta} \rangle \right) \nonumber\\
          & \quad \nonumber  + i t_1 \sum_{\substack{\alpha = \pm x; \\ \alpha = \pm y;}}\, \sum_{\substack{j+\alpha \in \mathcal{R}_{r_0} \\ j \in \mathcal{O}_{r_0,r'}}} (-1)^j C_j C_{j+\alpha } \,  \hat{b}^{\dagger}_{j + \alpha} \hat{b}^{\phantom{\dagger}}_j \\
&\quad  + t_2 \sum_{\alpha = \pm x \pm y} \, \sum_{\substack{j+\alpha \in \mathcal{R}_{r_0} \\ j \in \mathcal{O}_{r_0,r'}}} C_j C_{j+\alpha } \,  \hat{b}^{\dagger}_{j + \alpha} \hat{b}^{\phantom{\dagger}}_j \quad .
\end{align}

One is therefore left with the following self-consistency equations for the mean-field parameters $\langle \hat{n}_{j}\rangle$ with\\ 
$j\in \mathcal{R}_{r_0}$
\begin{align} \label{MF_iteration}
\langle \phi^{(m)} | \hat{b}_{j}^{\dagger}\hat{b}^{\phantom{\dagger}}_{j} | \phi^{(m)} \rangle  &= \langle \hat{n}^{\phantom{\dagger}}_{j}\rangle^{(m)}  \\
\hat{H}_{\rm eff}^{\rm (mf)} \left(\langle \hat{n}_{j}\rangle^{(m)}\right) |\phi^{(m)} \rangle &= \epsilon^{(m)} |\phi^{(m+1)}\rangle \quad ,
\end{align}
which are solved iteratively. The index $m$ refers to the $m$-th iteration step. Let us stress that the one-body problem for the particle in a region is effectively taken place in the finite Hilbert space $\mathcal{R}_{r_0}$ because all other degrees of freedom are considered to be frozen. Consequently, the diagonal elements $\langle j| \hat{H}_{\rm eff}^{\rm (mf)} | j\rangle $ with $j \in \mathcal{R}_{r_0}$ of the resulting matrix only depend on the two-body interactions. Here $|j\rangle$ refers to the one-particle state where the particle in region $\mathcal{R}_{r_0}$ is located on dimer $j$ with $\langle \hat{n}_{j}\rangle = 1$ while all other particles remain in their (delocalized) state of the previous iteration. For the first iteration ($m=0$), we choose a purely classical state for the particles inside the different regions, i.e.~the density distribution of a particle in one region is $\langle \hat{n}_{j} \rangle = 1$ for one specific dimer $j$. The classical energy $E_{\rm cl}$ corresponds to the energetically lowest configuration of the potential terms $V_\delta$. Then the triplon sits on its classical position $j_{\rm cl}$ with $\langle \hat{n}_{j_{\rm cl}} \rangle = 1$. Thus one obtains the classical energy per dimer $E_{\rm cl}/N = (N_r/N)\langle j_{\rm cl}| \hat{H}_{\rm eff}^{\rm (mf)} | j_{\rm cl} \rangle$ where $N_r/N$ just represents the density of the system.

The non-vanishing off-diagonal matrix elements are either $\propto\pm i t_1$ or $\propto t_2$ depending on the dimers which are involved. Since all matrices are small, they can be solved readily 
by exact diagonalization. The resulting lowest eigenvalue $E^{(\rm mf)}/(JN)$ corresponds to the mean-field energy per dimer of the whole system. The associated eigenvector contains the information how all periodically equivalent triplons delocalize inside the Wigner crystal.

Up to now, we have considered an effective one-body problem where every periodically equivalent region displays the identical behaviour. We therefore need to calculate the mean-field parameters respectively the expectation values of the density operator only once per iteration. This is surely right for the $1/8$-diamond plateau (see Fig.~\ref{fig:classical_plateaux}). But if we consider other plateau structures having more than one triplon per unit cell, we have to take care that not every particle is related by symmetry (e.g. the structure $2/15$-rhomb shown in Fig.~\ref{fig:classical_plateaux}). We therefore have to iterate additionally over all different regions using the results of the last iteration for the particles of the other regions. 

We therefore obtain the plateau energies and the wave function out of the mean-field procedure. We calculate the energies and wave functions with ten different starting configurations. Eight of them are randomly initialized, one uses the classical solution as input, and one is initialized by setting particles in a diagonal fashion into the unit cell. The latter is motivated by the observation that diagonal configurations often converge rather rapidly to the global energy minimum of the mean-field equations. The resulting wave function can be used to calculate the delocalization of the triplons inside the Wigner crystals by taking the density expectation value of the mean-field ground-state wave function. Let us stress that this delocalization is a consequence of the quantum fluctuations induced by the kinetic processes of the effective Hamiltonian. Additionally, one has the quantum fluctuations arising from the fact that the effective hardcore boson model (except the contribution from the DM interaction) has been derived as a high-order series expansion from the Shastry-Sutherland model, i.e. observables like the density or the local magnetization has to be evaluated in the same basis \cite{knetter00,knetter03} as already done for the purely classical solution of the effective model \cite{dorier08}. Let us mention that the spin-density profile has been also calculate for various structures using exact diagonalizations \cite{miyahara03}. 

\begin{figure}
   \begin{center}
\subfigure[$\,$Energy gain $E_{\rm gain}/(J N)$ per dimer]{
    \includegraphics[width=0.87\columnwidth]{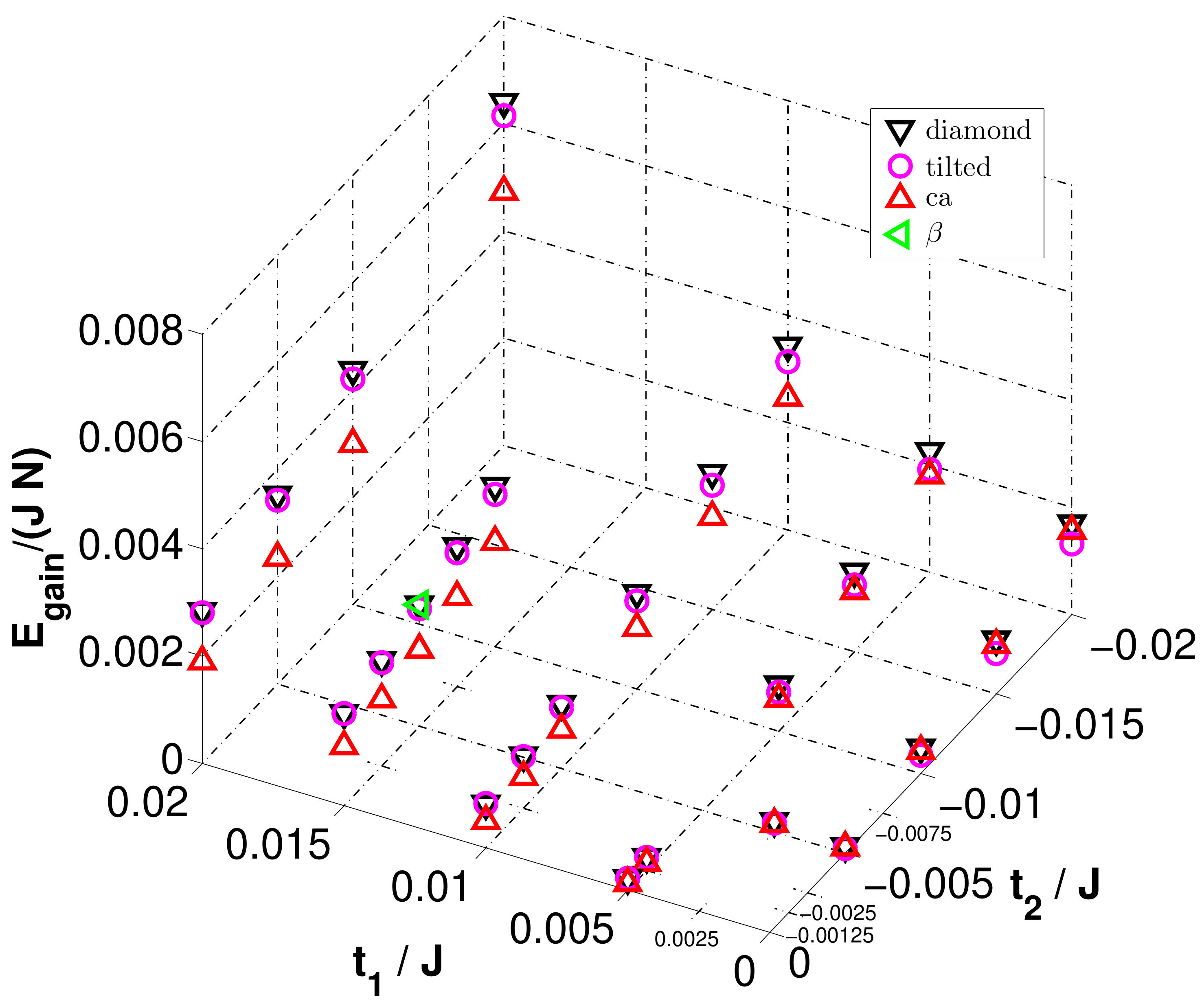}
\label{fig:energygain_J65_1_8}
}\\
\subfigure[$\,$Mean-field energies $E_{\rm mf}/(J N)$ per dimer]{
    \includegraphics[width=0.87\columnwidth]{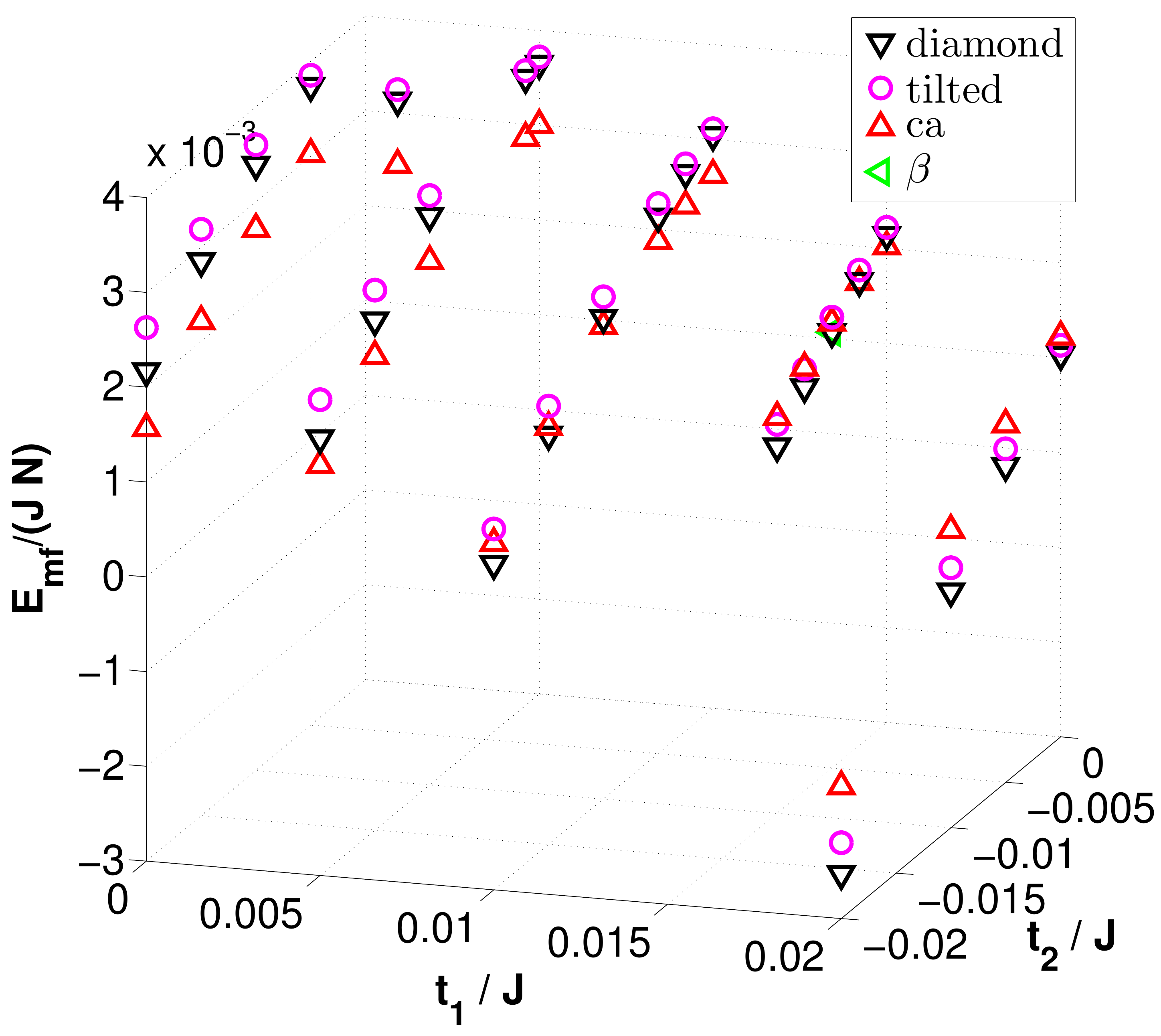}
\label{fig:energies_J65_1_8}
}
   \end{center}
   \caption{(a): The energy gain per dimer $E_{\rm gain}/(J N)$ is plotted as a function of $t_1/J$ and $t_2/J$ for density $1/8$ at $J'/J = 0.65$. The three considered structures are $1/8$-diamond (triangle down), $1/8$-tilted (circle), and $1/8$-ca (triangle up). The combination of both kinetics terms favors the structure $1/8$-diamond with a diamond unit cell. For other $J'/J$ values (not shown), the energy gain $E_{\rm gain}/(J N)$ has a very similar behaviour. 
(b): The mean-field energy per dimer $E_{\rm mf}/(J \, N)$ is plotted for $1/8$-diamond (triangle down), $1/8$-tilted (circle), and $1/8$-ca (triangle up) as a function of $t_1/J$ and $t_2/J$. Note that in the shown parameter regime always one of these three structures corresponds to the global minimum of the mean-field energy.}
    \label{fig:Gain+Energies_1_8_J65}
\end{figure} 

Here we are especially interested in the local magnetization on each spin $\hat{S}_{1,j}^{{\rm eff},z}$ and $\hat{S}_{2,j}^{{\rm eff},z}$ of the Shastry-Sutherland lattice, because this is the experimentally relevant quantity for NMR measurements. We therefore have calculated the following effective observables
\begin{align}
 \label{fig:obs}
 \hat{S}_{1,j}^{{\rm eff},z} &\simeq \sum_\alpha A_\alpha \hat{n}_{j+\alpha} \\
 \hat{S}_{2,j}^{{\rm eff},z} &\simeq \sum_\alpha B_\alpha \hat{n}_{j+\alpha} \, ,
\end{align}
as it has already been done in Ref.~\onlinecite{dorier08} for the pure Shastry-Sutherland model. The coefficients $A_\alpha$ and $B_\alpha$ are determined up to order 10 in $J'/J$ and we have extrapolated the series using Pad\'{e} extrapolation. The local magnetization is then obtained on each spin-site as illustrated in Fig.~\ref{fig:magnetScheme}. Physically, there are two mechanism leading to a delocalization of the triplons inside the magnetization plateaux. First, the quantum fluctuations of the Shastry-Sutherland model itself already present for the purely classical Wigner crystals \cite{dorier08} and, second, the quantum fluctuations induced by the kinetic processes of triplons about their classical positions inside the magnetization plateaux.

The just presented mean-field theory is expected to work well for plateau structures which are not too far from  
classical plateaux of frozen single triplons. But there are obvious limitations of our approach which are of two 
different types: $\alpha$) The kinetic terms are effectively too large compared to the involved repulsive 
interactions. Then a superfluid (supersolid) solution is expected and any crystalline solution is unphysical. Clearly, this
 has to happen in the dilute low-density limit. In our approach targetting Mott insulating phases, the involved
 repulsive interactions typically stabilizing the Wigner crystals are negligibly small because the distance between
 triplons becomes large at low densities. As a consequence, already small values of the kinetic terms are expected 
to melt any formation of a Wigner crystal. In the mean-field calculation, one then observes a negative mean-field 
energy $E_{\rm mf}<0$ at $\mu=\mu_0$ which we consider as an indicator for the presence 
of superfluid phases. The mean-field solutions having $E_{\rm mf}<0$ are unphysical and we will denote such solutions
 in the following to be of $\alpha$-type. $\beta$) Our mean-field treatment is expected to breakdown when the reduction 
 to an effective one-body problem fails. Later, we observe in certain situations that the mean-field solution 
corresponds to a Wigner crystal where triplons inside their region completely delocalize among various dimers. Such 
a behaviour is likely an artefact of our mean-field solution and it is therefore unphysical. In any case, one cannot
 expect that such strongly delocalized plateaux are well described on the mean-field level. We will denote such 
problematic solutions in the following to be of $\beta$-type. To be specific, we classify a mean-field solution as 
$\beta$ if the second-largest density of a triplon on a dimer inside a region exceeds the value $0.25$.       
   
\section{Results} 
\label{Sect:Results}

In this section we present the results obtained by the mean-field calculation. The numerical setup we use is to iterate over all unit cells up to 100 dimers getting 257 different plateau densities for $n \leq 1/6$. We focus on the range $0.6 \leq J'/J < 0.7$ with $0 \leq t_1/J \leq 0.02$ and $0 \leq -t_2 /J \leq 0.02$, which is expected to be the relevant regime for SrCu$_2$(BO$_3$)$_2$. 

In the following, we study first the favored plateau structures for a fixed density. This is done for $n=1/8$, $n=1/6$, $n=2/15$, and $n=1/9$. Afterwards we discuss the full low-density phase diagram of the extended Shastry-Sutherland model in Eq.~\ref{eq:Heff}. For all cases we give a detailed comparison to the physics of the frustrated quantum magnet SrCu$_2$(BO$_3$)$_2$. 

\begin{figure}
   \begin{center}
\subfigure[$\,1/8$-diamond]{
    \includegraphics[width=0.495\columnwidth]{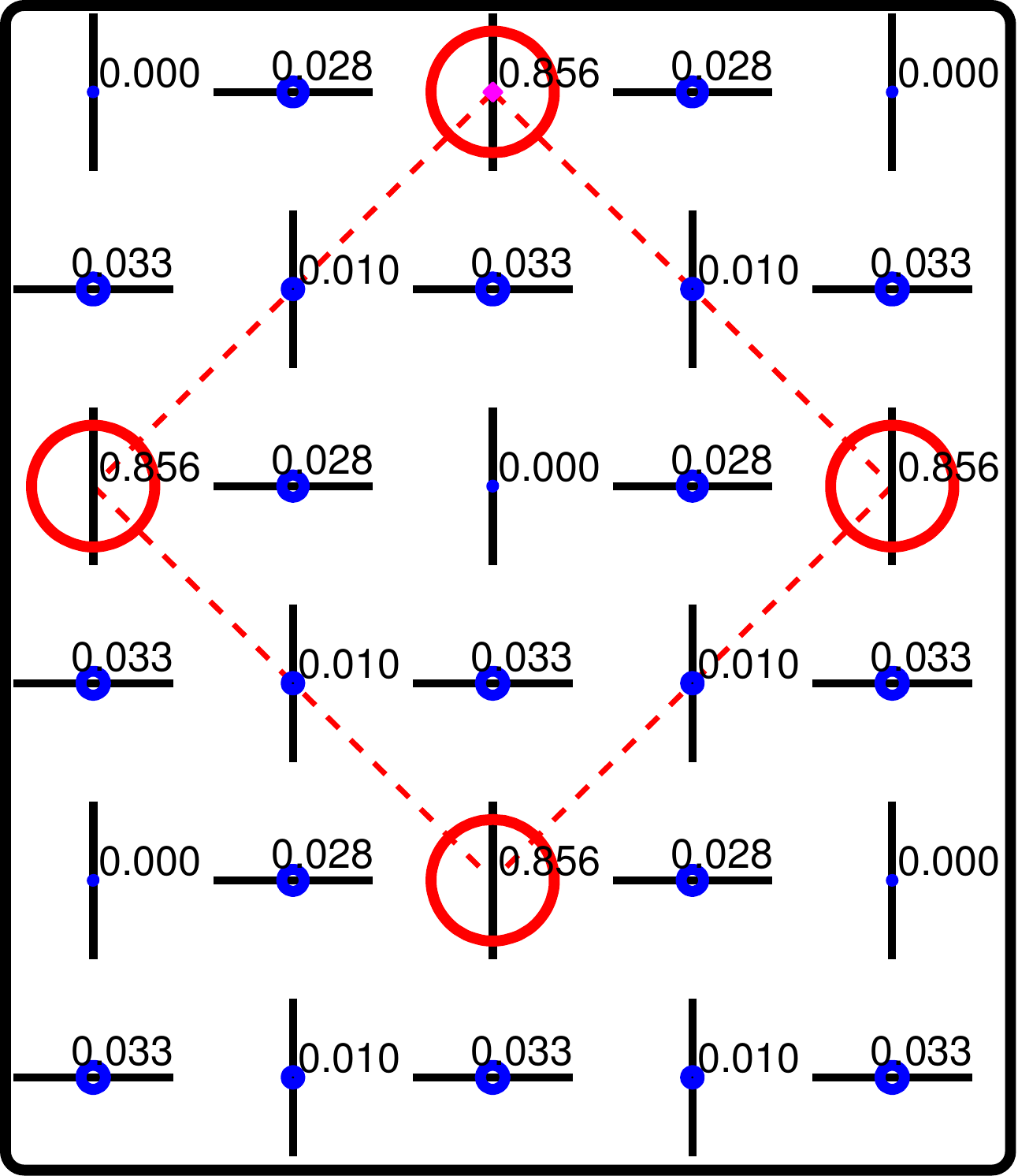}
\label{fig:eightFluct_A}
}
\subfigure[$\,1/8$-tilted]{
   \includegraphics[width=0.41\columnwidth]{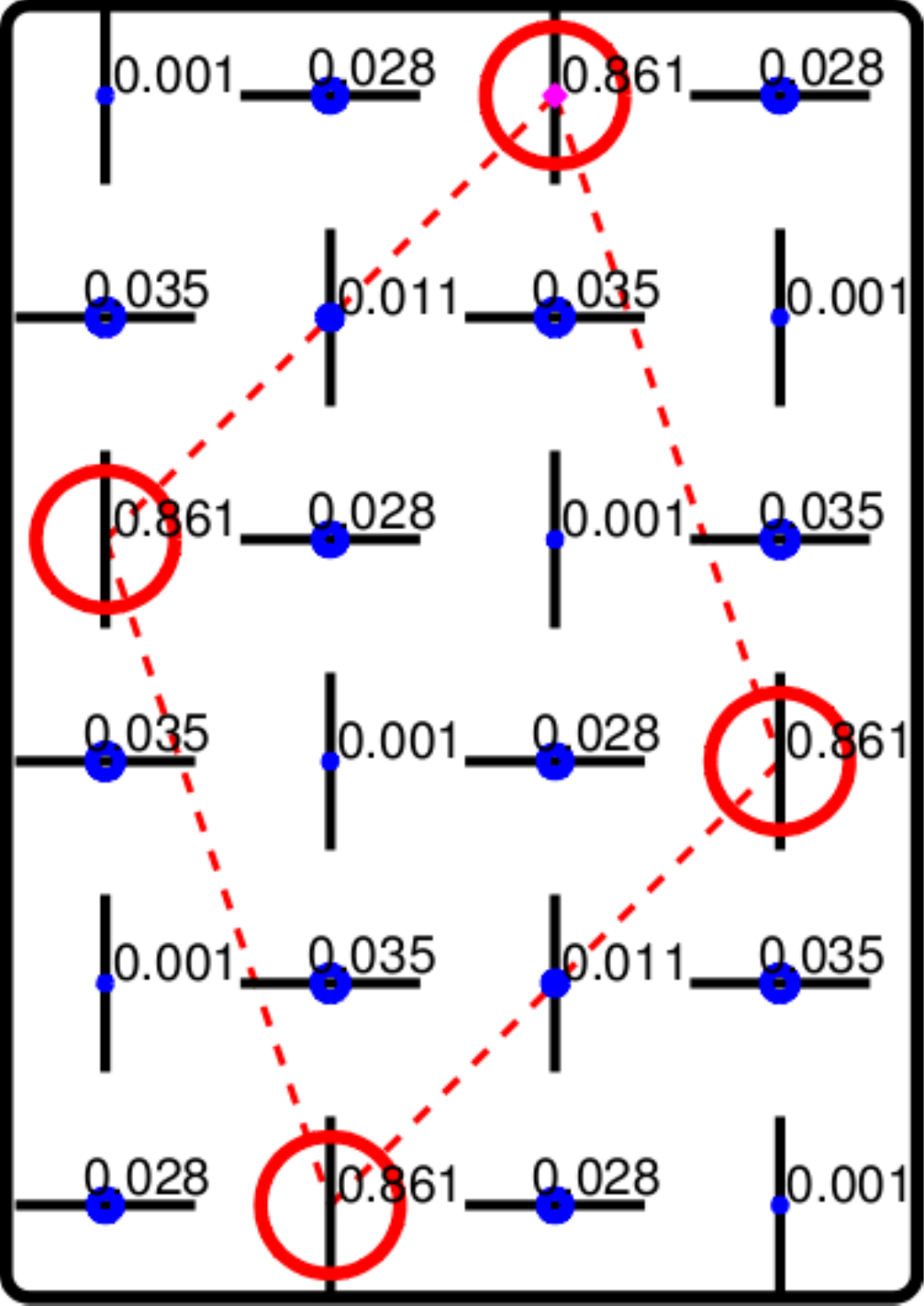}
\label{fig:eightFluct_B}
}\\
\subfigure[$\,$$1/8$-ca]{
    \includegraphics[width=0.9\columnwidth]{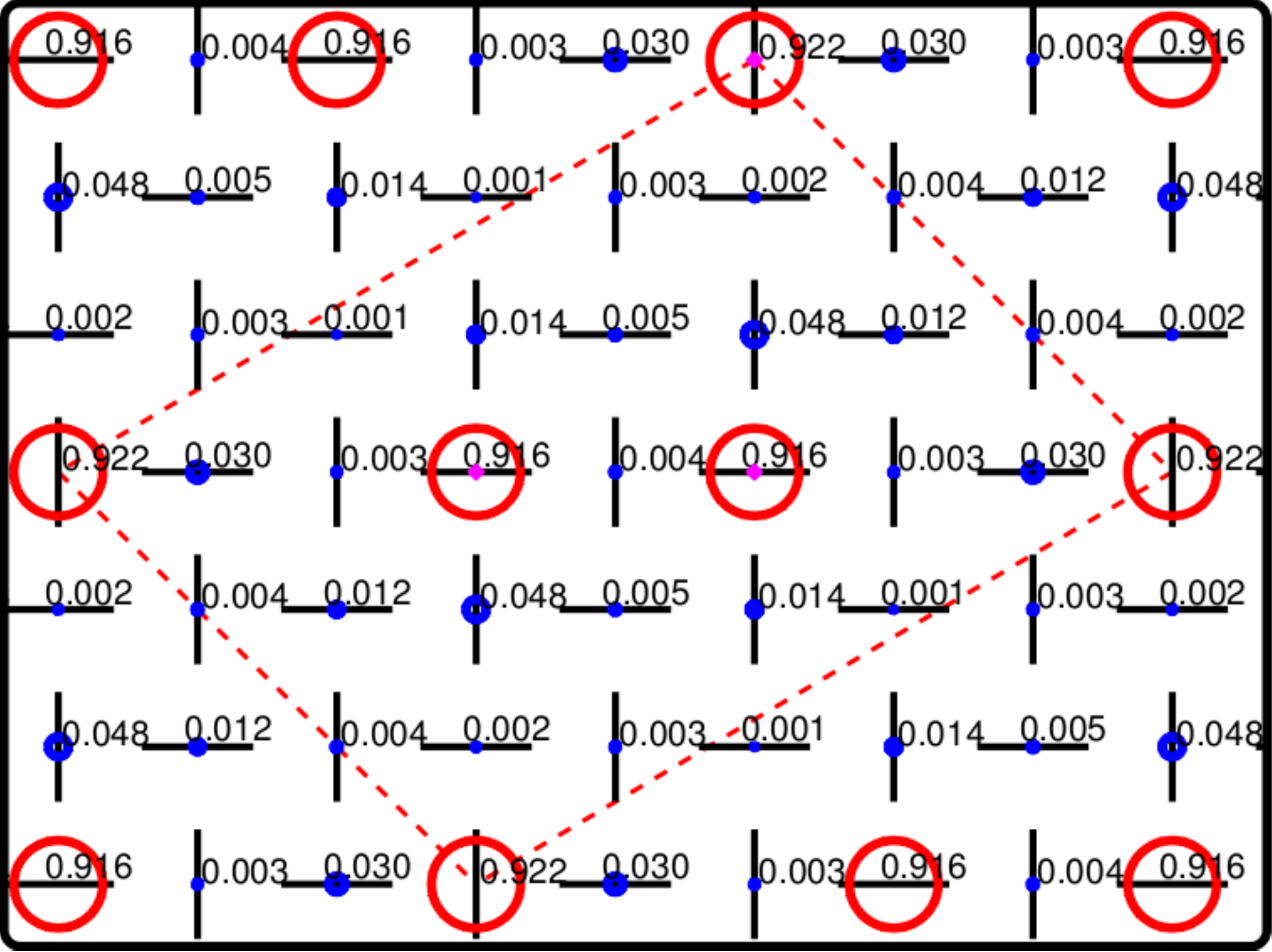}
\label{fig:eightFluct_C}
}
   \end{center}
   \caption{Density distribution of the mean-field solution for density $1/8$ at $J'/J$ = 0.65, $t_1/J$ = 0.01, and $t_2/J$ = -0.005. Note that the densities of all dimers are rounded to the third decimal digit. The suppression of certain fluctuation channels are clearly visible. The radius of the plotted circles is proportional to the square root of the density on each dimer. The red circles denote densities which are larger then 0.5.}
    \label{fig:eightFluct}
\end{figure} 

\subsection{$1/8$ plateau} 
\label{SSect:eight}

A triplon can well delocalize to neighboring dimers if the potential difference is small. The largest  
repulsive two-body interactions are $V_1$ and $V_3$. All dimer sites which involve these two interactions
represent therefore a sizable energy barrier and fluctuations to these dimers are well suppressed. 
To understand which kind of $1/8$ plateau is favored, let us start by investigating the different structures at fixed density $n=1/8$ in our numerical setup. To be concrete, we compare the structures $1/8$-diamond, $1/8$-tilted, and $1/8$-ca (see Fig.~\ref{fig:classical_plateaux}) as well as possibly other structures in case they have the lowest energy. 

\begin{figure}
   \begin{center}
\subfigure[$\,$$J'/J = 0.6$]{
    \includegraphics[width=0.61\columnwidth]{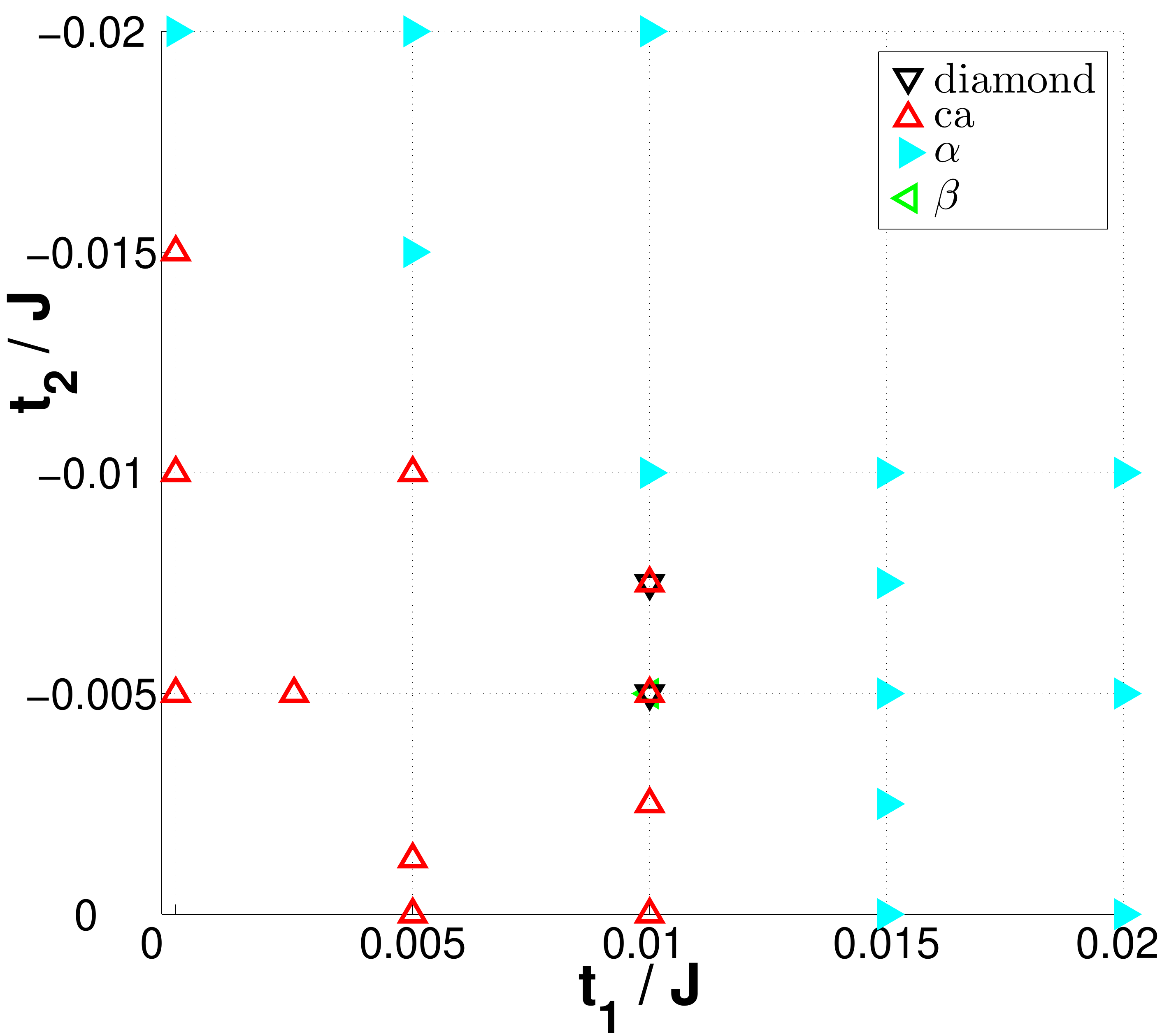}
\label{fig:EightLowArea_A}
}
\subfigure[$\,$$J'/J = 0.65$]{
   \includegraphics[width=0.61\columnwidth]{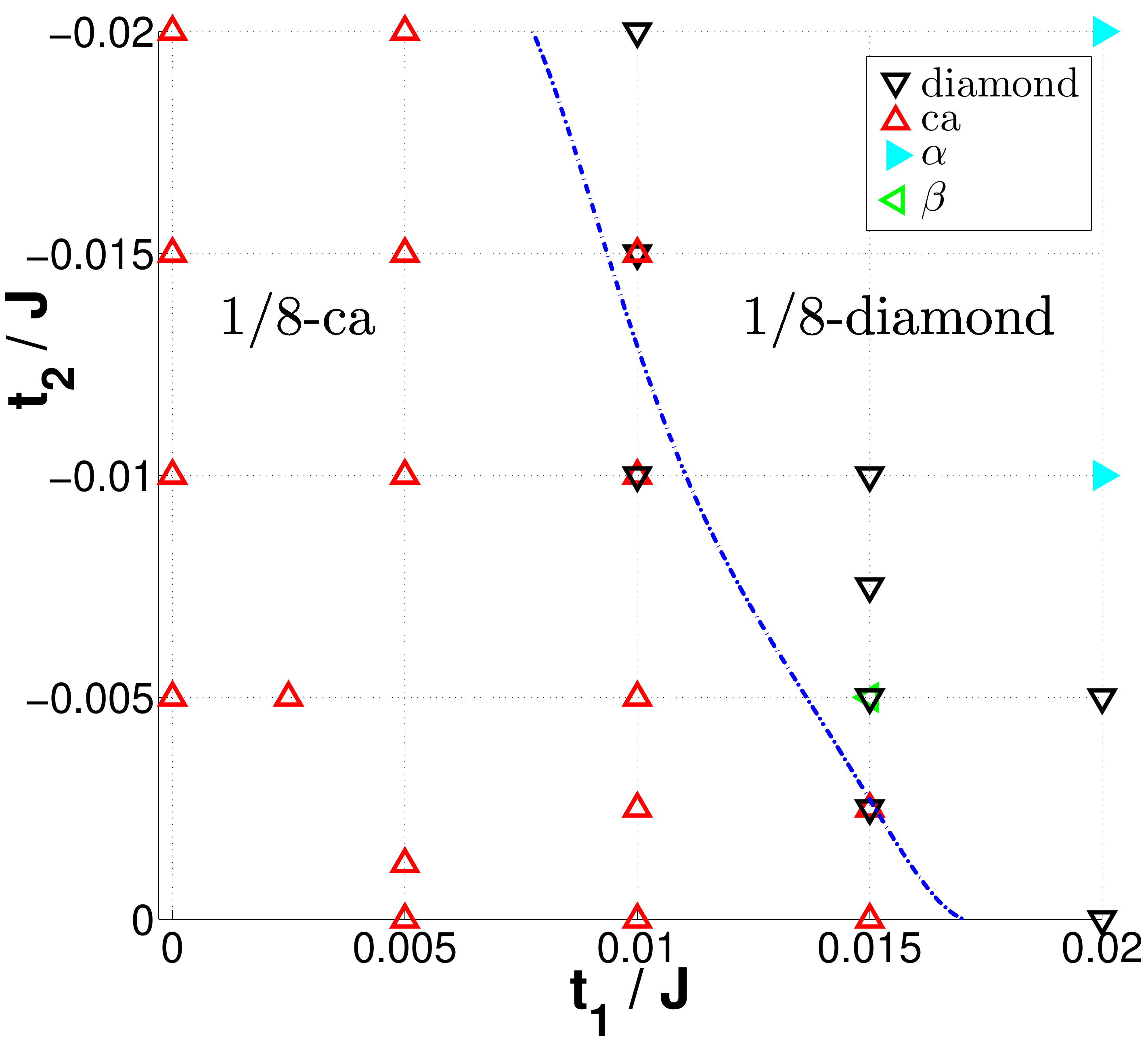}
\label{fig:EightLowArea_B}
}
\subfigure[$\,$$J'/J = 0.68$]{
    \includegraphics[width=0.61\columnwidth]{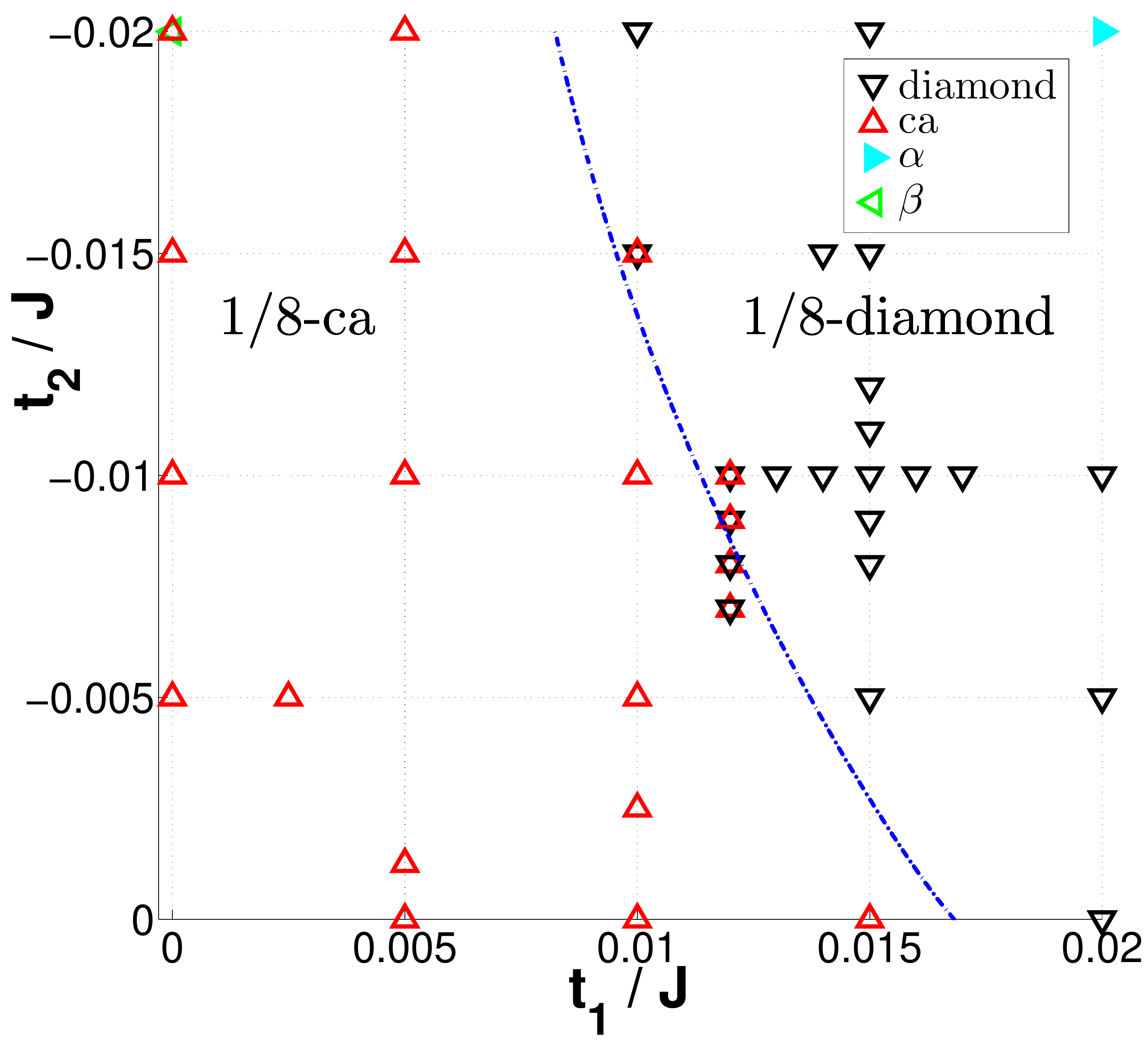}
\label{fig:EightLowArea_C}
}
   \end{center}
   \caption{Minima of the mean-field energy are  plotted as a function of $t_1/J$ and of $t_2/J$. The dashed blue line separates the parameter regime where the classical structure $1/8$-ca is favored from the regime where the $1/8$-diamond structure is realized. This line is obtained by fitting the lowest energies of $1/8$-diamond and $1/8$-ca along the displayed grid. Symbols $\alpha$ ($\beta$) represent mean-field solutions of $\alpha$-type ($\beta$-type). Mean-field solutions which differ by less than $10^{-4}$J are considered to be degenerate. In this case all such structures are displayed.}
    \label{fig:EightLowArea}
\end{figure} 

We consider first the energy gain $E_{\rm gain}$ for these three structures defined by the difference between the classical energy $E_{\rm cl}$ and the converged mean-field energy $E_{\rm mf}$ including quantum fluctuations
\begin{align}
 \frac{E_{\rm gain}}{J\, N} &= \frac{1}{J \, N} \left(E_{\rm cl} - E_{\rm mf}\right) \, .
\end{align} 
It is remarkable that the $1/8$-diamond structure has in most cases the highest energy gain (see Fig.~\ref{fig:energygain_J65_1_8}). Qualitatively, we find that increasing $t_1/J$ gives basically the same high energy gain for the $1/8$-diamond and for the $1/8$-tilted structure. In contrast, the energy gain for the classical plateau structure $1/8$-ca is rather small. Furthermore, increasing $t_2/J$ results in an energy splitting between the preferred plateau with a diamond unit cell $1/8$-diamond and the other two structures at density $n=1/8$.

These findings are well understood by looking at the potential differences between the classical position of the triplons inside the Wigner crystals and the dimers which can be reached via one hopping $t_1$ or one hopping $t_2$ (see Fig.~\ref{fig:eightFluct}). For a finite $t_1$ and $t_2=0$, only the two plateaux $1/8$-diamond and $1/8$-tilted are such that all four neighboring dimers do not involve the large repulsive interactions $V_1$ or $V_3$ (compare Figs.~\ref{fig:eightFluct}(a-b)). In contrast, the classical structure $1/8$-ca is stabilized by $V^\prime_{3}$ interactions, and therefore two out of the four nearest neighbors involve the largest repulsive interaction $V_1$ and quantum fluctuations to these dimers are suppressed. One then expects that a finite $t_1$ favors the $1/8$ structures with diamond and tilted unit cells in a similar fashion. This is different for the other limiting situation, i.e.~a finite $t_2$ and a vanishing $t_1$. Here it is only the $1/8$-diamond structure where all four (equal) fluctuation channels do not involve large interactions terms. In contrast, for the other two structures minimally one out of four fluctuation channels are suppressed due to the interaction $V_3$. This explains the energy splitting between the tilted and the diamond structure triggered by $t_2$ as discussed above (see Fig.~\ref{fig:energygain_J65_1_8}). 

Next we compare the ground-state energies $E_{\rm mf}$ of the different plateau structures at density $1/8$ which are shown for $J'/J=0.65$ in Fig.~\ref{fig:energies_J65_1_8}. Interestingly, it is in principle only the 1/8 structure with diamond unit cell which is stabilized besides the already expected classically realized 1/8-ca structure. The unphysical structure denoted by $\beta$ for $J'/J = 0.68$ is of $\beta$-type, i.e.~triplons are delocalized completely on two different dimers. Furthermore, it is the hopping to nearest-neighbors $t_1$ originating from the DM interaction which is the driving force for the 1/8-diamond structure found within the phenomenological theory for SrCu$_2$(BO$_3$)$_2$ \cite{takigawa12}. Indeed, we checked explicitly that the next-nearest neighbor hopping $t_2$ alone is not sufficient to stabilize the structure with a diamond unit cell. These trends are present for all ratios of $J'/J$ as long as $J'/J$ is large enough (see Fig.~\ref{fig:EightLowArea}). Indeed, even for a sizable $J'/J=0.6$ only a tiny region of 1/8-diamond is found. Instead, we observe that in most cases mean-field solutions of $\alpha$-type have the lowest mean-field energy.    

As stated above, solutions of $\alpha$-type indicate a melting of the Wigner crystals which is expected once the kinetic terms are large enough to win over the potential terms. In our mean-field treatment focusing on crystalline phases, an indication for such a melting transition to superfluid phases is a negative mean-field energy $E_{\rm mf}$  at $\mu=\mu_0$ as can be seen for example in Fig.~\ref{fig:energies_J65_1_8} for $t_1/J=0.02$ and $t_2/J=-0.02$. The melting transitions of the different Wigner crystals have to occur for smaller values of $t_{1}$ and $t_2$ when the ratio of $J'/J$ is reduced, because then also the repulsive interactions stabilizing the magnetization plateaux are smaller. This is exactly what we observe in the behaviour of the mean-field energy which becomes negative at $\mu=\mu_0$ for smaller values of the kinetic terms if $J'/J$ is reduced (see Fig.~\ref{fig:EightLowArea}). The presence of a 1/8 plateaux with diamond unit cell in the frustrated quantum magnet SrCu$_2$(BO$_3$)$_2$ points therefore to a ratio $J'/J\approx 0.65$ very close to the phase transition point in the Shastry-Sutherland model. The latter finding is further confirmed in the next subsection when studying the magnetization plateau at fixed density $1/6$.

\begin{figure}
   \begin{center}
   \includegraphics[width=0.95\columnwidth]{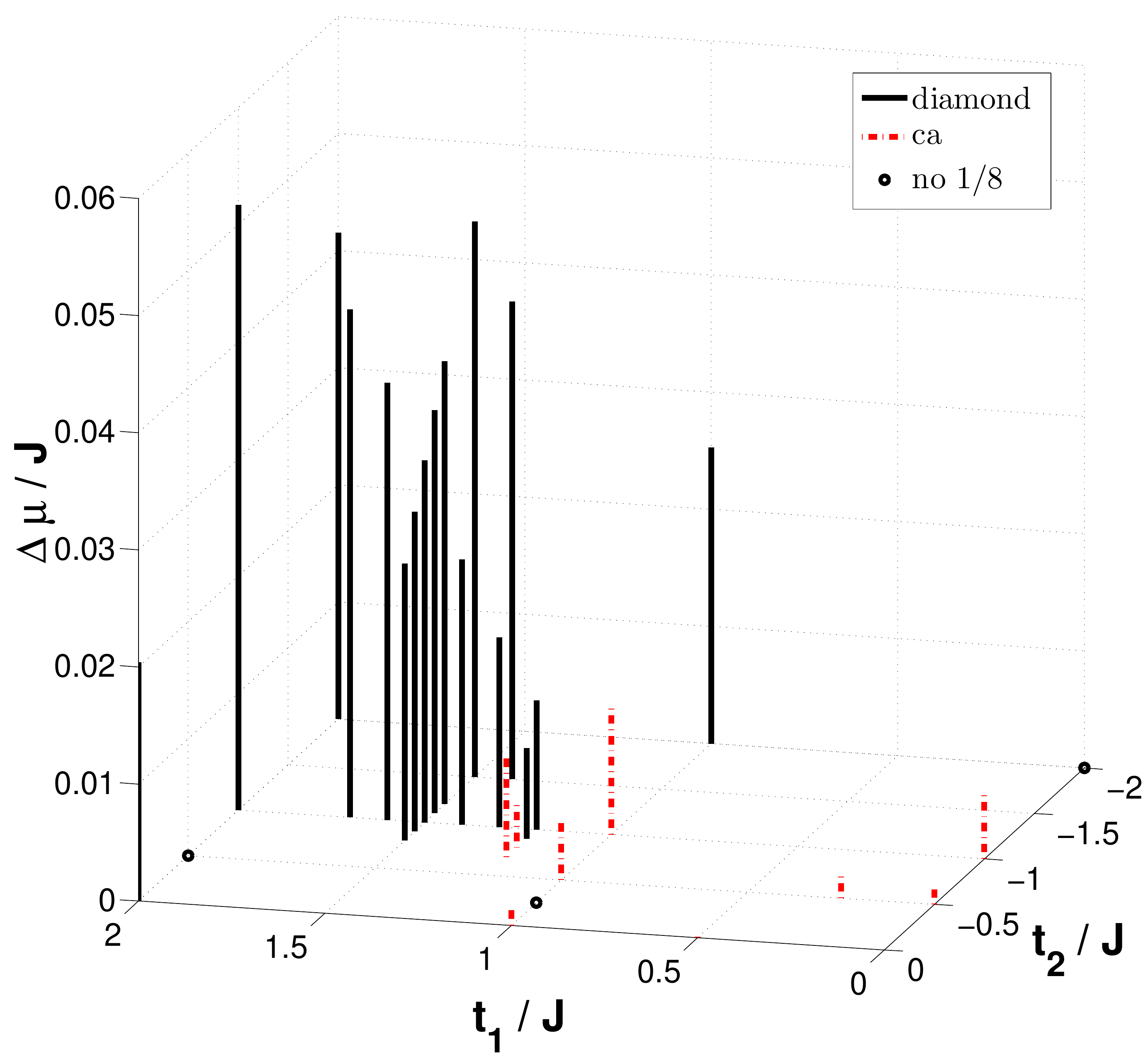}
   \end{center}
   \caption{The figure displays for $J'/J = 0.68$ the range in the chemical potential $\Delta\mu/J$ where a $1/8$ plateau is realized in the full low-density phase diagram compared to all other plateau structures at different densities.
 The stabilized $1/8$-plateau changes its structure from the classical $1/8$-ca structure (dashed lines, present for small values of $t_1/J$ and $-t_2/J$) to the $1/8$diamond structure (shown as solid black line) when both hoppings $t_1/J$ and $-t_2/J$ are of the order $0.01$.} 
    \label{fig:muRange_1_8_J68}
\end{figure} 

Finally, we want to see whether the above discussed 1/8 structures (especially the one with a diamond unit cell) are indeed realized in the phase diagram. To do so we compare all different plateaux at different densities as a function of the chemical potential $\mu/J$ and we plot in Fig.~\ref{fig:muRange_1_8_J68} the $\mu$-range 
\begin{align}
\frac{\Delta\mu}{J} = \frac{\mu_{\rm end} - \mu_{\rm start}}{J}
\end{align}
 where a $1/8$-plateau is realized as a function of $t_1/J$ and $t_2/J$ for $J'/J=0.68$. Here $\mu_{\rm end}$ ($\mu_{\rm start}$) denotes the end (beginning) of the plateau at density $1/8$ in the low-density phase diagram. The quantity $\Delta\mu/J$ corresponds therefore to the width of the plateau. Let us mention again that the $1/8$ plateau is not favored in the classical limit, i.e. for $t_1=t_2=0$. It is therefore remarkable that already for small kinetic hopping terms a $1/8$ plateau with the $1/8$-ca structure is present in the mean-field phase diagram. But one should keep in mind that the transition point between the plateau at $1/9$ and at $2/15$ in the classical limit is highly degenerate, i.e. many structures including the $1/8$-ca have the same classical enegy for this specific chemical potential. Thus small kinetic terms are able to lift this degeneracy and to favor the $1/8$-ca plateau in a tiny $\mu$-range depending on the size of the kinetic terms.
 
More importantly, we find that the $1/8$ plateau with diamond unit cell is present in the phase diagram for a wide range of hopping parameters including the expected values for SrCu$_2$(BO$_3$)$_2$: $t_1/J \approx 0.015i$ and $t_2/J \approx -0.01$. It is indeed the combination of both kinetic terms which is responsible for the stabilization of the $1/8$-diamond plateau. 

Altogether, our mean-field results for density $1/8$ are in very good agreement with the experimental findings and the phenomenological theory for SrCu$_2$(BO$_3$)$_2$ \cite{takigawa12}. Quantum fluctuations induced by the kinetic hopping terms $t_1$ and $t_2$ are essential to obtain a sizable 1/8 plateau with a diamond unit cell. Furthermore, the coupling ratio $J'/J$ must be rather large $J'/J\approx 0.65$ in order to prevent a melting of the Wigner crystal due to the kinetic processes. 

\subsection{$1/6$ plateau}
\label{SSect:six}

The other prominent low-density plateau observed experimentally in the compound SrCu$_2$(BO$_3$)$_2$ is at density $1/6$ which we focus on in this subsection. Interestingly, the phenomenological interpretation of the NMR data on SrCu$_2$(BO$_3$)$_2$ \cite{takigawa12} yields a structure for the $1/6$ plateau which is in disagreement with the plateau $1/6$-ca found in the classical limit for $J'/J=0.5$ \cite{dorier08}. The latter structure is stabilized by $V'_3$ interactions. This has to be contrasted with the phenomenologically deduced $1/6$ plateau which is built out of $V_4$ and $V_5$. As stated above, we find exactly such structures ($1/6$-stripe, $1/6$-square, and $1/6$-new) naturally already in the classical limit for ratios $J'/J \geq 0.67$. So again, our results indicate a rather large ratio $J'/J$ in SrCu$_2$(BO$_3$)$_2$. 

The three plateau structures $1/6$-stripe, $1/6$-square, and $1/6$-new have exactly the same classical energy. In the following we want to study how this degeneracy is lifted in our mean-field theory if the kinetic processes $t_1$ and $t_2$ are turned on. To this end we discuss first the dominant fluctuation channels for these plateaux (plus the structure $1/6$-ca) in an analogue fashion as we did for the plateaux at density $1/8$. One therefore has to check for each  particle on each structure whether the potential barrier on the eight dimers which can be reached from the classical position by one $t_1$- or one $t_2$-hopping is large or not.

\begin{figure}
   \begin{center}
   \includegraphics[width=0.95\columnwidth]{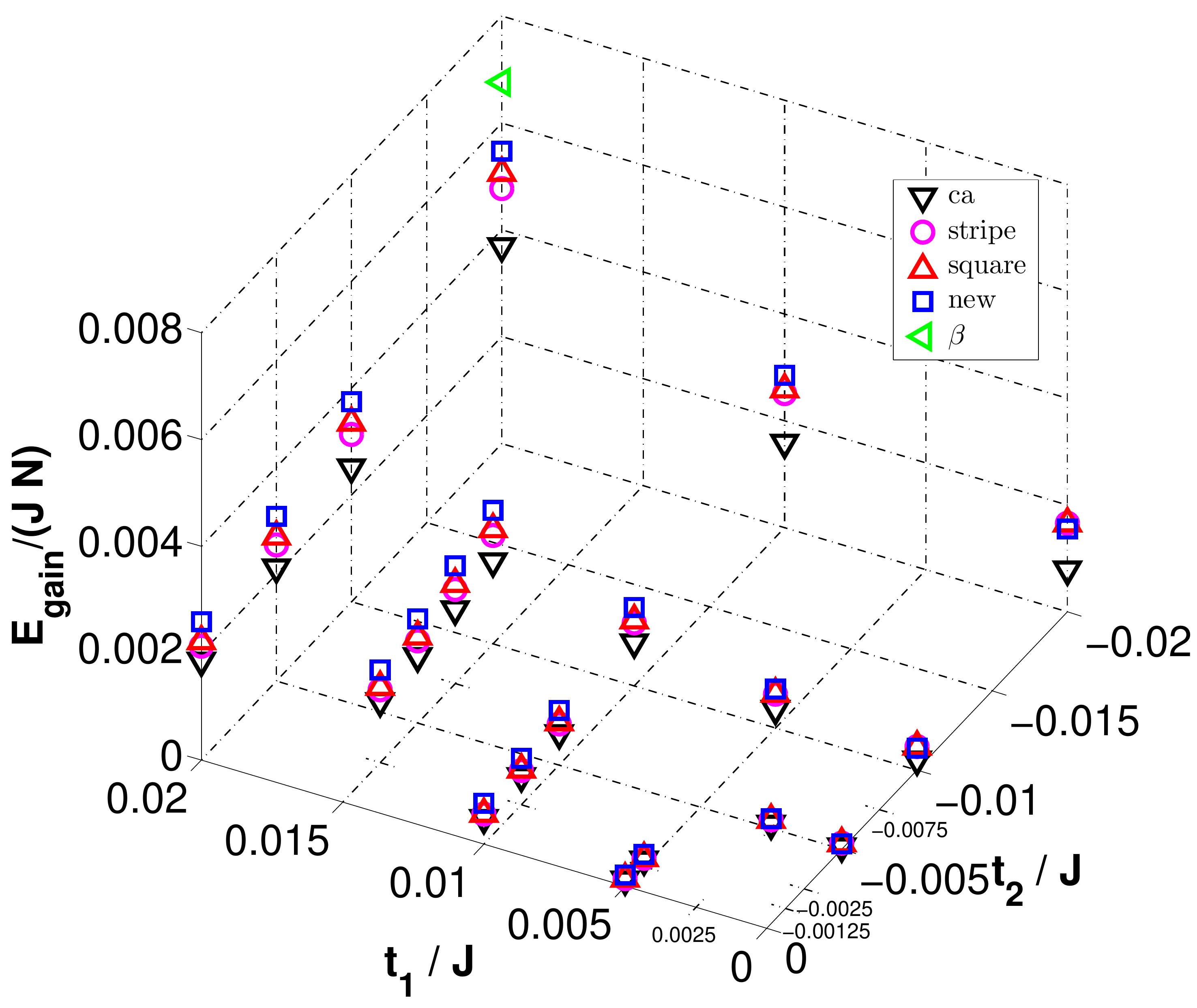}
   \end{center}
   \caption{The energy gain per dimer $E_{\rm gain}/(J N)$ is plotted as a function of $t_1/J$ and $t_2/J$ for fixed density $n=1/6$ at $J'/J = 0.65$. In contrast to the structure $1/6$-new (squares), the classical structure $1/6$-ca (triangles down) has a lower energy gain for all considered parameter values. The structure $1/6$-square ($1/6$-stripe) is denoted by triangles up (circles). If the kinetic terms become large $t_1/J \approx 0.02$ and $t_2/J \approx -0.02$, an unphysical plateau of $\beta$-type has the lowest energy.}
    \label{fig:energygain_J65_1_6}
\end{figure}

In the $1/6$-ca structure all particles have the same environment. For each particle, there are only two dimers which can be reached via one $t_1$-hopping from the classical position that do not involve the large $V_1$ or $V_3$ interactions. All other six fluctuation channels are suppressed. This is different for the other structures. The two structures $1/6$-stripe and $1/6$-square stay degenerate when the quantum fluctuations of the particles are restricted to the eight dimers around the classical position. For both structures one finds that four out of the eight fluctuation channels do not involve $V_1$ or $V_3$. To be concrete, for each particle inside the Wigner crystal there are two out of four dimers (for $t_1$ and $t_2$) where the largest potential term to be paid is $V_2$. Finally, the $1/6$-new plateau has very similar quantum fluctuations as the latter two structures. The key difference is the existence of one additional fluctuation channel for the triplons having their classical location on the vertical dimers inside the Wigner crystal (see Fig.~\ref{fig:classical_plateaux}). This additional channel is reached via one $t_1$ hopping to a dimer where one has to pay only $V_4$ interactions. The structure $1/6$-new is therefore expected to gain most from quantum fluctuations. This is exactly what we get (see Fig.~\ref{fig:energygain_J65_1_6}). We therefore find that the $1/6$-new plateau is favored for most combinations of kinetic hopping terms $t_1$ and $t_2$ for $J'/J=0.68$. This is only different for small values of $t_1$ where the structures $1/6$-square and $1/6$-stripe have a slightly smaller energy (see Fig.~\ref{fig:energygain_J65_1_6}).

\begin{figure}
   \begin{center}
\subfigure[$\,$$J'/J = 0.65$]{
    \includegraphics[width=0.61\columnwidth]{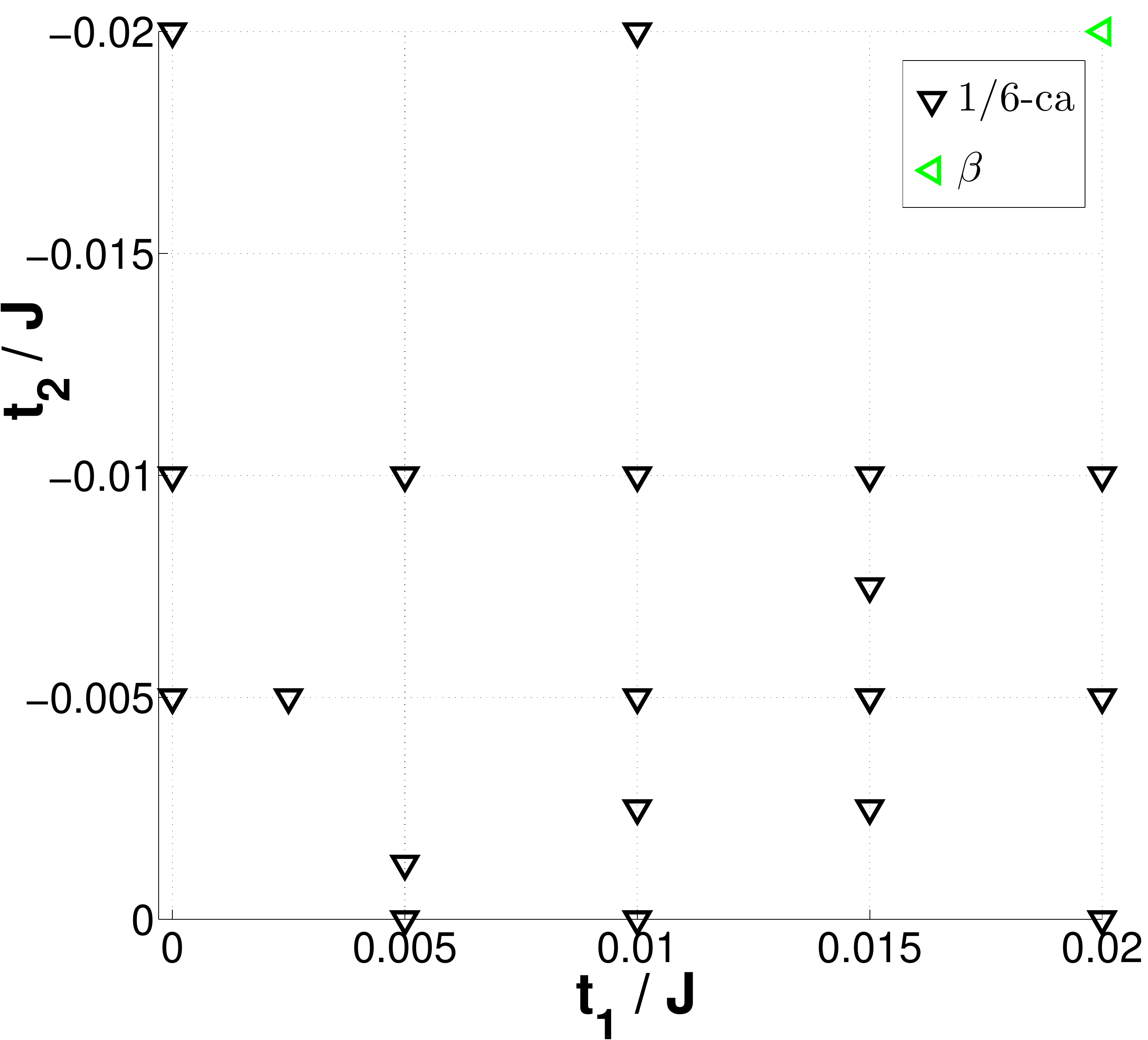}
\label{fig:SixLowArea_A}
}
\\
\subfigure[$\,$$J'/J = 0.68$]{
   \includegraphics[width=0.61\columnwidth]{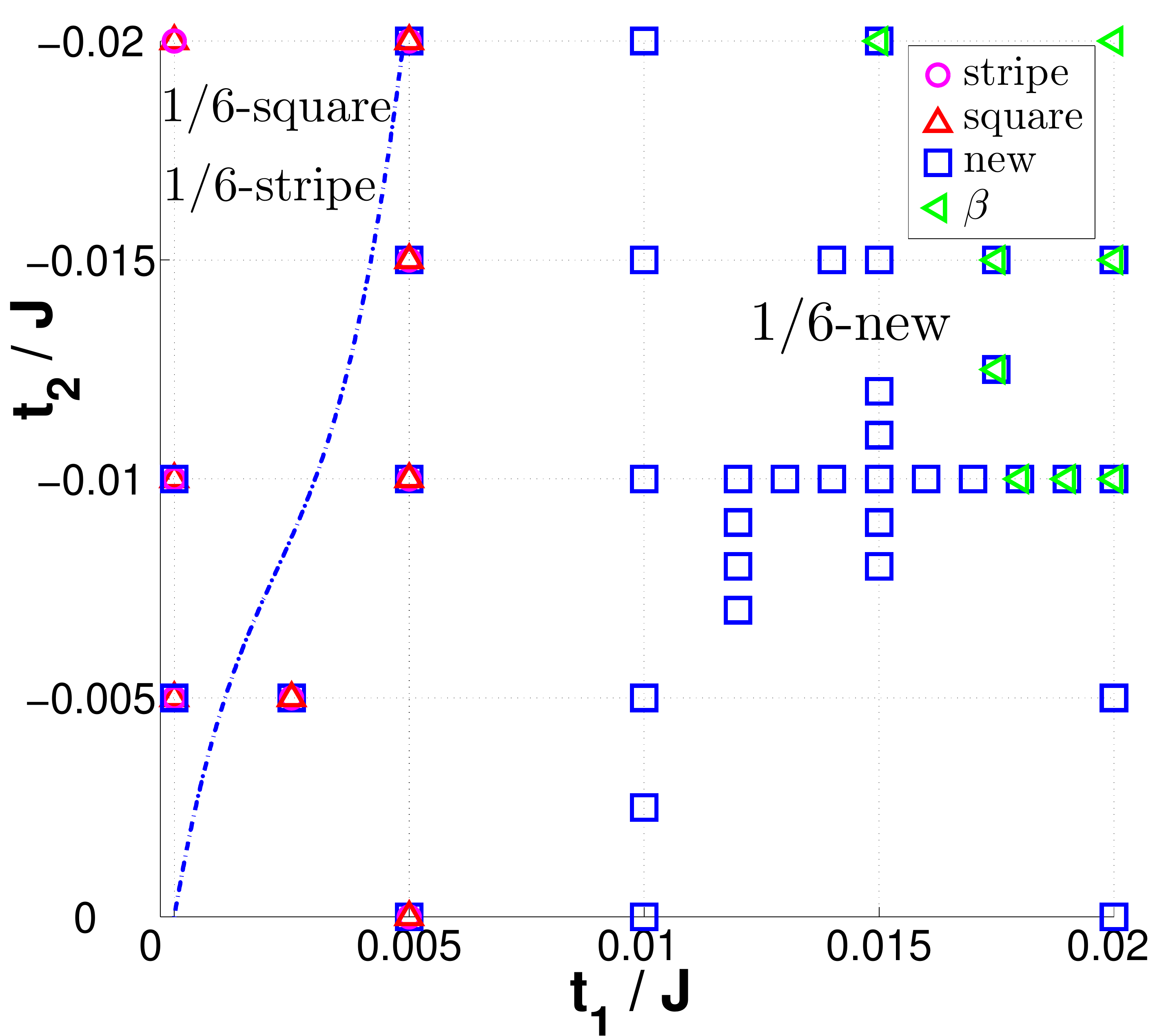}
\label{fig:SixLowArea_B}
}
   \end{center}
   \caption{The structures at fixed density $n=1/6$ having the lowest mean-field energy $E_{\rm mf}$ are plotted as a function of $t_1/J$ and $t_2/J$ for (a) $J'/J$= 0.65 and for (b) $J'/J$ = 0.68. The most important difference between $J'/J$ = 0.65 and $J'/J$ = 0.68 is the change from the $1/6$-ca structure to the $1/6$-new structure. Symbols $\beta$ represent mean-field solutions of $\beta$-type. Mean-field solutions which differ by less than $10^{-4}$J are considered to be degenerate. In this case all such structures are displayed.}
    \label{fig:SixLowArea}
\end{figure} 

Let us stress that for too large values of the kinetic hopping terms a different Wigner crystal has the lowest mean-field energy. As already mentionned above, this novel structure is likely an unphysical artefact of our mean-field treatment. To be specific, we find a mean-field solution where the triplons are not anymore well localized at their classical position. In contrast, the particles get completely delocalized among different dimers in their regions which is not the behaviour where our mean-field treatment is expected to work well. Such mean-field solutions are of $\beta$-type as introduced in Sect.~\ref{Sect:MF}.     

Let us compare our mean-field results with the findings deduced from the NMR data of SrCu$_2$(BO$_3$)$_2$ \cite{takigawa12}. The latter militates in favor of the $1/6$-stripe structure. Interestingly, we find that this plateau is indeed among the ones with the lowest energy as long as the ratio $J'/J$ is sufficiently large. In this respect our results are in better agreement with the experimental data as the purely classical results for $J'/J=0.5$ \cite{dorier08}. But the structure $1/6$-new has a slightly lower energy which one can understand due to the larger number of fluctuation channels as discussed above. This mismatch between microscopic and phenomenological theory originates most likely from subleading effects not contained in our calculation. Indeed, the experimental data do not find evidences for rather larger unit cells like for example the structure $1/6$-new. One possible explanation of this discrepancy is the relevance of three-dimensional couplings or the renormalization of the two-body interactions $V_\delta$ due to the additional couplings. 
 
Nevertheless, our mean-field results as discussed so far for densities $1/8$ and $1/6$ are quite promising. The findings for $1/8$ are in full agreement with experiments and its phenomenological interpretation. Additionally, the analysis for both densities point to a large ratio $J'/J \geq 0.67$ and one clearly sees the relevance of the quantum fluctuations we have introduced in our microscopic calculation.
               
\subsection{$2/15$ plateau} 
\label{SSect:stripeMech}
The NMR data on SrCu$_2$(BO$_3$)$_2$ reveals the existence of a third low-density plateau at density $2/15$ as predicted theoretically for $J'/J=0.5$ by the classical solution of the effective hardcore boson model \cite{dorier08}. But as for the plateau at $1/6$, the classical structures $2/15$-rhomb and $2/15$-rect are inconsistent with the experimental data. Interestingly, the phenomenological theory point to the stripe structure $2/15$-big \cite{takigawa12}, i.e.~this Wigner crystal is built by a pattern of three stripes of the structure $1/8$-tilted and one stripe of the structure $1/6$-stripe. Let us mention that the phenomenological interpretation of the NMR data for the $2/15$ plateau is very complicated, because the experimental signal is very complex due to the large unit cell.

In the following we compare the results of our microscopic theory with the above predictions. Most importantly, our mean-field theory naturally gives a stripe structure for the density $2/15$. The microscopic reason is rather simple. As we have seen in the last sections, the $1/8$ structure with a diamond unit cell is one of the very rare Wigner crystals where all eight fluctuation channels to dimers reached by one $t_1$ or one $t_2$ hopping do not involve the large repulsive interactions $V_1$ and $V_3$. As a consequence, it seems preferable to construct stripe structures above $1/8$ which contain the $1/8$-diamond Wigner crystal as a substructure. One relevant example for such a stripe structure with density $2/15$ is the structure $2/15$-b$_2$ illustrated in Fig.~\ref{fig:classical_plateaux}. One can see that the $2/15$-b$_2$ plateau is formed by a subpart of three $1/8$-diamond stripes (gray shaded) and one $1/6$ subpart which is built by $V_4$ and $V_5$ potential terms. It is clear that such stripe structures will mainly benefit from the $1/8$ component since the fluctuation channels for the thin $1/6$ stripe are mainly suppressed.

\begin{figure}
   \begin{center}
\subfigure[$\,$$J'/J = 0.65$]{
    \includegraphics[width=0.61\columnwidth]{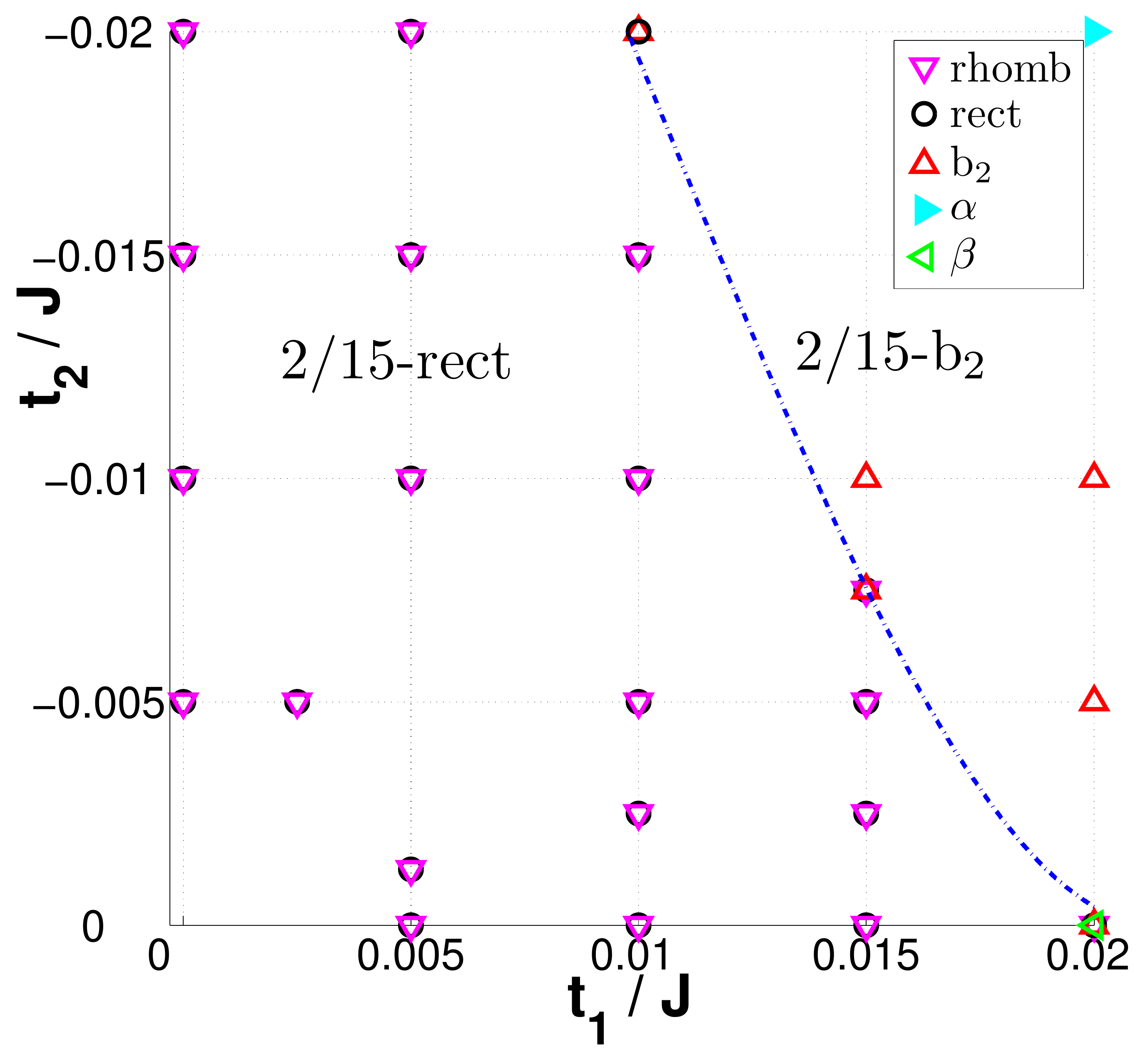}
\label{fig:FiveLowArea_A}
}
\\
\subfigure[$\,$$J'/J = 0.68$]{
   \includegraphics[width=0.61\columnwidth]{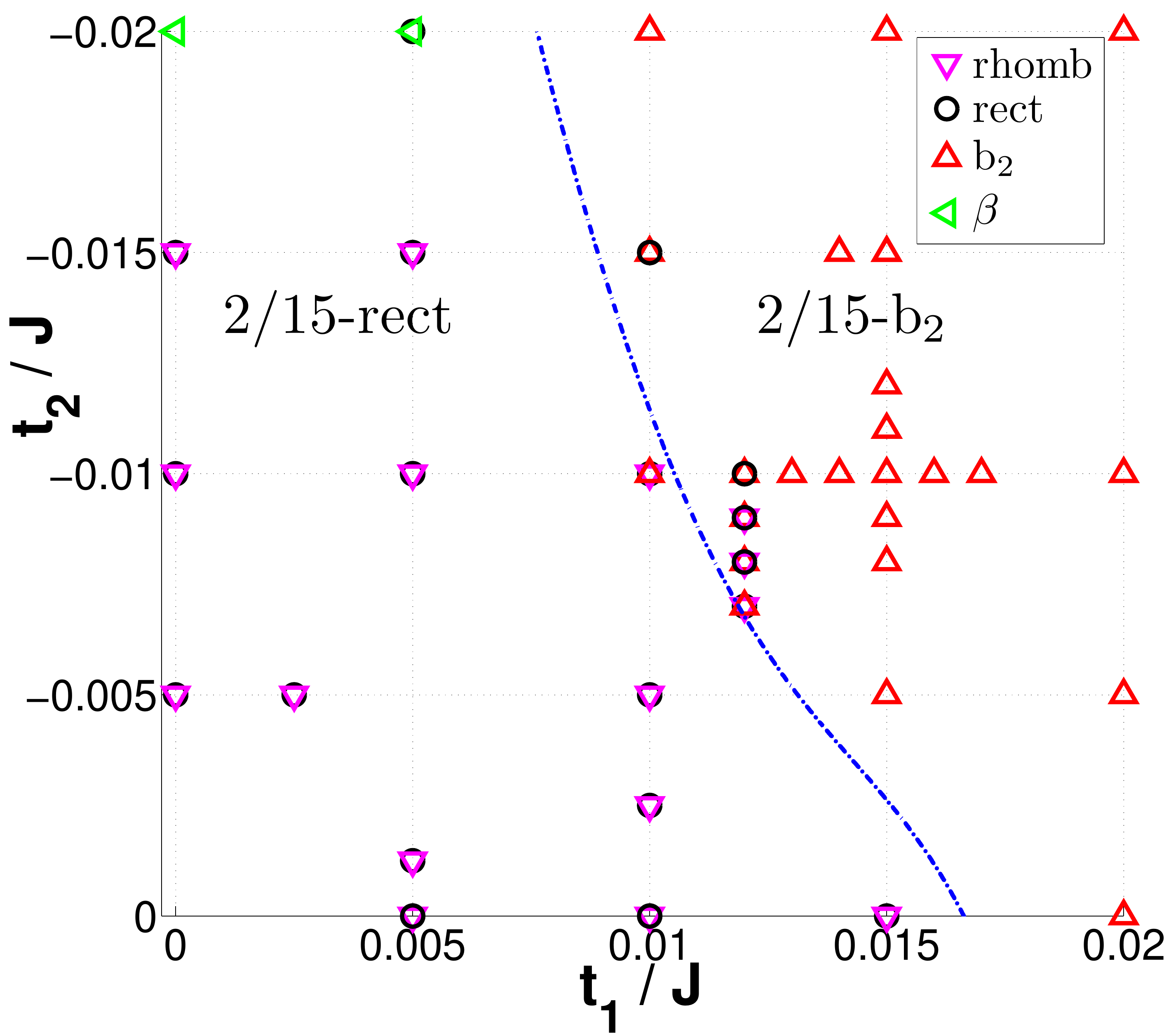}
\label{fig:FiveLowArea_B}
}
   \end{center}
   \caption{The structures at fixed density $n=2/15$ having the lowest mean-field energy $E_{\rm mf}$ are plotted as a function of $t_1/J$ and $t_2/J$ for (a) $J'/J$= 0.65 and for (b) $J'/J$ = 0.68. The dashed blue line separates the region favoring the structure $2/15$-rect present for small values of the kinetic terms from the regime where the structure $2/15$-b$_2$ is realized. This line is obtained by fitting the lowest energies of $2/15$-rect and $2/15$-b$_2$ along the displayed grid. Symbols $\alpha$ ($\beta$) represent mean-field solutions of $\alpha$-type ($\beta$-type). Mean-field solutions which differ by less than $10^{-4}$J are considered to be degenerate. In this case all such structures are displayed.}
    \label{fig:FiveLowArea}
\end{figure}

After this discussion, it is obvious that it is less advantegeous to form stripes by taking $1/8$-tilted substructures instead of $1/8$-diamond for all combinations of $J'/J$, $t_1/J$, and $t_2/J$. The $2/15$ striped crystal formed by $1/8$-diamond substructures has always a lower mean-field energy than the one consisting of tilted substructures. We therefore find that the striped $2/15$ structure containing $1/8$-diamond substructures are favored nearly in the same region as the $1/8$-diamond crystal itself (see Figs.~\ref{fig:EightLowArea} and \ref{fig:FiveLowArea}).

\subsection{1/9 plateau}

Before we present the resulting phase diagram for realistic values of the kinetic hopping terms, let us first discuss our results for fixed density $n=1/9$. Neither Ref.~\onlinecite{kodama02} nor Ref.~\onlinecite{takigawa12} find experimental indications for a plateau at density $1/9$. In contrast, Sebastian et al have interpreted their high-field torque measurements as evidence for a plateau at $1/9$ \cite{sebastian07} which is clearly in conflict with the other measurements.

On the theoretical side, the CA \cite{dorier08} finds at $J'/J$=0.5 a stable $1/9$ plateau and not the one with density $1/8$. Furthermore, we find a $1/9$ plateau with this classical structure (see Fig.~\ref{fig:classical_plateaux}) for a wide range of kinetic couplings for all investigated $J'/J$. One nevertheless expects that this $1/9$ plateau should melt already for rather small values of the induced quantum fluctuations triggered by the kinetic hopping processes due to the fact that the $1/9$ plateau is stabilized by the weak repulsive interaction $V_6$ appearing only in order 8 perturbation theory. 

\begin{figure}
   \begin{center}
\subfigure[$\,$$J'/J = 0.65$]{
    \includegraphics[width=0.61\columnwidth]{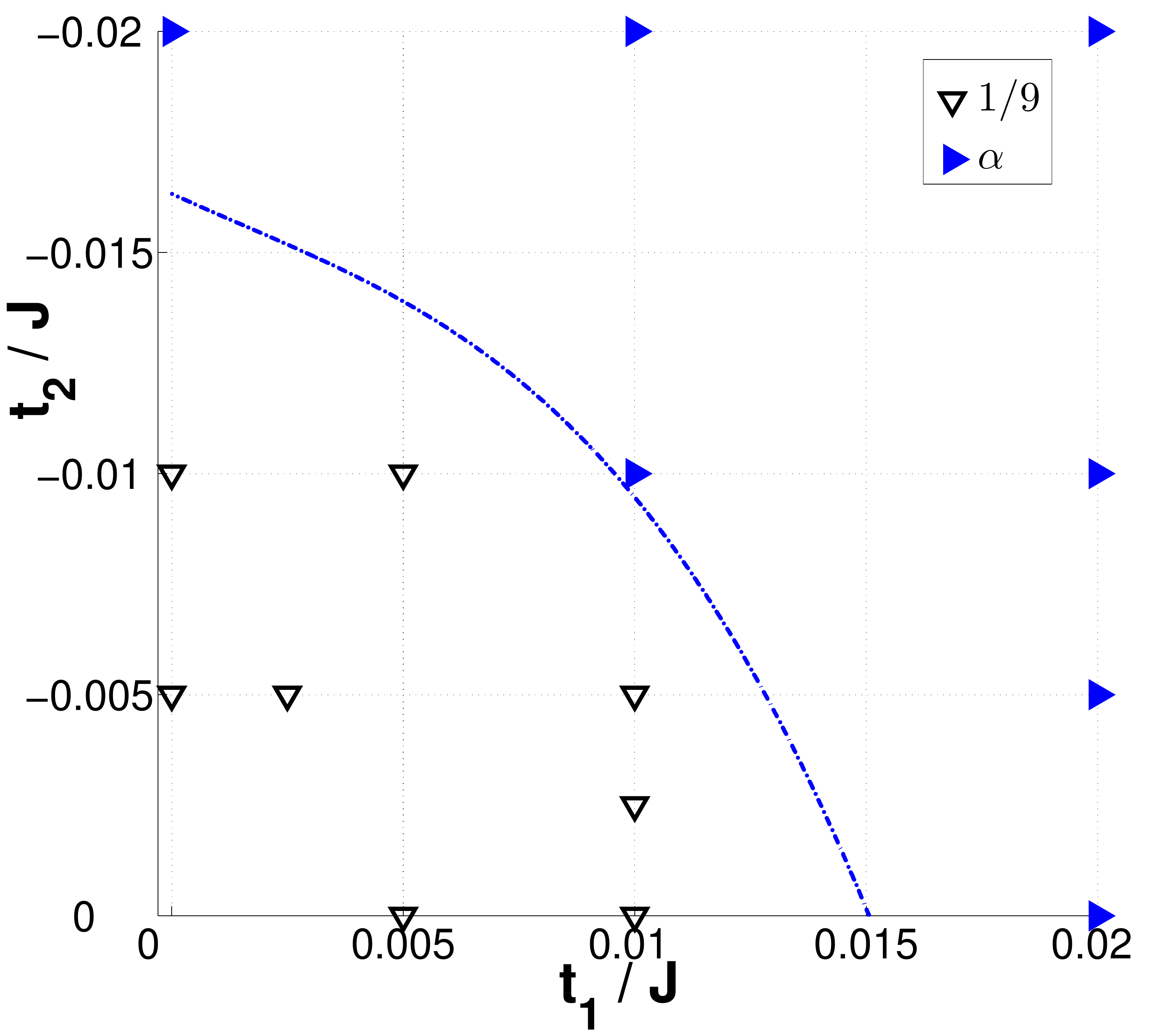}
\label{fig:NineLowArea_A}
}
\\
\subfigure[$\,$$J'/J = 0.68$]{
   \includegraphics[width=0.61\columnwidth]{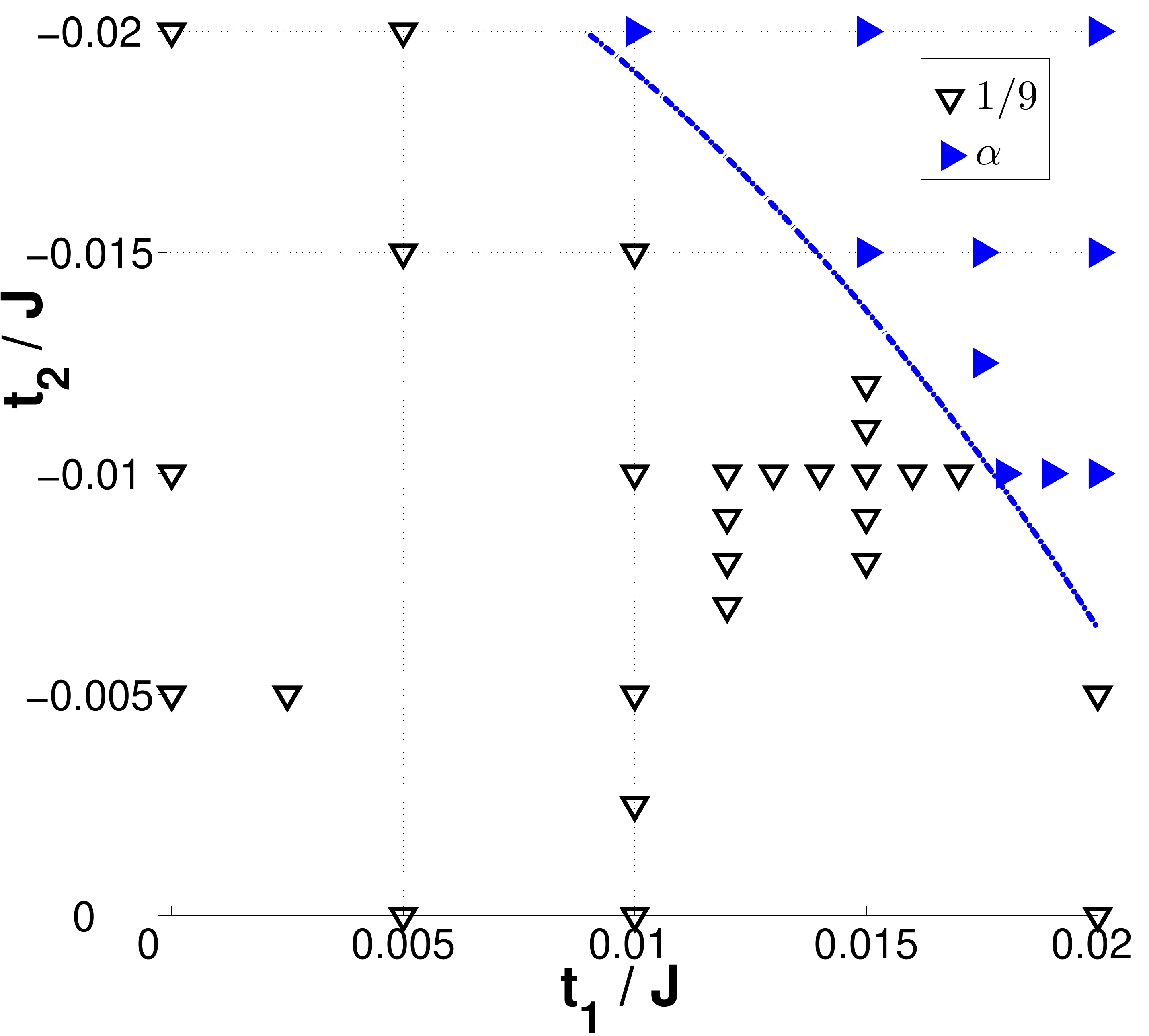}
\label{fig:NineLowArea_B}
}
   \end{center}
   \caption{The structures at fixed density $n=1/9$ having the lowest mean-field energy $E_{\rm mf}$ are plotted as a function of $t_1/J$ and $t_2/J$ for (a) $J'/J$= 0.65 and for (b) $J'/J$ = 0.68. The dashed blue line separates the parameter regime where the $1/9$-plateau has a positive mean-field energy at $\mu=\mu_0$ from the parameter region where the mean-field energy becomes negative at $\mu=\mu_0$ which is clearly unphysical. This unphysical behaviour is denoted by $\alpha$.}
    \label{fig:NineLowArea}
\end{figure}

An upper bound for this melting transition can be obtained in our mean-field calculation by determining the sign of the mean-field energy at $\mu=\mu_0$. This corresponds to mean-field solutions of $\alpha$-type which is shown in Fig.~\ref{fig:NineLowArea} for $J/J=0.65$ and $J'/J=0.68$. One clearly sees that the $1/9$ plateau is about to melt for realistic values of the kinetic hopping processes $t_1$ and $t_2$. It is therefore likely that the $1/9$ plateau is not a stable Wigner crystal for realistic parameters because quantum fluctuations destroy the Mott insulator. But it is beyond the validity of our mean-field approach to answer this question in a quantitative fashion.

\subsection{Phase diagram} 
\label{SSect:phase diagram}
In the last paragraphs we have concentrated on comparing different Wigner crystals at fixed densities $1/8$, $1/6$, $2/15$, and $1/9$ which is motivated by the recent experimental NMR data on SrCu$_2$(BO$_3$)$_2$ and their phenomenological interpretation \cite{takigawa12}. Our microscopic mean-field theory of the effective hardcore boson model plus additional kinetic terms confirms many aspects of the phenomenological theory. Altogether, all our results point to the following coupling ratios for a microscopic description of SrCu$_2$(BO$_3$)$_2$: $J'/J\geq 0.67$, $t_1/J\approx 0.015$, and $t_2/J\approx -0.01$. Interestingly, the magnitude of the kinetic terms are in very good agreement with estimates for the pure Shastry-Sutherland model (see discussion on $t_2$ above) as well as estimates for the DM interaction $D_z/J = 2\,t_1\approx 0.02$ \cite{cepas01,cheng07,mazurenko08,takigawa10,romhanyi11}. The latter is estimated for $J'/J \approx 0.635$. Obviously, such estimates give the correct order of magnitude but do not pinpoint the value quantitatively. Additionally, our findings of a rather large ratio $J'/J \geq 0.67$ might result in slightly renormalized fitting values. In this subsection we study the full low-density phase diagram in this most realistic coupling regime and we compare our results to the experimental findings. 

\begin{figure}
   \begin{center}
\subfigure[$\,$$J'/J = 0.6; t_1/J = 0; t_2/J = -0.0025$]{
    \includegraphics[width=0.75\columnwidth]{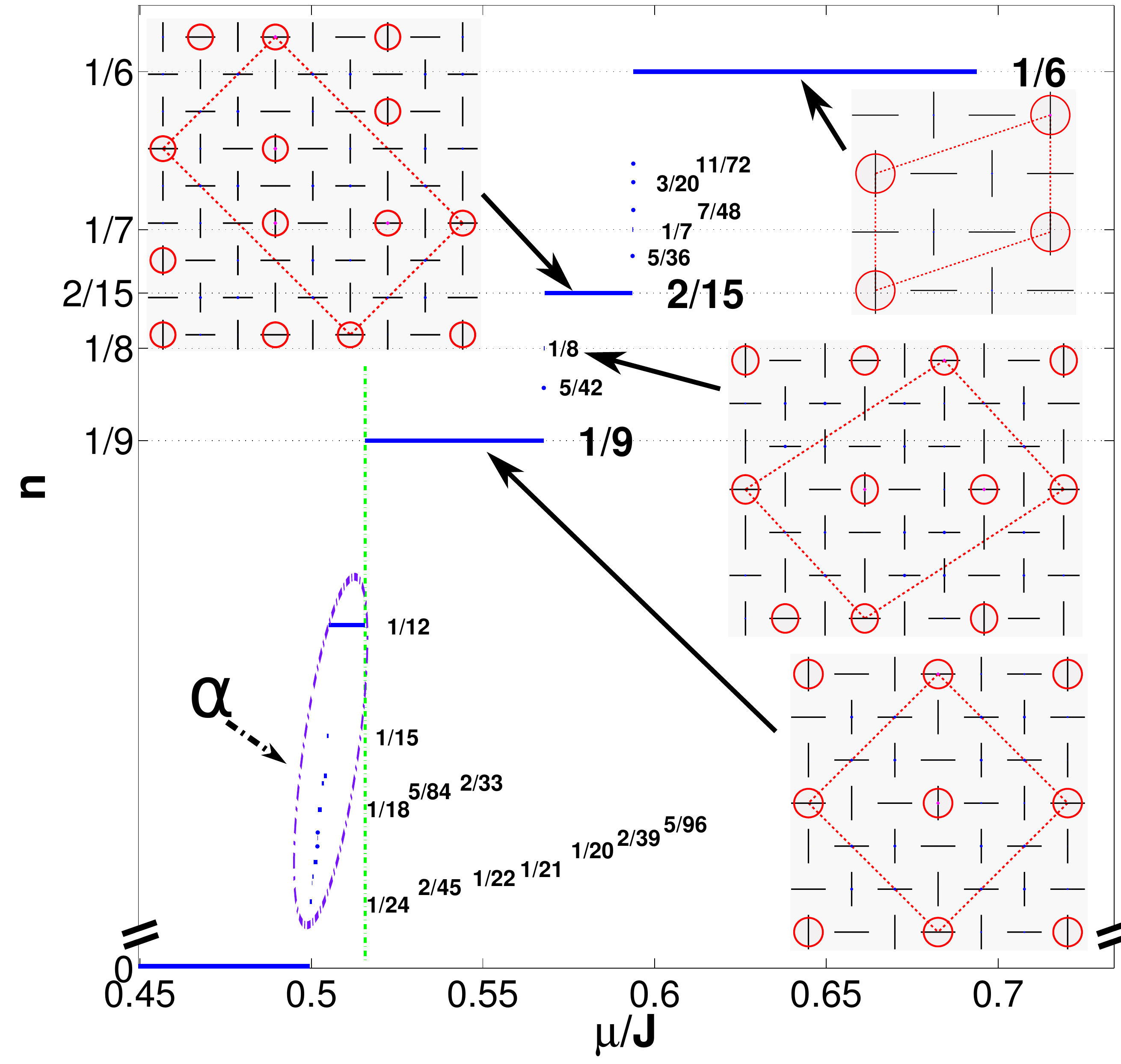}
\label{fig:MagnetJ60_no_t1}
} \\
\subfigure[$\,$$J'/J = 0.65; t_1/J = 0; t_2/J = -0.005$]{
    \includegraphics[width=0.75\columnwidth]{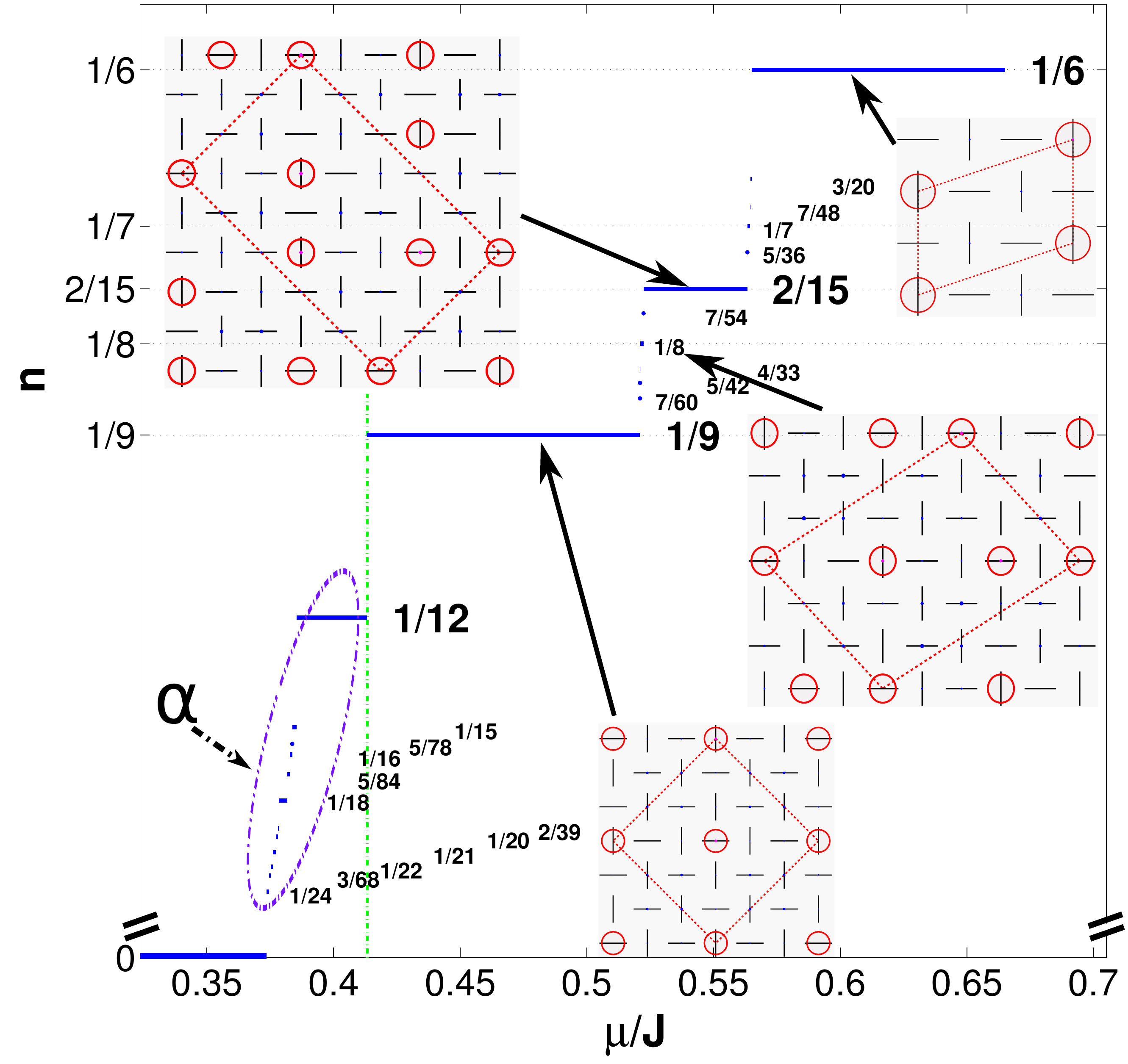}
\label{fig:MagnetJ65_no_t1}
} \\
\subfigure[$\,$$J'/J = 0.68; t_1/J = 0; t_2/J = -0.01$]{
    \includegraphics[width=0.75\columnwidth]{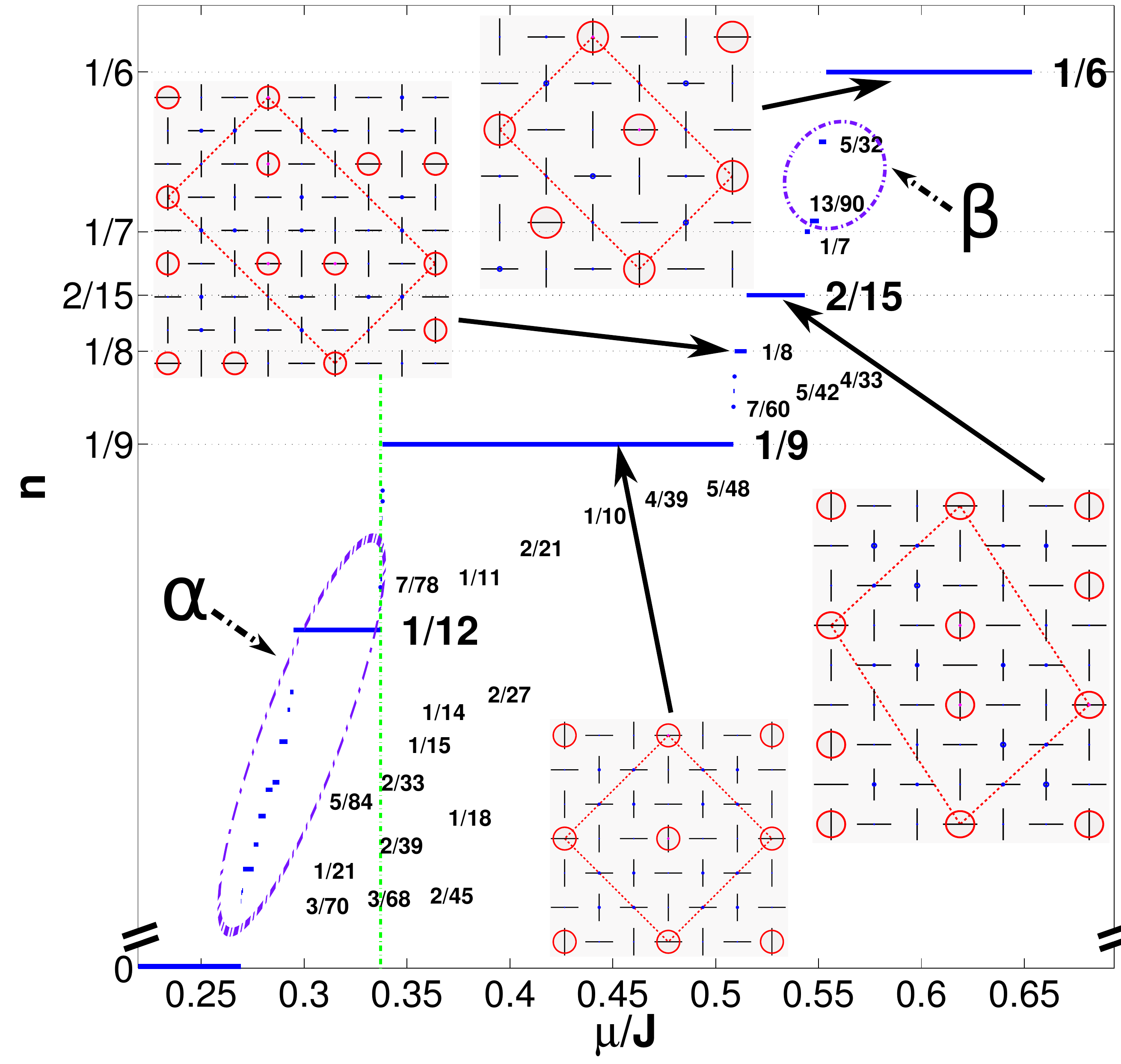}
\label{fig:MagnetJ68_no_t1}
} 
\end{center}
   \caption{The density $n$ is plotted as a function of $\mu/J$ for a vanishing nearest-neighbor hopping $t_1 = 0$ for different values of $J'/J$ and $t_2/J$.  The value of the diagonal hopping $t_2$ is consistent with the extrapolated order 17 series of the pure Shastry-Sutherland model. Additionally, the most important structures are shown.}
    \label{fig:Magnet_no_t1}
\end{figure} 

We start by discussing the phase diagram of the pure Shastry-Sutherland model which contains intrinsically the diagonal hopping term $t_2$. It is therefore already interesting to compare our mean-field results treating quantum fluctuations with the CA used in Ref.~\onlinecite{dorier08} giving a sequence of magnetization plateaux at densities $1/9$, $2/15$, and $1/6$ for $J'/J=0.5$. 

\begin{figure}
   \begin{center}
\subfigure[$\,$$J'/J = 0.6; t_1/J = 0.01; t_2/J = -0.0025$]{
    \includegraphics[width=0.75\columnwidth]{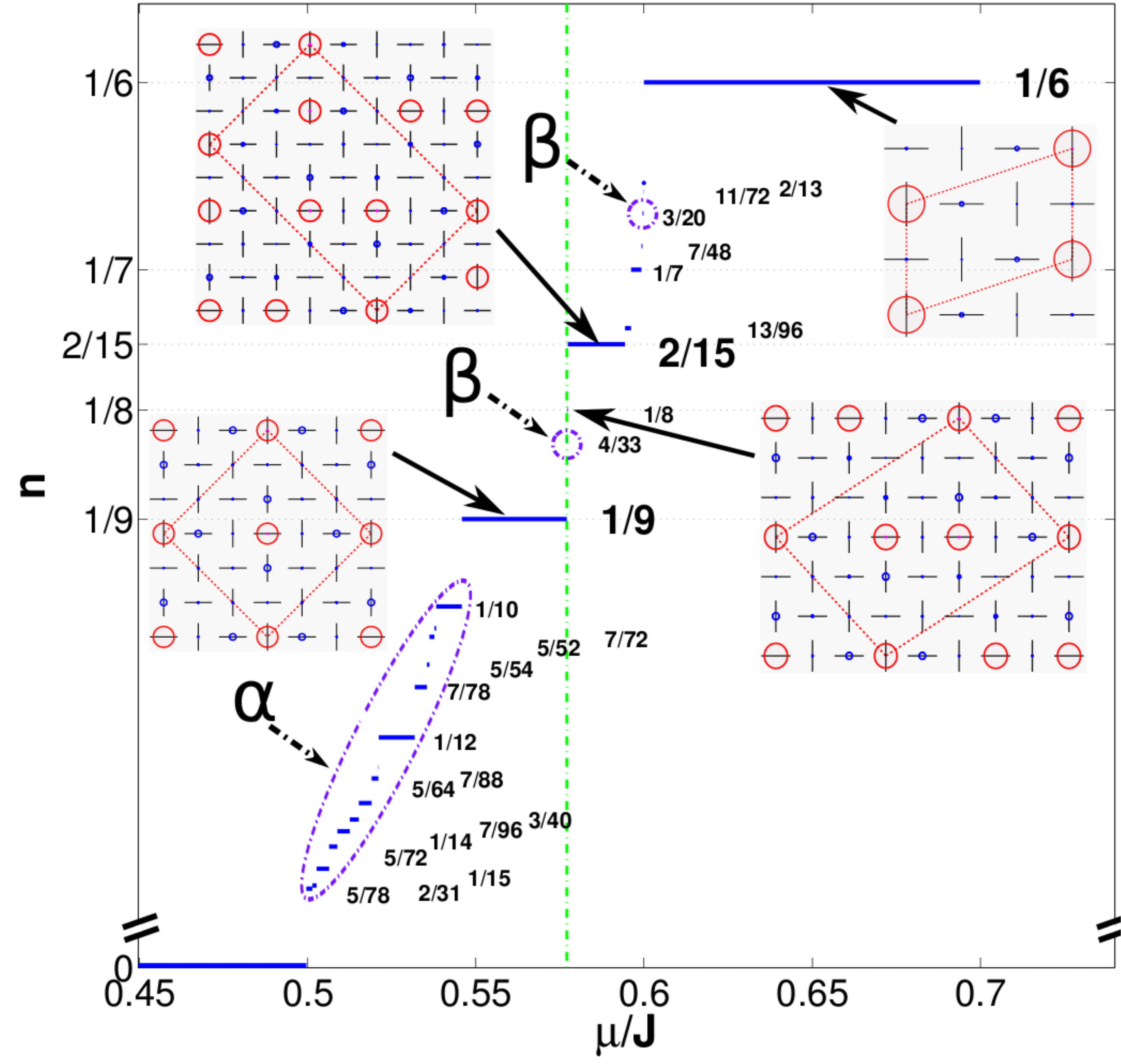}
\label{fig:MagnetJ60_with_t1}
} \\
\subfigure[$\,$$J'/J = 0.65; t_1/J = 0.01; t_2/J = -0.005$]{
    \includegraphics[width=0.75\columnwidth]{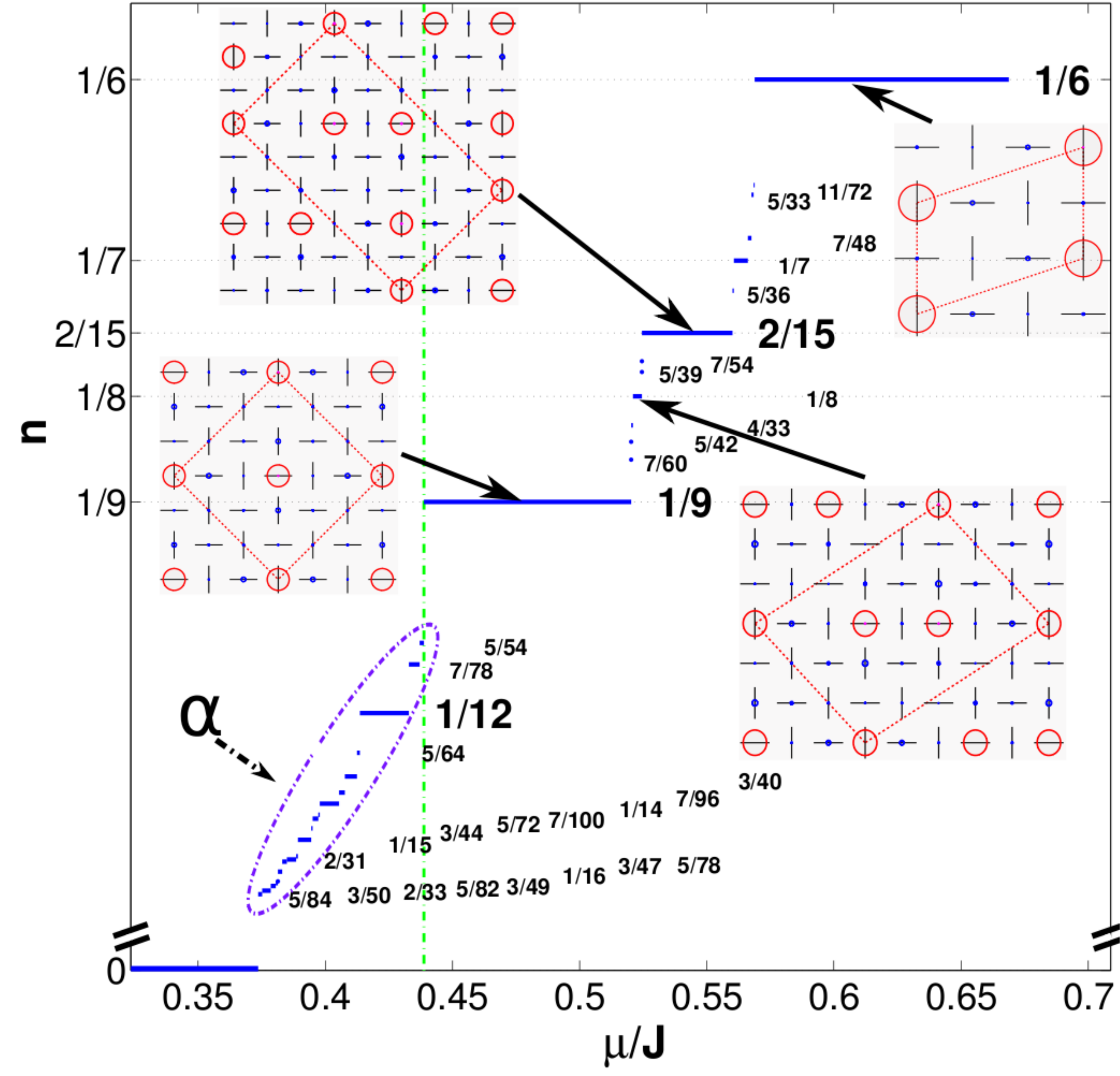}
\label{fig:MagnetJ65_with_t1}
} \\
\subfigure[$\,$$J'/J = 0.68; t_1/J = 0.01; t_2/J = -0.01$]{
    \includegraphics[width=0.75\columnwidth]{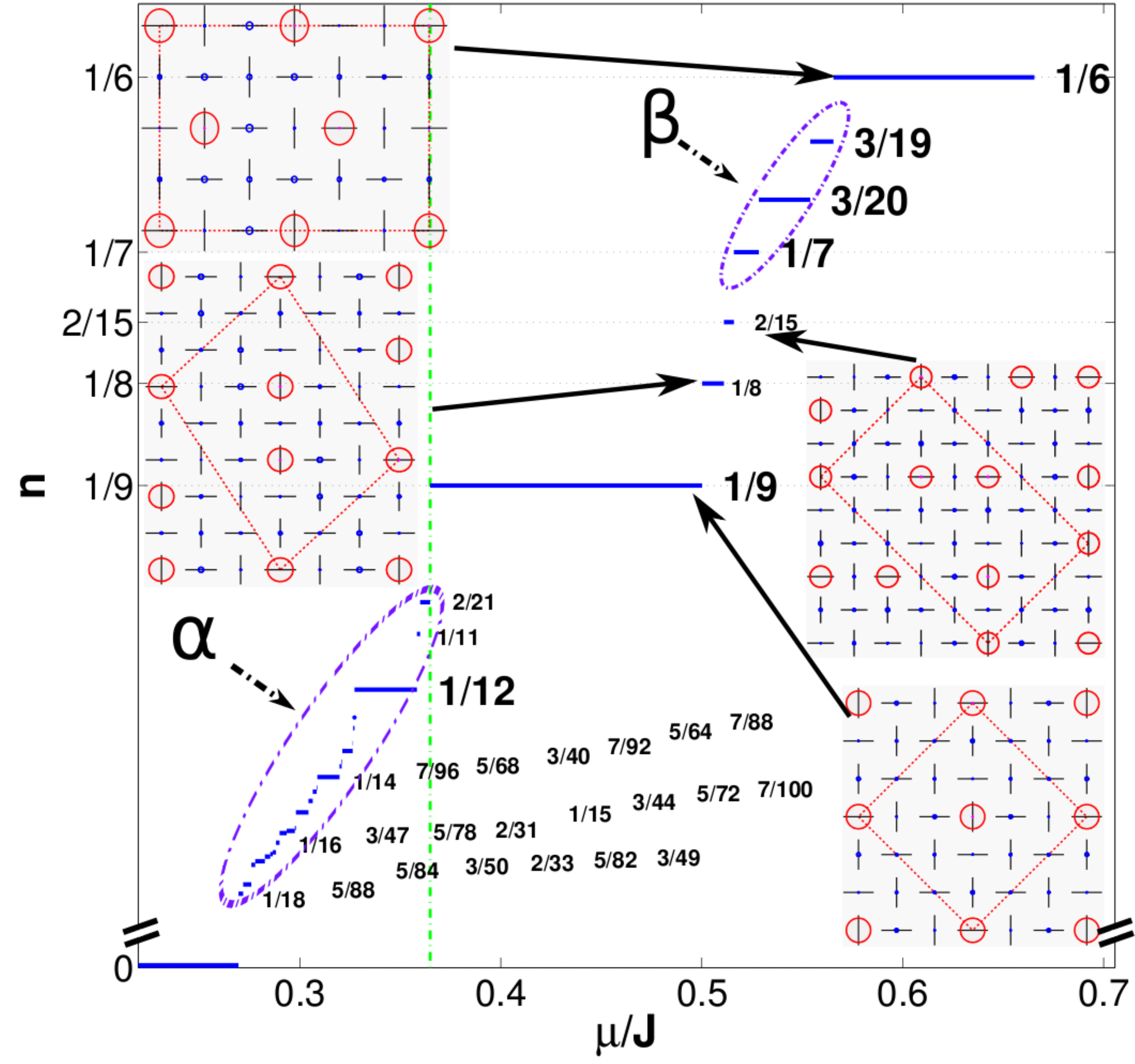}
\label{fig:MagnetJ68_with_t1}
} 
\end{center}
   \caption{The density $n$ is plotted as a function of $\mu/J$ for a nearest-neighbor hopping $t_1 = 0.01$ and different values of $J'/J$ and $t_2/J$. The value of the diagonal hopping $t_2$ is consistent with the extrapolated order 17 series of the pure Shastry-Sutherland model for $J'/J\geq 0.65$. Additionally, the relevant structures are shown.}
    \label{fig:Magnet_with_t1}
\end{figure}

The density curves for $J'/J \in\{ 0.6, 0.65, 0.68\}$ and different values for $t_2/J$ are shown in Fig.~\ref{fig:Magnet_no_t1}. One clearly sees that the density displays basically the same sequence $1/9$, $2/15$, and $1/6$ as the CA \cite{dorier08}.  The only major change is for the $1/6$ plateau whose structure changes for $J'/J \geq 0.67$ as discussed above which is consistent with the phenomenological theory \cite{takigawa12}. Furthermore, it is not surprising that already small kinetic terms lift the degeneracy of the multi-intersectional points present in the classical phase diagram (see Fig.~\ref{fig:ClassR2}). Most importantly, the diagonal hopping $t_2$ alone is not responsible for the experimentally stabilized $1/8$ plateau. These findings are well seen in Fig.~\ref{fig:Magnet_no_t1}: other plateaux distinct from the classical structures at densities $1/9, 2/15$, and $1/6$ are basically not realized.

Next, we introduce a finite value for the nearest-neighbor hopping $t_1/J = 0.01$ and we study again the low-density phase diagram shown in Fig.~\ref{fig:Magnet_with_t1}. Still, the sequence of plateaux is close to the one already observed in the classical limit for $J'/J \leq 0.65$. The first real change happens for $J'/J = 0.68$. Here the $1/8$ plateau is realized with an appropriate width. Nevertheless the structure of the $1/8$ plateau is still incompatible with the one seen in experiments. This suggests that one should consider slightly larger values for the nearest-neighbor hopping $t_1$ in our mean-field calculation.

\begin{figure}
   \begin{center}
\subfigure[$\,$$1/8$ plateaux ($J'/J = 0.68$)]{
    \includegraphics[width=0.61\columnwidth]{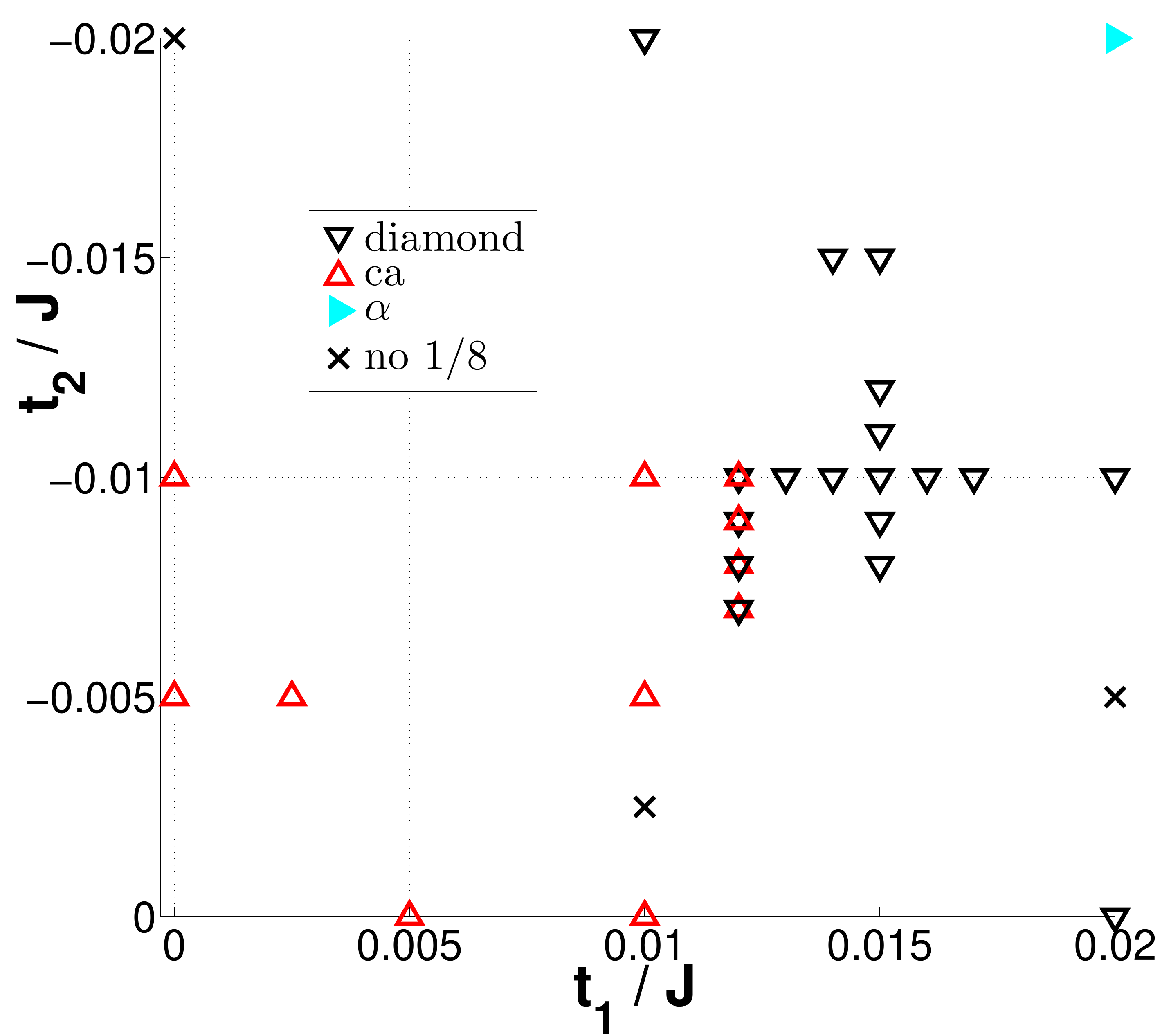}
\label{fig:EightInComp}
}
\\
\subfigure[$\,$$2/15$ plateaux ($J'/J = 0.68$)]{
   \includegraphics[width=0.61\columnwidth]{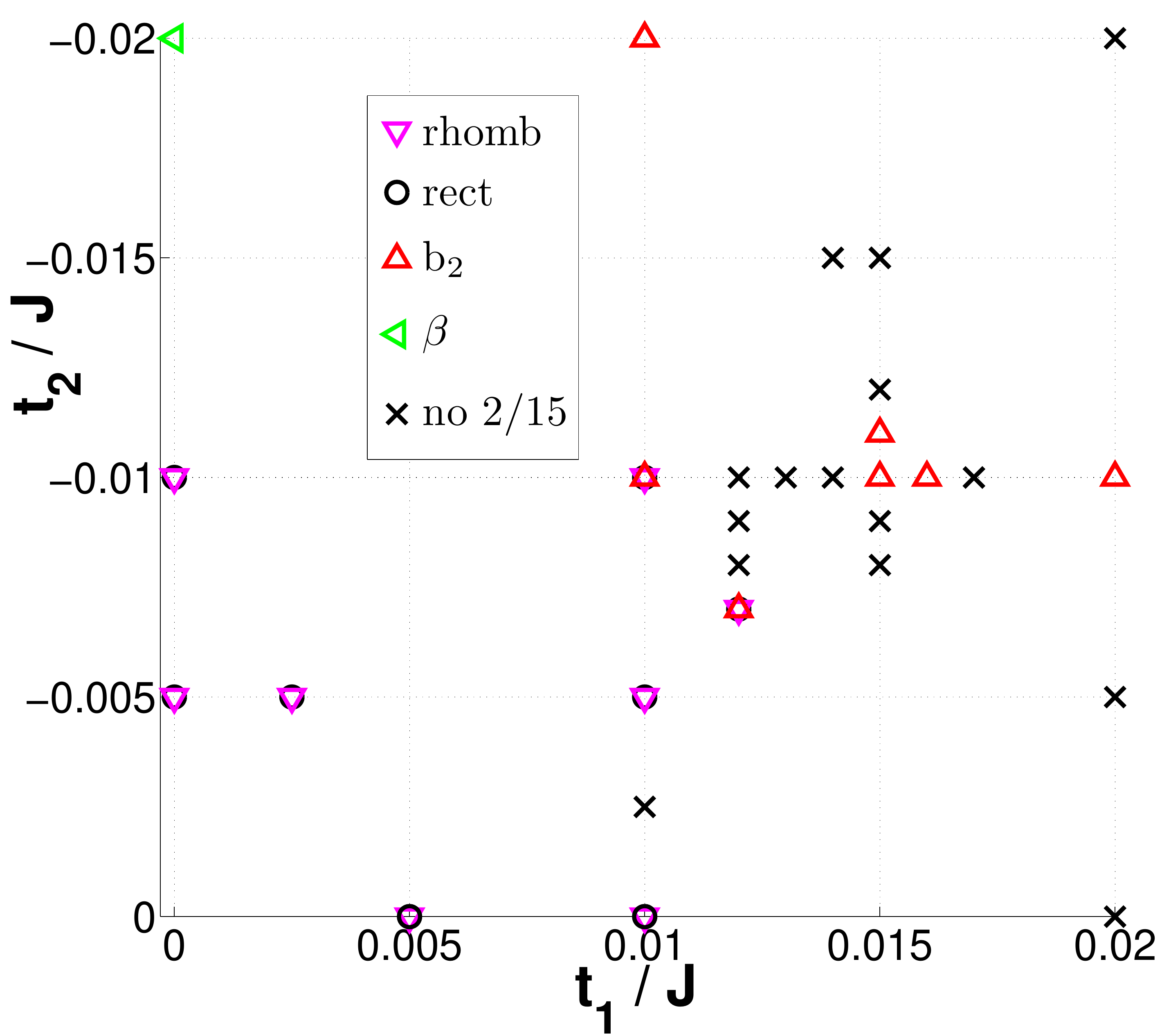}
\label{fig:FiveInComp}
}
   \end{center}
   \caption{Structures with the lowest mean-field energy are plotted in comparison to all other densities as a function of $t_1/J$ and $t_2/J$ for the density (a)$1/8$ and for the density (b)$2/15$. The black symbol x marks a parameter regime where both plateaux are realized in the phase diagram. In the relevant regime $t_1/J \approx 0.015$ and $t_2/J \approx -0.01$, the $2/15$b$_2$ plateau is only realized in a small region in contrast to the $1/8$-diamond. Symbols $\alpha$ ($\beta$) represent mean-field solutions of $\alpha$-type ($\beta$-type). Mean-field solutions which differ by less than $10^{-4}$J are considered to be degenerate. In this case all such structures are displayed.}
    \label{fig:EnergyInComparison}
\end{figure} 

\begin{figure}
   \begin{center}
\subfigure[$\,$Density $n$ as a function of the chemical potential for $J'/J = 0.68$, $t_1/J=0.016$, and $t2/J=-0.01$. All plateaux left of the green dashed line have a negative mean-field energy at $\mu = 0$.]{
    \includegraphics[width=0.90\columnwidth]{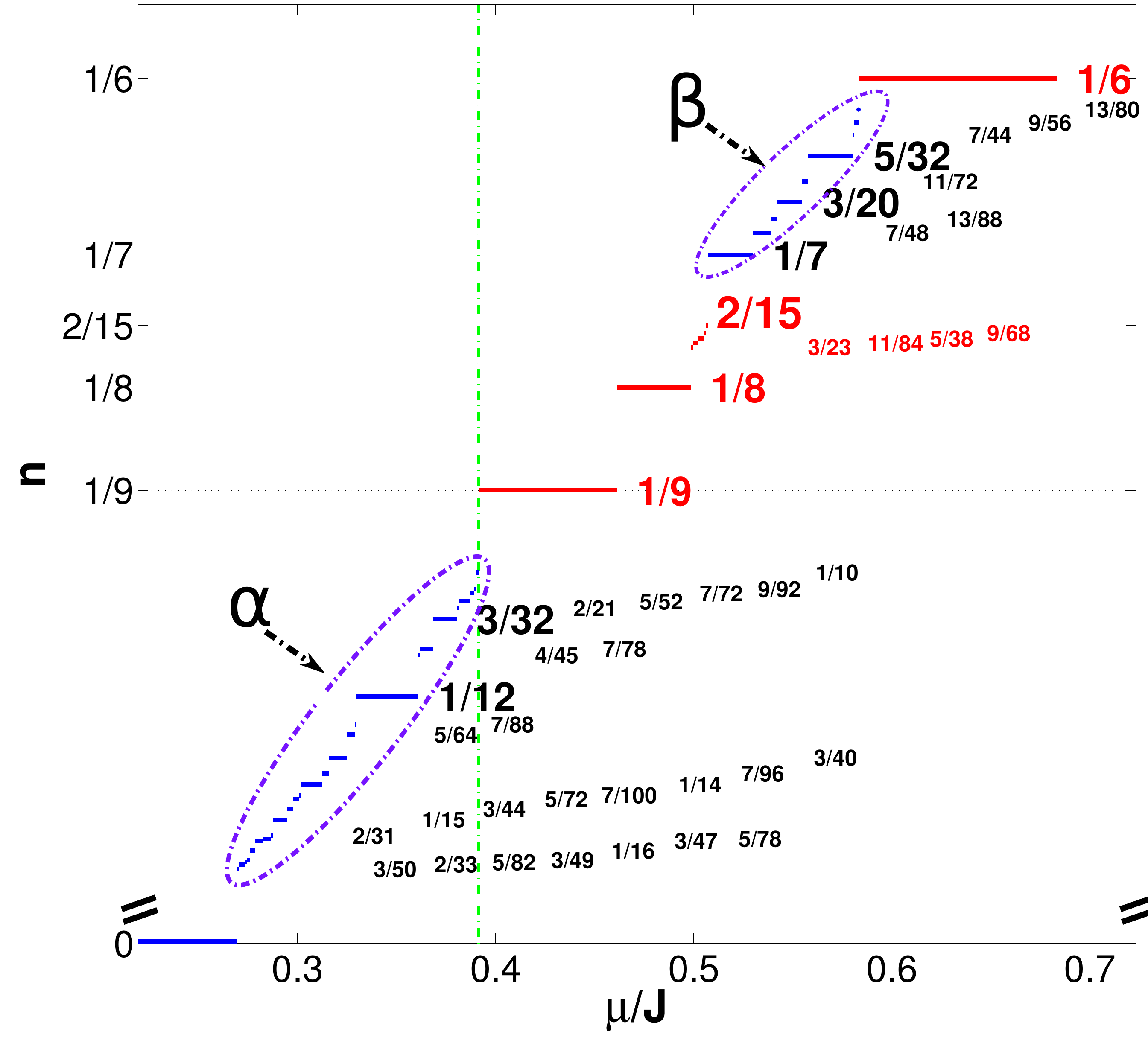}
\label{fig:Magnet_J68_2_15_3_32}
}\\
\subfigure[$\,$$1/9$ structure]{
   \includegraphics[width=0.44\columnwidth]{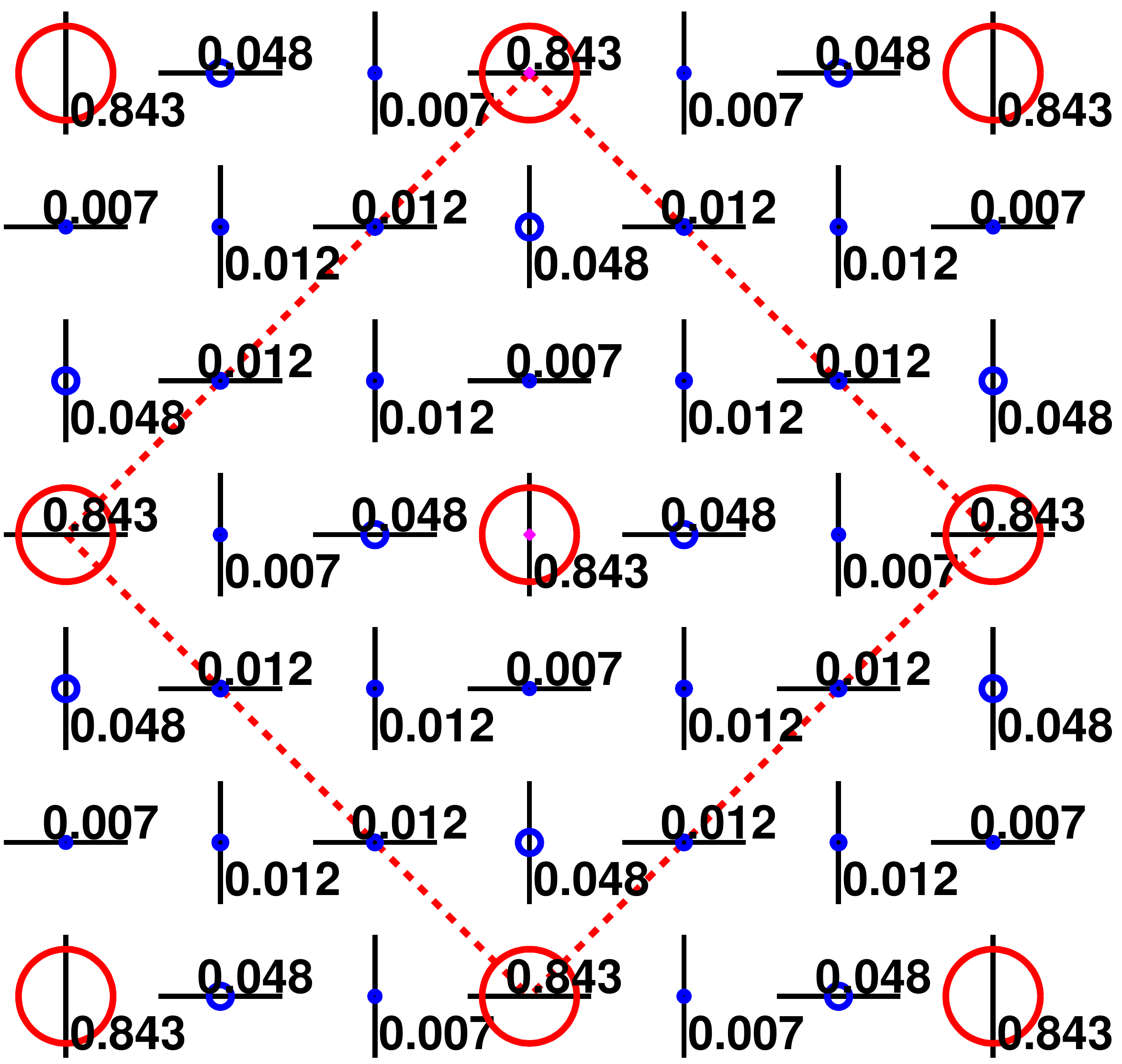}
\label{fig:Nine_Plat}
}
\subfigure[$\,$$1/9$ magnetization]{
   \includegraphics[width=0.45\columnwidth]{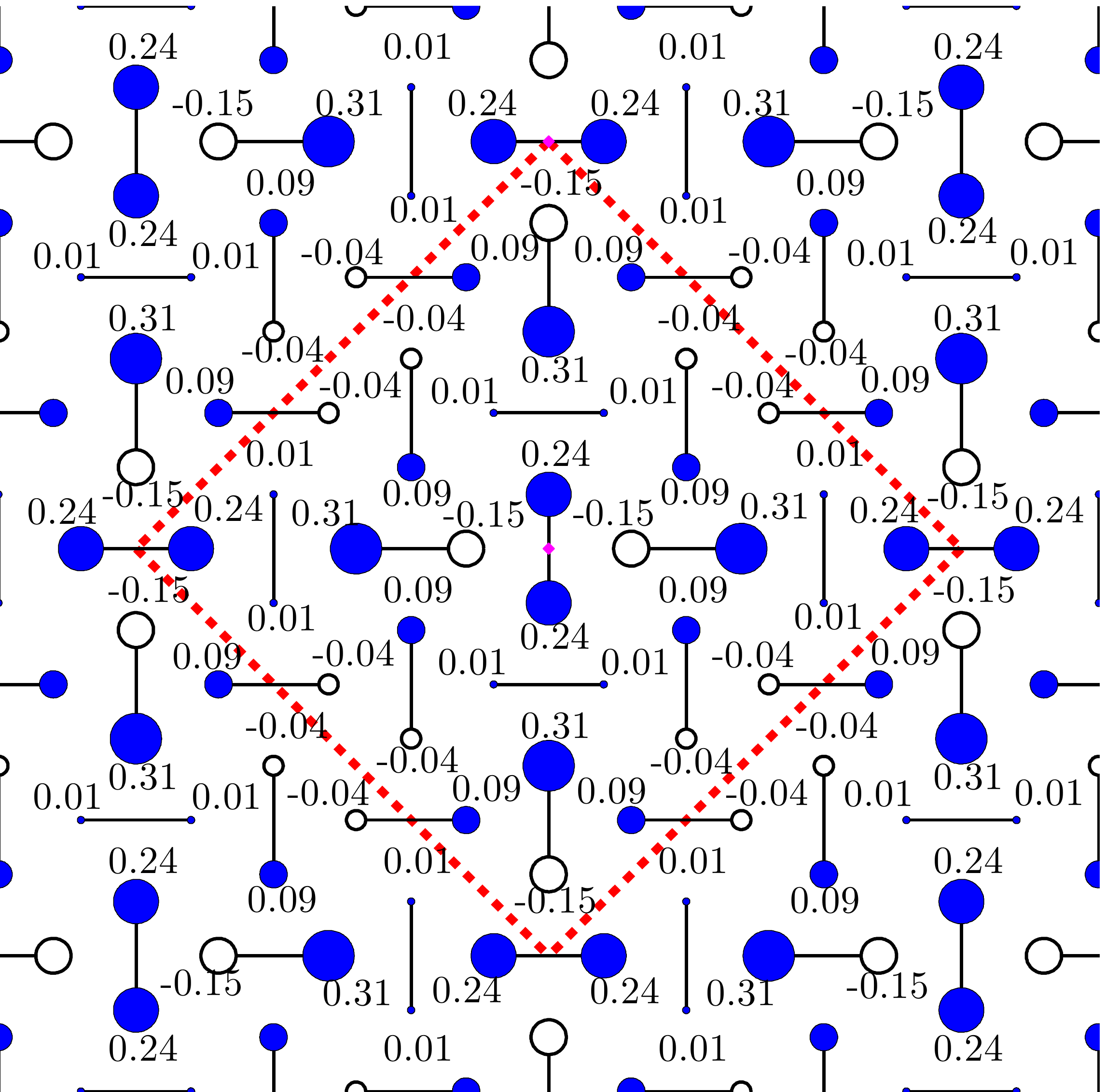} 

}
\\
\subfigure[$\,$$1/8$-diamond structure]{
   \includegraphics[width=0.43\columnwidth]{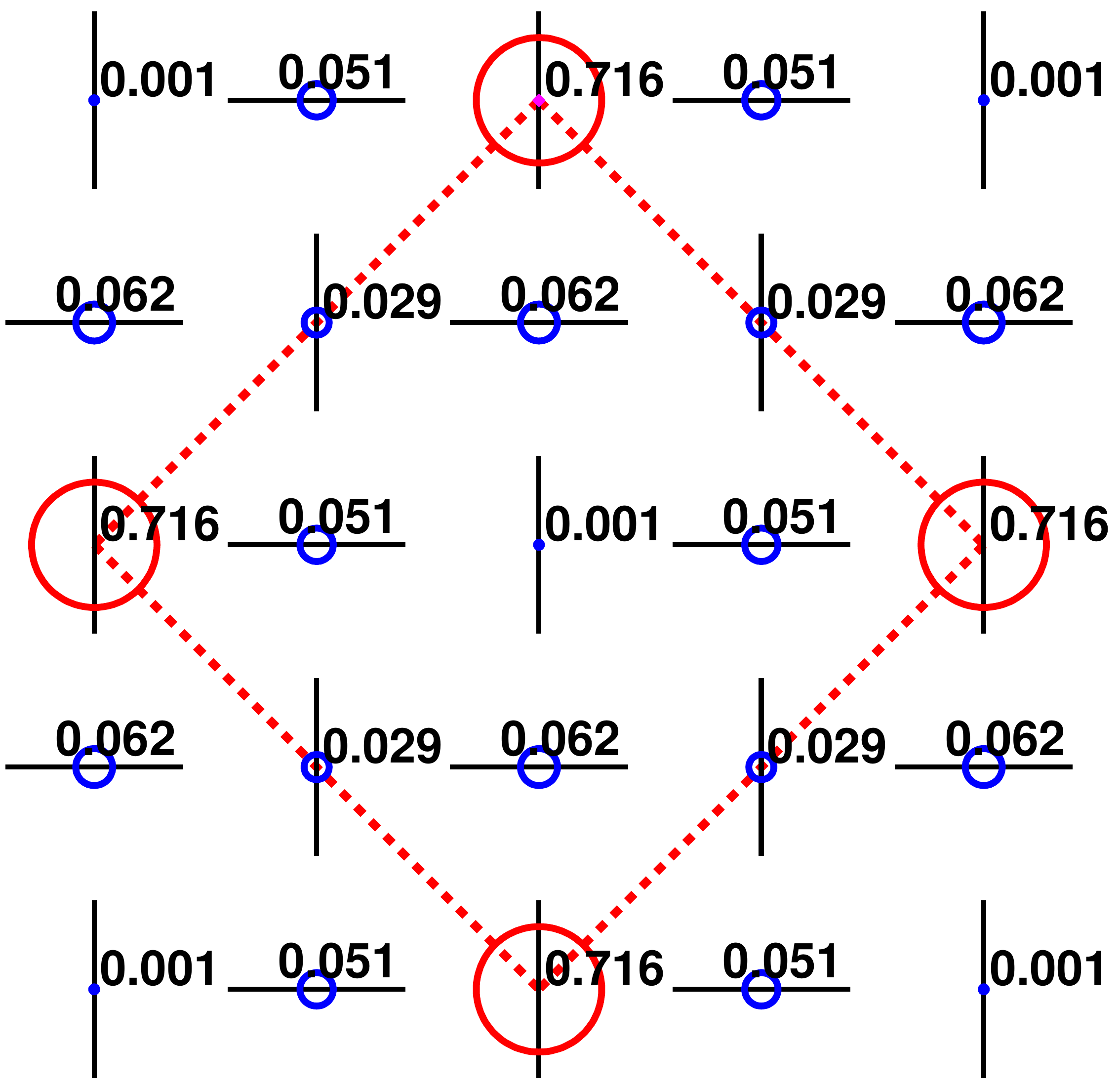}
\label{fig:Eight_Plat}
}
\subfigure[$\,$$1/8$-diamond magnetization]{
   \includegraphics[width=0.49\columnwidth]{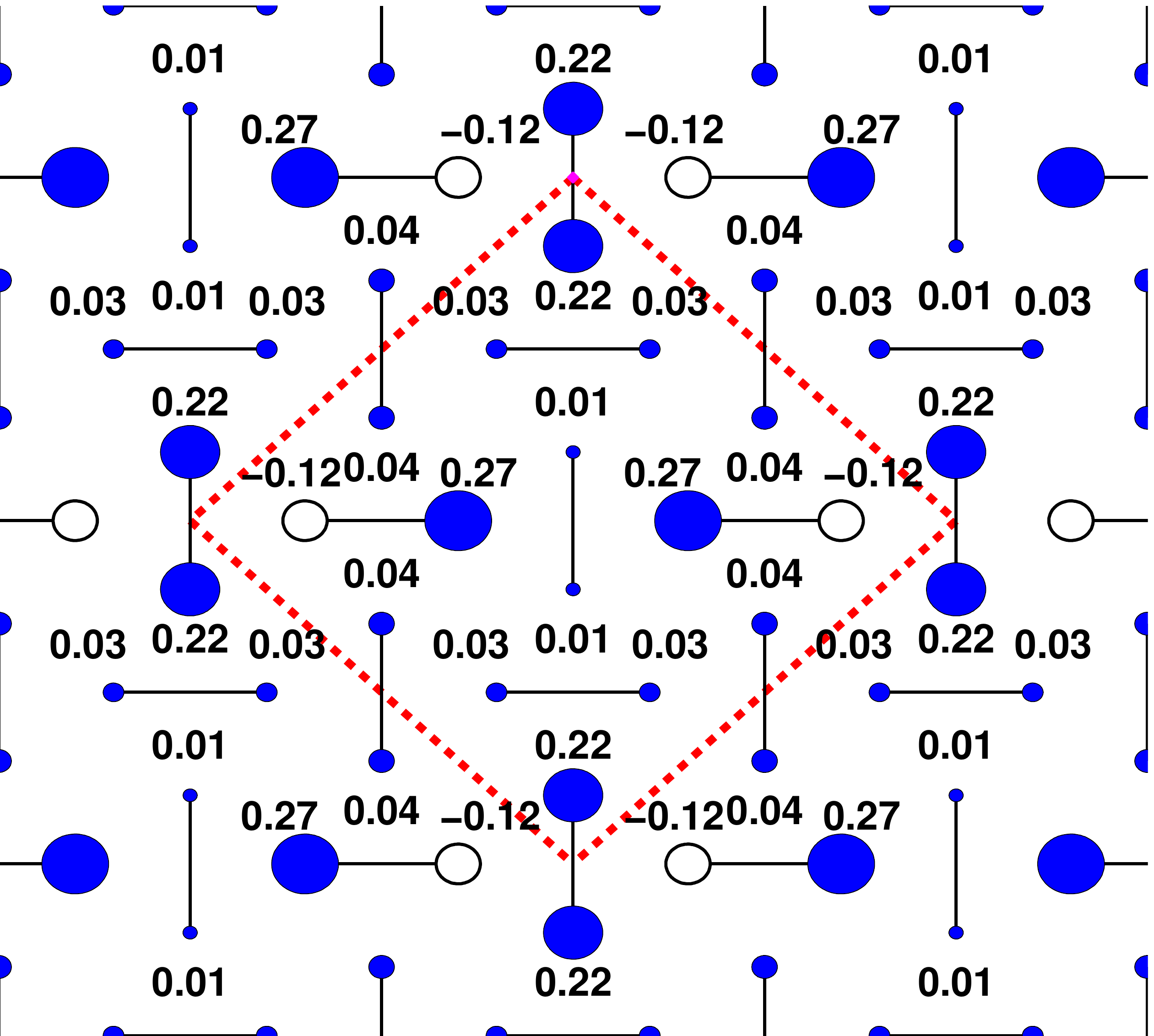}
\label{fig:Eight_Mag}
}\\
\subfigure[$\,$$1/6$-new structure]{
   \includegraphics[width=0.43\columnwidth]{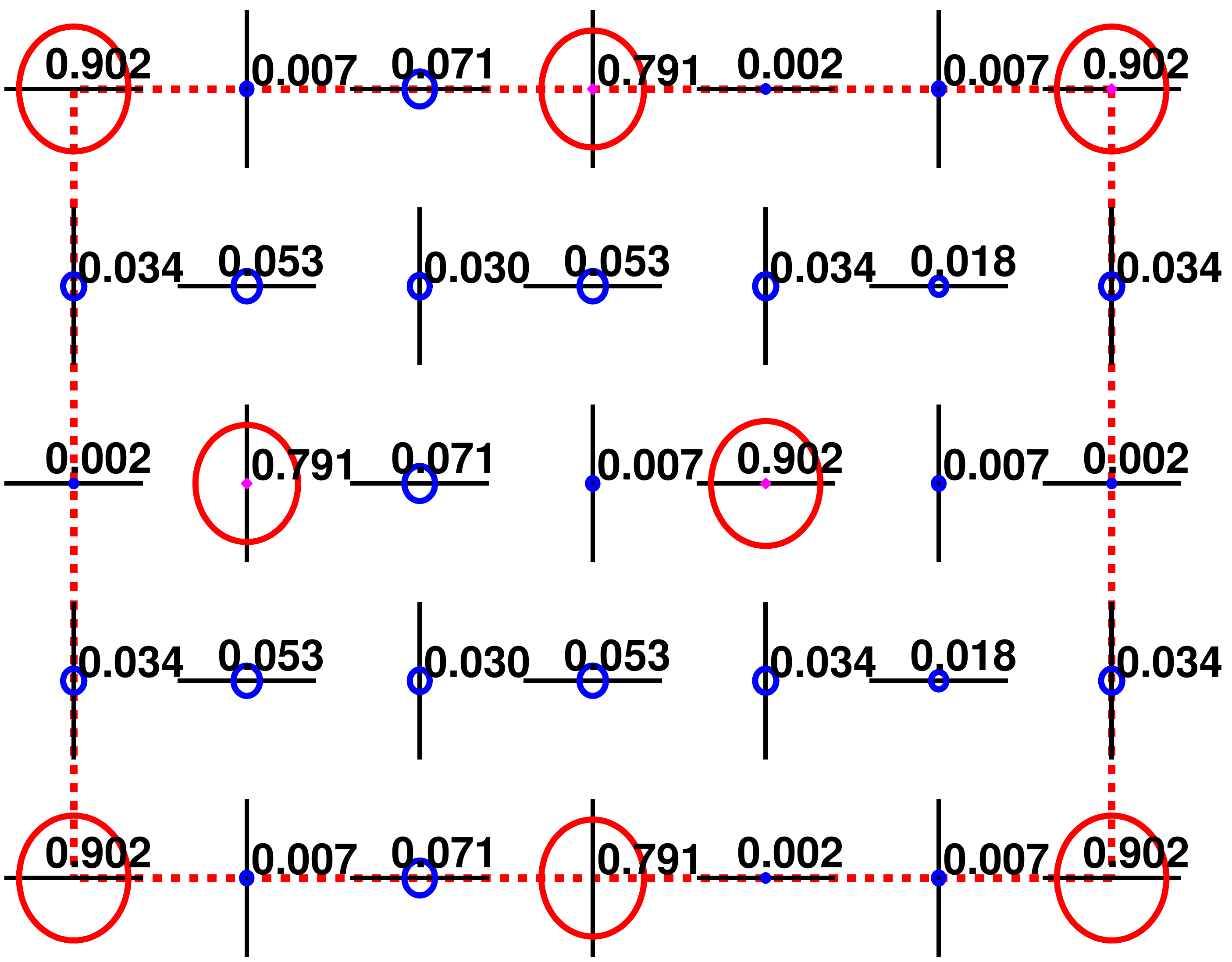} 
\label{fig:Six_Plat}
}
\subfigure[$\,$$1/6$-new magnetization]{
   \includegraphics[width=0.47\columnwidth]{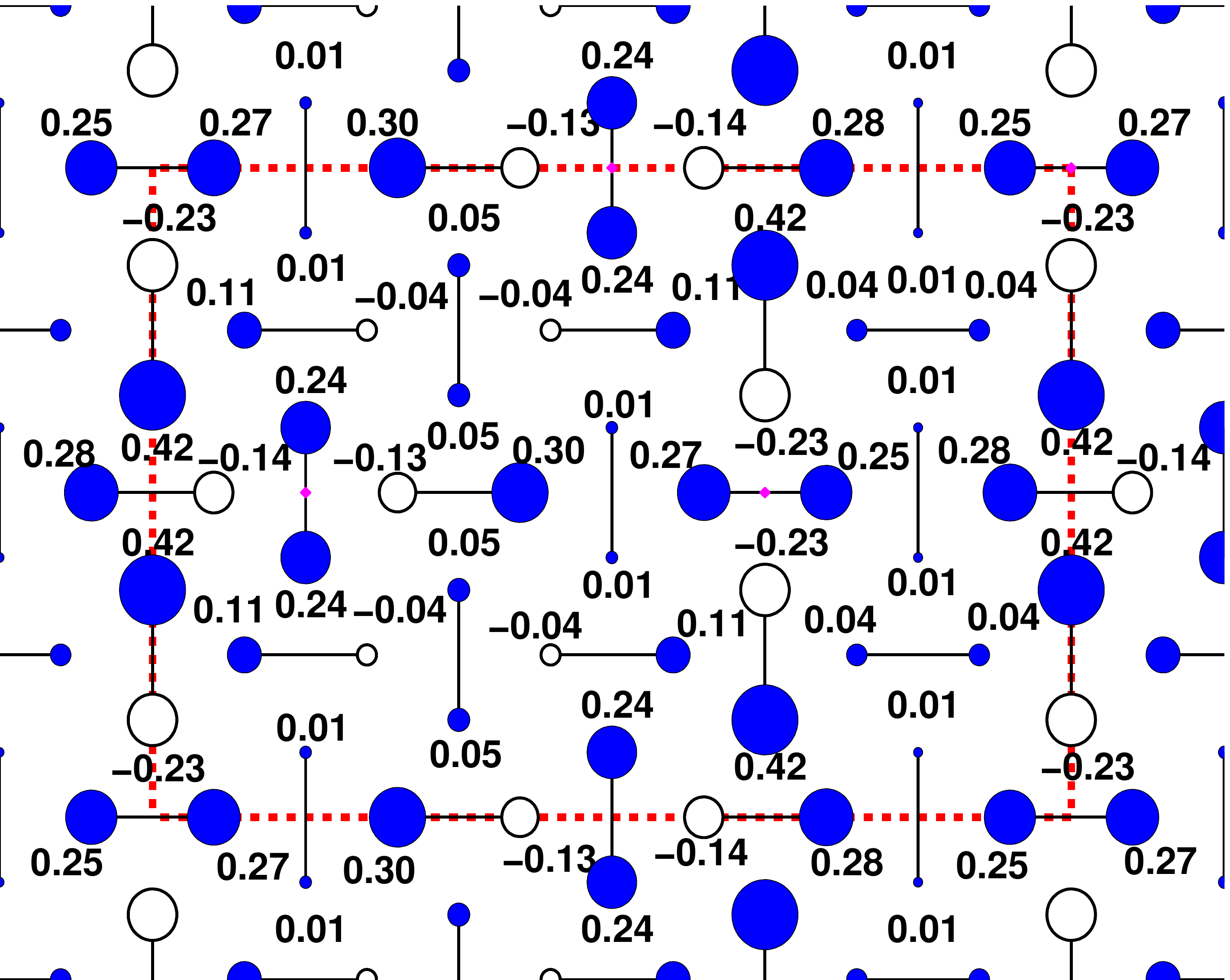} 
\label{fig:Six_Mag}
}
   \end{center}
   \caption{(a) Density $n$ is plotted as a function of $\mu/J$ for $J'/J = 0.68$, $t_1/J=0.016$, and $t_2/J=-0.01$. Plateaux shown in red correspond to valid Wigner crystals. Mean-field solutions in the regimes $\alpha$ and $\beta$ are unphysical: In $\alpha$ the mean-field energies at $\mu=\mu_0$ are negative. In $\beta$ particles are completely delocalized among various dimers inside the crystalline solution. The structure and the magnetization of the plateau at $1/9$ are given in (b)-(c). The corresponding figure for $1/8$ and $1/6$ are shown in (d)-(e) and (f)-(g).}
    \label{fig:Magnet}
\end{figure} 

We therefore checked first whether a (realistic) parameter range exists for $t_1$ and $t_2$ where a $1/8$ plateau with diamond unit cell and a $2/15$ plateau with a striped structure is realized simultaneously in the phase diagram, i.e.~where both plateaux are present in the density curve. This is shown in Fig.~\ref{fig:EnergyInComparison}. We find that the nearest-neighbor hopping $t_1$ should be $t_1/J \approx 0.015$. Let us stress that it is well possible that such absolute values slightly shift due to uncertainties in the extrapolation of the potential terms $V_\delta$. But it is already encouraging that the value for $t_1$ deduced from our mean-field theory is so close to the one proposed in the literature \cite{cepas01,cheng07,mazurenko08,takigawa08,romhanyi11}.

Summarizing the above results, we should focus on a ratio $J'/J \geq 0.67$ and we should consider kinetic couplings $t_1/J \approx 0.015$ and $t_2/J \approx -0.01$. Consequently, we plot the density as a function of $\mu/J$ for $J'/J = 0.68$, $t_1/J = 0.016$ and $t_2/J = -0.01$ in Fig.~\ref{fig:Magnet_J68_2_15_3_32}. Additionally, we present our results for the density pattern (Fig.~\ref{fig:Magnet_J68_2_15_3_32}b-d-f) and for the magnetization (Fig.~\ref{fig:Magnet_J68_2_15_3_32}c-e-g). The latter is calculated according to Eqs.~\ref{fig:obs} (see also Fig.~\ref{fig:magnetScheme}). Most prominently, a $1/8$ plateau with diamond unit cell as well as the $1/6$-new plateau is realized in the phase diagram. The just discussed $2/15$ striped plateau is also realized but its width is very tiny. 

\begin{figure}
   \begin{center}
\subfigure[$\,1/7$ structure]{
   \includegraphics[width=0.95\columnwidth]{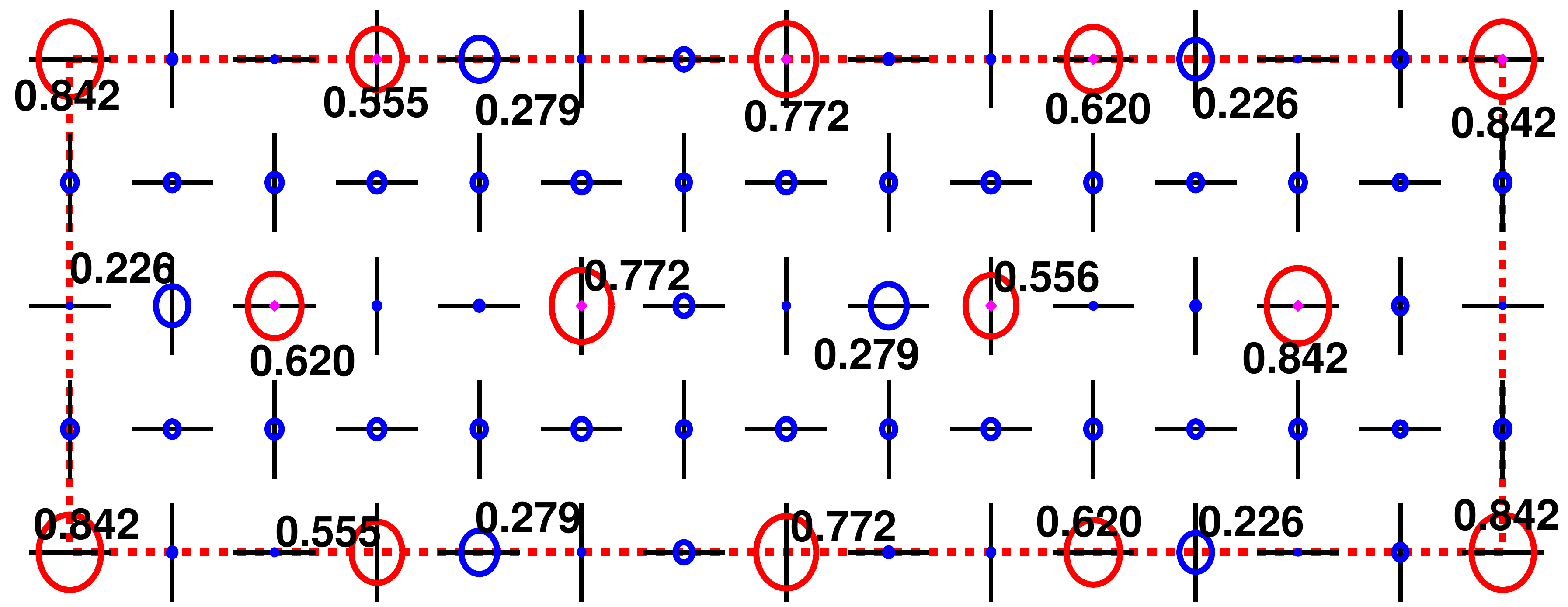}
\label{fig:Seven_Plat}
}
\\
\subfigure[$\,3/20$ structure]{
   \includegraphics[width=0.95\columnwidth]{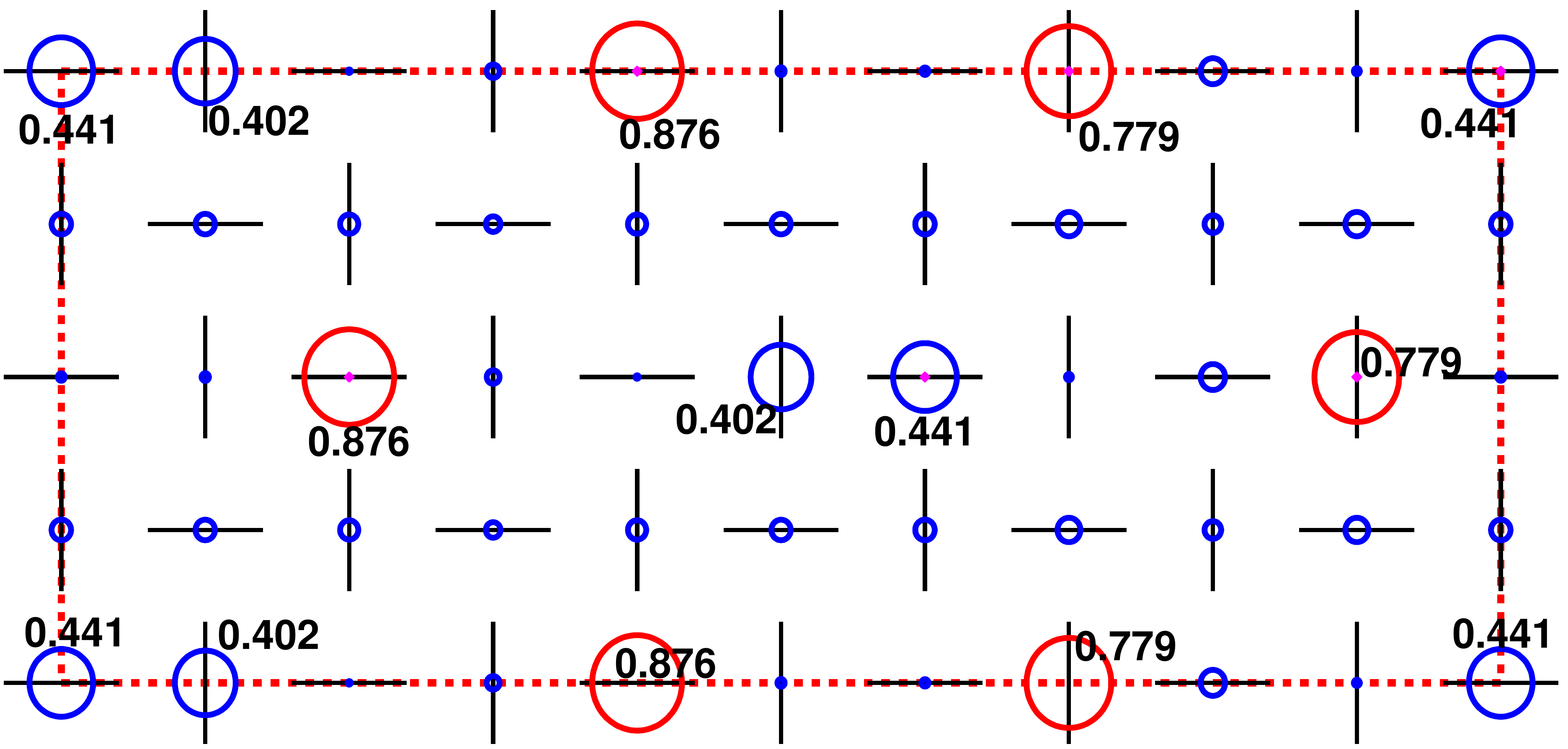}
\label{fig:twenty_Plat}
}
   \end{center}
   \caption{Figure illustrates the local density of two unphysical mean-field states classified as $\beta$ at densities $1/7$ (a) and $3/20$ (b) shown in Fig.~\ref{fig:Magnet_J68_2_15_3_32}.}
    \label{fig:Magnet2}
\end{figure}

Furthermore, there are additionally three interesting density regimes which one should discuss: i) densities below $1/9$, ii) densities $1/8< n \leq 2/15$, and iii) densities $2/15< n < 1/6$. We start the discussion with the regimes i) and iii) which are denoted by $\alpha$ and $\beta$ in Fig.~\ref{fig:Magnet_J68_2_15_3_32}. We believe that our mean-field theory gives unphysical results for both density regions, but due to different physical reasons. In the low-density region $\alpha$, we find a negative mean-field energy $E_{\rm mf}$ at $\mu=\mu_0$ for all plateau structures below $1/9$. As explained above, it is very likely that these magnetization plateaux melt due to the introduced quantum fluctuations if one allows also superfluid mean-field solutions (or treat the problem beyond mean-field theory). In contrast, for the regime $\beta$ between $2/15$ and $1/6$, all plateau structures are such that at least one particle is completely delocalized between at least two different dimers. Two examples of such structures for densities 1/7 and 3/20 are displayed in Fig.~\ref{fig:Magnet2}. These are therefore clearly structures which are extremely far away from perturbed classical plateau structures where our mean-field theory is expected to work well. In total, we do not trust our mean-field results in the regions $\alpha$ and $\beta$. 

\begin{figure}[t]
  \begin{center}
\subfigure[$\,$Stripe structures]{
    \includegraphics[width=0.85\columnwidth]{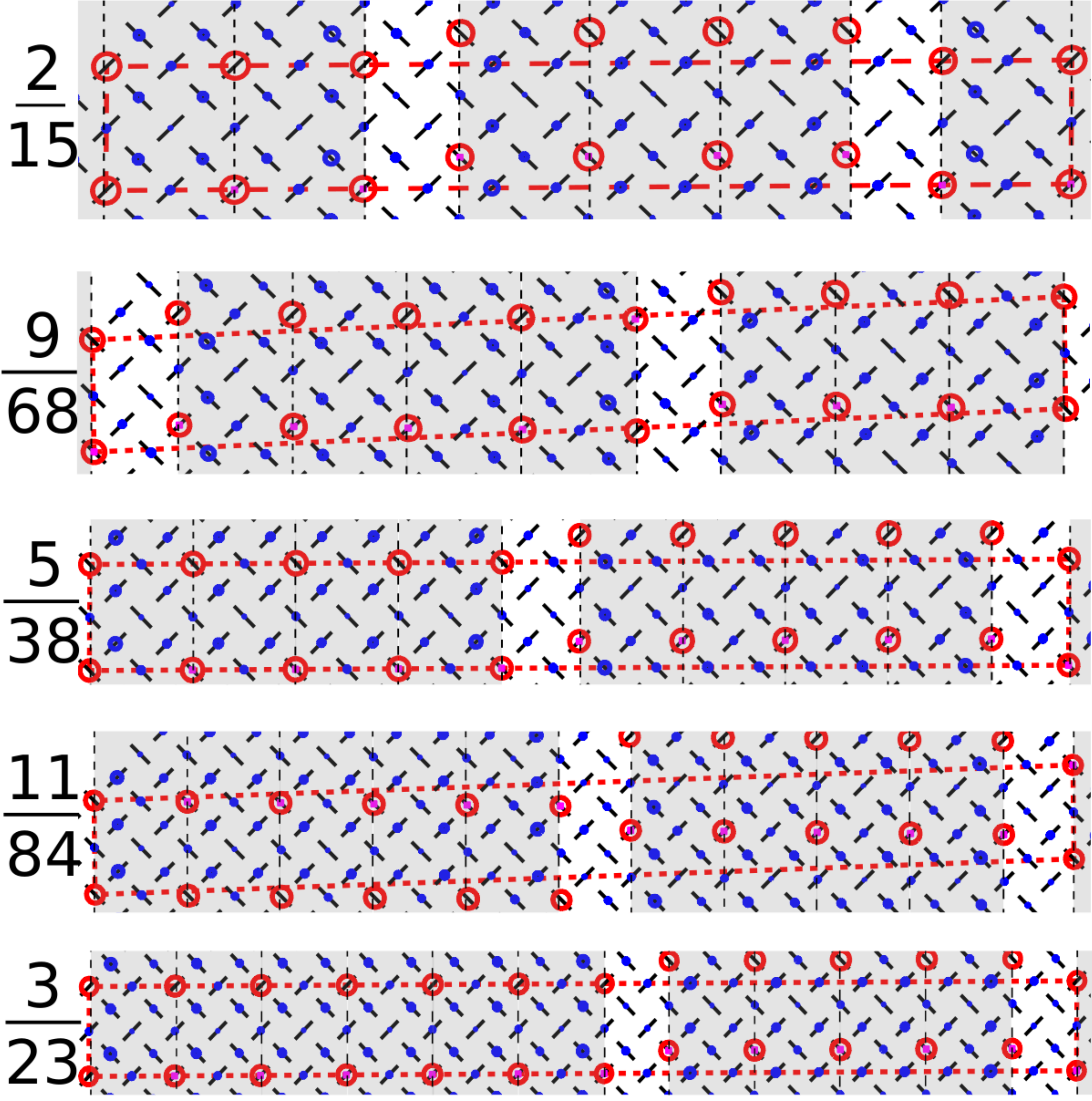}
\label{fig:AllStripeMech_PLATEAUX}
}
\subfigure[$\,$Magnetization of the stripe structures]{
    \includegraphics[width=0.85\columnwidth]{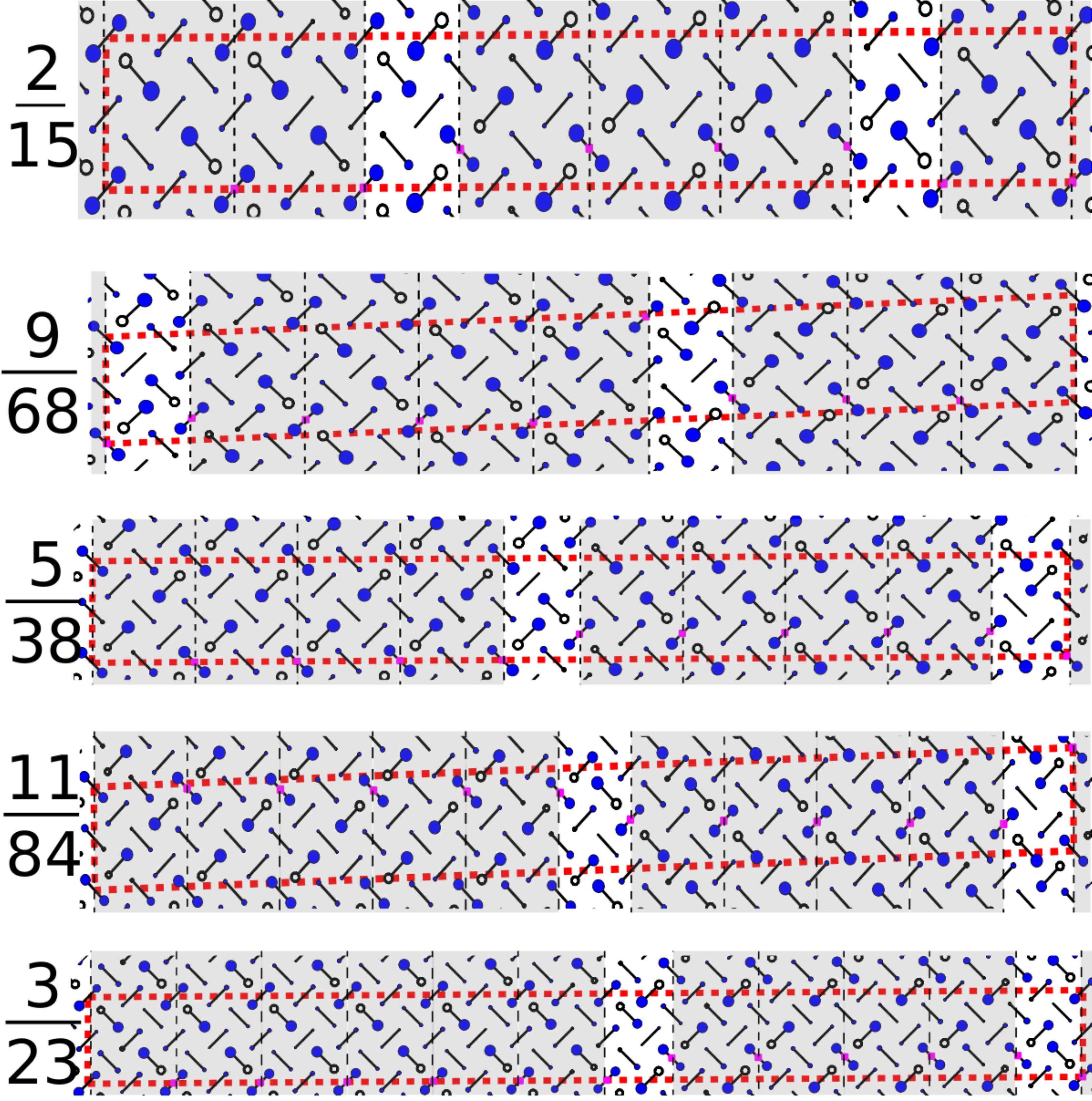}
\label{fig:AllStripeMech_MAGNETIZATION}
}
   \end{center}
   \caption{Figure shows all stripe structures present in the low-density phase diagram for $J'/J=0.68$, $t1/J=0.016$,  and $t2/J=-0.01$ as displayed in Fig.~\ref{fig:Magnet_J68_2_15_3_32}. All Wigner crystals contain stripes build by the $1/8$-diamond structure (shaded in grey).}
    \label{fig:AllStripeMech}
\end{figure}

The second regime ii) which includes the above discussed striped $2/15$ structure is different. In fact, the same microscopic mechanism giving rise to the $2/15$ plateau, applies to an (in principle) infinite number of striped plateaux above $1/8$. The specific stripe structures together with their magnetizations are shown in Fig.~\ref{fig:AllStripeMech}. Most importantly, our findings agree well with the experimental data of SrCu$_2$(BO$_3$)$_2$ \cite{takigawa08,takigawa12}: One finds that the translational symmetry stays broken above the $1/8$ plateau.

\subsection{Correlated hopping}

Let us finally discuss to what extent so-called correlated hopping terms can alter the above findings. We have computed all correlated hopping terms starting at most in order 6 in $J'/J$ up to order 14 in $J'/J$. As already mentionned, such couplings allow a hopping of a particle when another particle is present and remains static. The quantum fluctuations induced by correlated hopping depends therefore on the density. The higher the density, the more important is the effect of correlated hopping for the formation of plateau structures. Let us remark that correlated hopping has been also identified as a possible driving force for the realization of so-called pair superfluids \cite{bendjama05,schmidt06}. If realized, such phases are expected to be present at very low densities in the density range which we have classified as $\alpha$. The description of such phases goes well beyond our mean-field considerations. 

Formally, a correlated hopping process is written as  $t'_{i,j,k} \hat{b}^\dagger_i \hat{b}^{\phantom{\dagger}}_j \hat{n}^{\phantom{\dagger}}_k $, i.e.~a particle hops from dimer $j$ to dimer $i$ with an amplitude $t'_{i,j,k}$ if another particle is present on dimer $k$. The importance of such a correlated hopping term for a specific Wigner crystal depends on one hand on the strength $t'_{i,j,k}$ and on the other hand on the potential barrier on dimer $i$ on which the particle hops. A correlated hopping process induces therefore large quantum fluctuations whenever its amplitude is large and the involved repulsive interactions in intermediate states are small. 

Let us discuss the most important correlated hopping terms for the relevant structures at densities $1/9$, $1/8$, and $1/6$ as discussed above. For $n=1/9$, there is no correlated hopping process starting up to order 6 which induce quantum fluctuations on the classical 1/9-plateau. One can therefore conclude that correlated hopping has no real impact on the mean-field energy of this plateau. 

\begin{figure}
   \begin{center}
    \includegraphics[width=\columnwidth]{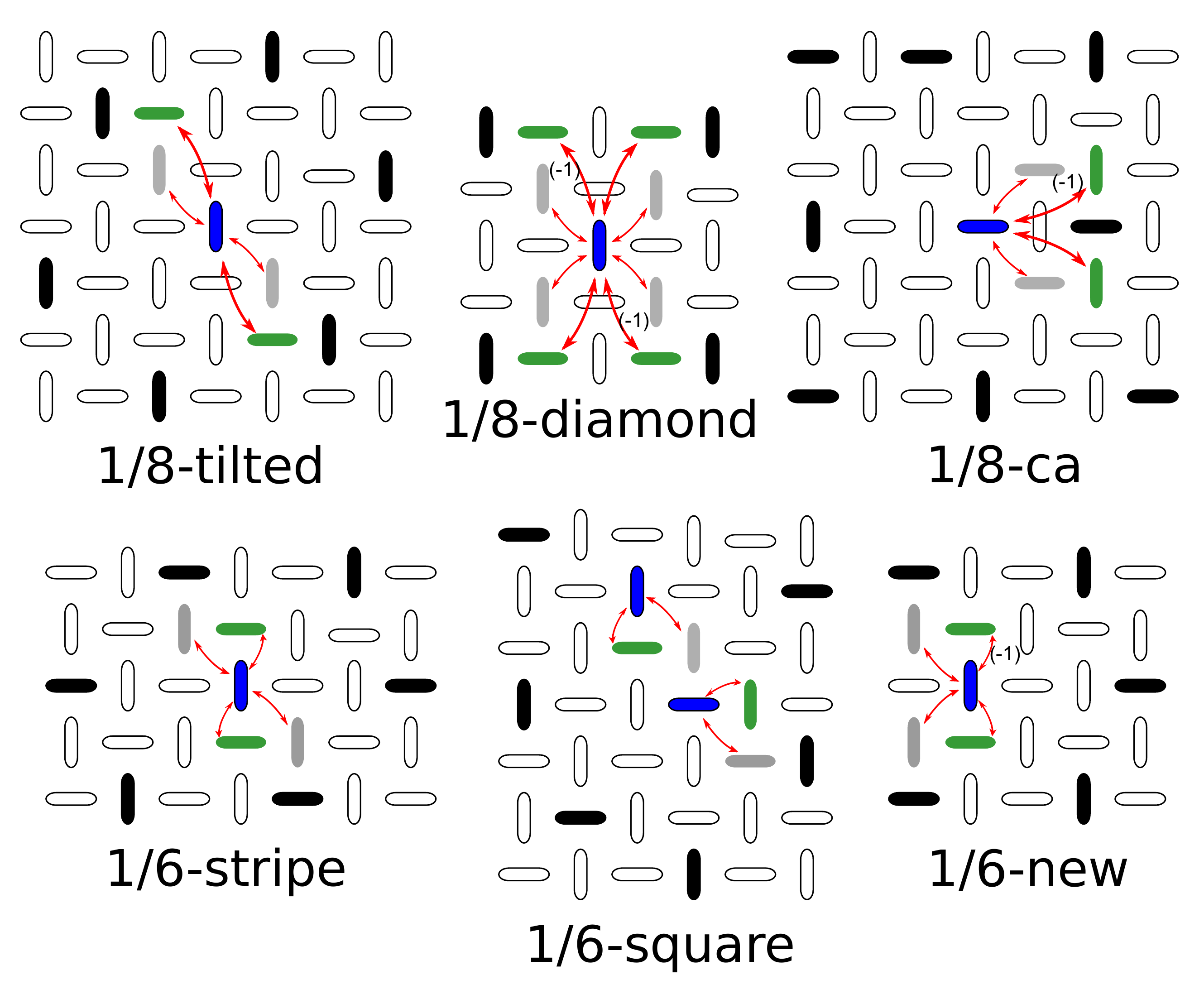}
   \end{center}
   \caption{Illustrations of the relevant correlated hopping processes for the structures at density $1/8$ (top) and density $1/6$ (down).}
    \label{fig:correlated_hopping}
\end{figure}

For the relevant structures at density $n=1/8$, there exist correlated hopping terms starting in order six perturbation theory having an amplitude of the order $J/1000$ in the regime $J'/J\approx 0.65$ (see Fig.~\ref{fig:correlated_hopping}). In our mean-field treatment, these terms renormalize slightly the values for $t_1$ and $t_2$. Furthermore, the induced quantum fluctuations are very similar for the structures $1/8$-diamond, $1/8$-tilted, and $1/8$-ca. Altogether, all our findings for density $1/8$ are expected to be unchanged when correlated hopping terms are included.

The situation is different for the structures at density $n=1/6$. Here correlated hopping processes exist which start already in order four perturbation theory. The corresponding amplitudes are of the order $J/100$ for $J'/J\approx 0.65$ (see Fig.~\ref{fig:correlated_hopping}). The most relevant hopping processes are those which do not contain any of the large repulsive interactions $V_1$ or $V_3$ in intermediate states because then in full analogy to the discussion above fluctuations from the classical position are not strongly suppressed. Interestingly, the induced quantum fluctuations due to these correlated hopping terms are again almost identical for the structures $1/6$-stripe, $1/6$-square, and $1/6$-new. We therefore expect that our results for density $1/6$ also hold when correlated hopping terms are taken into account.

The most subtle density regime is the one between $1/8$ and $1/6$ where our mean-field calculation predicts an infinite sequence of stripe structures. Again, there exist correlated hopping terms starting in order four perturbation theory which are of similar magnitude as the kinetic processes $t_1$ and $t_2$. Consequently, one expects that correlated hopping may have some impact on this hierarchy of plateau structures. This might be especially true for the density regime just below $1/6$ where our mean-field calculation stabilizes unphysical states classified as $\beta$ in the phase diagram. It would be therefore highly interesting to study this part of the phase diagram in more detail, because one might expect supersolid phases due to the increasingly important correlated hopping terms \cite{schmidt08}.        

\section{Summary}
\label{Sect:Summary}
The main motivation of this work is the discrepancy between experiment and theory on the understanding of 
the magnetization curve of the frustrated quantum magnet SrCu$_2$(BO$_3$)$_2$. The most prominent magnetization
 plateau at $M=1/8$ is currently not understood microscopically. Instead a sequence of plateaux at $M=1/9$, $M=2/15$,
 and $M=1/6$ is found at low magnetization \cite{dorier08}. Interestingly, additional low-density plateaux at $M=2/15$ and
 $M=1/6$ have been recently observed experimentally \cite{takigawa12} but their structures disagree with theory.  

Microscopically, it is believed that the material SrCu$_2$(BO$_3$)$_2$ is well described by the Shastry-Sutherland model. In Ref.~\onlinecite{dorier08}, an appropriate effective low-energy theory for the Shastry-Sutherland model in an external magnetic field is derived which is then solved in the classical limit. The classical approach is believed to work well because quantum fluctuations are strongly suppressed due to the strong frustration in the Shastry-Sutherland model. The latter is certainly true as long as one is not too close to the phase transition point $J'/J\approx 0.7$ or as long as additional magnetic couplings are negligible. Additionally, one assumes that bound states of triplons are not essential for the formation of plateaux at low densities \cite{manmana11}.

Here we have studied the effect of quantum fluctuations on the sequence of magnetization plateaux at low densities $n \leq 1/6$ which originate either from additional magnetic couplings like the DM-interaction or from the one-particle hopping already contained in the Shastry-Sutherland model. Both couplings are expected to be of the order $J/100$ at strong coupling $J'/J\approx 0.65$ which is the relevant parameter regime for SrCu$_2$(BO$_3$)$_2$. Physically, the plateaux are formed by freezing hardcore bosons in a regular fashion in order to avoid the strong repulsive interactions dominating the effective low-energy description. Quantum fluctuations correspond then to kinetic processes of the hardcore bosons in the effective model. Here we have treated the kinetic terms on a mean-field level by reducing the complex many-body problem to an effective one-body problem which is solved self-consistently. We have concentrated on crystalline solutions of the mean-field equations.

It is the central result of this article that quantum fluctuations are essential to understand the low-density phase diagram of the compound SrCu$_2$(BO$_3$)$_2$. Furthermore, we find several indications that the ratio $J'/J$ is very large $J'/J\geq 0.65$ placing the frustrated quantum magnet SrCu$_2$(BO$_3$)$_2$ in the most complex parameter regime. Both aspects together result in a low-density phase diagram which is different to the classical sequence $M=1/9$, $M=2/15$, and $M=1/6$ but much closer to the experimental results and their phenomenlogical interpretation \cite{takigawa12}: a) We find that quantum fluctuations stabilize in a natural way a 1/8 plateau with a diamond unit cell consistent with experiments. The quantum fluctuations induced by the DM interactions play a central role. b) The structure of the $1/6$ plateau changes already on the classical level for $J'/J\geq 0.67$. Microscopically, the repulsive interaction $V'_3$ becomes comparable to $V_4$ in this parameter regime and, as a consequence, it becomes favorable to realize structures avoiding $V'_3$. This is likely the reason why the classical structures found in Ref.~\onlinecite{dorier08} for $J'/J=0.5$ are different to the ones observed experimentally. All classical structures for $J'/J=0.5$ depend strongly on $V'_3$. For $J'/J\geq 0.67$, one finds that the three classical structures $1/6$-stripe, $1/6$-square, and $1/6$-new have exactly the same classical energy. Interestingly, the structure $1/6$-stripe is proposed phenomenologically \cite{takigawa12}. In our mean-field theory, quantum fluctuations favor the structure $1/6$-new but we would like to stress that energy differences are very small. c) In the regime between $1/8$ and $1/6$ our mean-field theory predicts a sequence of stripe structures all containing substructures of densities $1/8$ and $1/6$. This sequence includes a $2/15$ plateau. Qualitatively, this is again consistent with the phenomenological interpretation of the NMR data \cite{takigawa12}. But in the mean-field treatment it is always the $1/8$ plateau with a diamond unit cell which is realized inside the stripes because it benefits most from the induced quantum fluctuations. d) The mean-field solution is still consistent with a plateau at density $1/9$ which is in disagreement with the NMR data on SrCu$_2$(BO$_3$)$_2$ but in accordance with high-field torque measurements by Sebastian et al\cite{sebastian07}. But we stress that it is well possible that the $1/9$ plateau melts completely when superfluid solutions are considered, when effects beyond the mean-field level are taken into account, or when additional magnetic couplings are present. Indeed, as we have shown above, already slightly larger values for the kinetic terms lead to a melting of the $1/9$ plateau.

Clearly, there are limitations of our microscopic approach. One has to distinguish the following three levels: i) limitations of the mean-field treatment, ii) limitations of the effective low-energy description, and iii) limitations of the studied microscopic Hamiltonian. 

i) We have focused on crystalline solutions of the mean-field equations because the magnetization curve of SrCu$_2$(BO$_3$)$_2$ is dominated by magnetization plateaux corresponding to gapped bosonic Mott insulators. Definitely, this approach is hard to justify in the dilute limit where one expects superfluid solutions to be present in the phase diagram. In our theory, we find a sequence of low-density plateau structures with a negative energy for $\mu=\mu_0$. This part of the phase diagram denoted by $\alpha$ in this work corresponds likely to a superfluid phase. Additionally, the mean-field treatment becomes problematic when the physics cannot be reduced to an effective one-body problem. We found solutions in the density regime between $1/8$ and $1/6$ denoted by $\beta$ where particles in the regions are completely delocalized between different dimers. This is clearly an artefact of the mean-field calculation and we do not trust the mean-field solution in this density regime. One possibility to go beyond might be to map the effective low-energy model in terms of hardcore bosons to a quantum spin model which is then solved by spin-wave theory on various large unit cells.

ii) The effective low-energy theory is limited by the extrapolations used for the various couplings and by the choice of operators included in the effective low-energy description. Qualitatively, we expect that all revelant tendencies are correctly described by the extrapolations up to the relevant regime $J'/J\geq 0.65$. It is nevertheless well possible that errors in the extrapolations lead to little changes, e.g.~the value $J'/J\geq 0.67$ where the structure of the $1/6$ plateau changes already on the classical level might shift slightly. One option to go beyond is to determine the effective couplings like the repulsive interactions non-perturbatively using continuous unitary transformation \cite{yang11,krull12,yang12} or contractor renormalization \cite{capponi08}. Indeed, this would be of great help in order to describe the physics quantitatively in the relevant coupling regime of large $J'/J$. The choice of operators to be treated in the effective low-energy models depends on the density. Here we have included two-body interactions and one-particle hopping. At low densities this is certainly a very good approximation. As discussed above, corrections like correlated hopping are negligible for $n\leq 1/8$. We therefore expect that all our findings for the most relevant $1/8$ plateau are quantitatively correct. In contrast, the same is less obvious for the density regime $1/8 < n \leq 1/6$. Here certain correlated hopping processes could be important. A correct treatment of the correlated hopping processes is most likely only possible if one goes beyond a mean-field description. At larger densities like $n=1/4$, correlated hopping is expected to be relevant \cite{foltin12}. 

iii) Our results demonstrate clearly that residual interactions of the order $J/100$ beyond the pure Shastry-Sutherland model are important for the magnetization of the frustrated quantum magnet SrCu$_2$(BO$_3$)$_2$. Here we have included the effects of DM interactions and of additional Heisenberg couplings to next-nearest neighbors by including the order-one contribution to the effective low-energy model. Higher orders in such couplings are expected to be unimportant. But let us stress that we have not considered effects linear in the additional couplings times a factor $(J'/J)^n$ with $n>0$. Most importantly, it would be interesting to study the feedback effect of the additional couplings on the two-body repulsive interactions \cite{schmidt09}. Finally, let us mention that the role of inter-plane exchange couplings for the formation of plateau structures is currently not understood.

Altogether, we have presented an important step towards a microscopic description of the frustrated quantum magnet SrCu$_2$(BO$_3$)$_2$. Quantum fluctuations induced by unfrustrated additional magnetic couplings like the DM interaction are very important in the low-density regime. A natural mechanism for the realization of the most prominent $1/8$ plateau with a diamond unit cell is discovered. Furthermore, several indications for a rather large ratio $J'/J\geq 0.65$ are found placing SrCu$_2$(BO$_3$)$_2$ in the most challenging parameter regime. 
\\

\section{Acknowledgements}
We acknowledge very useful discussions with F.~Mila and S.~Manmana. KPS acknowledges ESF and EuroHorcs 
for funding through his EURYI.

\section{Appendix}
\label{sect:Appendix}

\subsection{Classical plateau energies}
In the following we list the explicit expressions for the classical energy
 per dimer $\epsilon_{\rm cl}=E_{\rm cl}/(JN)$ of all considered plateaux from Fig.~\ref{fig:classical_plateaux}:
\begin{eqnarray*}
 \epsilon^{1/6}_{\rm cl, ca} &=& \frac{1}{6} \left( V^\prime_3 + 2V_7\right) \\
 \epsilon^{1/6}_{\rm cl,square} &=& \frac{1}{6} \left( V_4+V_5 +V_6\right) \\
 \epsilon^{1/6}_{\rm cl,stripe} &=& \frac{1}{6} \left( V_4+V_5 +V_6\right) \\
 \epsilon^{1/6}_{\rm cl,new} &=& \frac{1}{6} \left( V_4+V_5 +V_6\right) \\
 \epsilon^{2/15}_{\rm cl,rect} &=& \frac{1}{15} \left( V^\prime_3+V_7 +2V_6 + 2V_8\right) \\
 \epsilon^{2/15}_{\rm cl,rhomb} &=& \frac{1}{15} \left( V^\prime_3+V_7 +2V_6 + 2V_8\right) \\
\epsilon^{2/15}_{\rm cl, big} &=& \frac{1}{30} \left( V_4+4V_5+ V_6 + 3V^\prime_{7} + 3V_7 \right) \\
 \epsilon^{2/15}_{\rm cl, b_2} &=& \frac{1}{30} \left( 7V_5 + V_4 + V_6 \right) 
\end{eqnarray*}
\begin{eqnarray*}
\epsilon^{1/8}_{\rm cl,dia} &=& \frac{1}{4} V_5 \\
 \epsilon^{1/8}_{\rm cl,tilted} &=& \frac{1}{8} \left( V_5+V_7 \right) \\
 \epsilon^{1/8}_{\rm cl,ca} &=& \frac{1}{24} \left( V^\prime_{3} + 4V_6 + V7 + 2V_8 \right) \\
 \epsilon^{1/9}_{\rm cl} &=& \frac{1}{9} 2V_6 \quad .
\end{eqnarray*}

\subsection{Relevant series expansions}

In the following we show the series expansion for the two-body interactions $V_{\delta}$:
\onecolumngrid
\begin{align*}
V_1 &=  \frac{1}{2}  x + \frac{1}{2}  x ^2- \frac{1}{8}  x ^3- \frac{9}{16}  x ^4- \frac{3}{64}  x ^5+ \frac{809}{768}  x ^6+ \frac{2173}{3072}  x ^7- \frac{70543}{24576}  x ^8- \frac{37816411}{5308416}  x ^9- \frac{2055058321}{637009920}  x ^{10} \\
&\quad + \frac{335455501303}{25480396800}  x ^{11} +\frac{164557631263}{8153726976}  x ^{12}- \frac{31644100269296887}{1100753141760000}  x ^{13}- \frac{40411124068847011421}{308210879692800000}  x ^{14} \\
V_2 &=  \frac{1}{4}  x ^3+ \frac{3}{8}  x ^4+ \frac{23}{64}  x ^5- \frac{41}{128}  x ^6- \frac{337}{192}  x ^7- \frac{283327}{221184}  x ^8+ \frac{23684687}{5308416}  x ^9+ \frac{1362864853}{127401984}  x ^{10}- \frac{12420874729}{76441190400}  x ^{11} \\
&\quad - \frac{65984600196191}{1834588569600}  x ^{12}- \frac{52361880469040173}{1100753141760000}  x ^{13}+ \frac{18756837485942785969}{308210879692800000}  x ^{14}+ \frac{64803177532673441622343}{258897138941952000000}  x ^{15} \\
V_3 &=  \frac{1}{2}  x ^2+ \frac{3}{4}  x ^3- \frac{1}{8}  x ^4- \frac{49}{64}  x ^5- \frac{289}{768}  x ^6+ \frac{4019}{9216}  x ^7+ \frac{77609}{110592}  x ^8+ \frac{243991}{1327104}  x ^9- \frac{73855279}{79626240}  x ^{10}- \frac{1584489421}{1061683200}  x ^{11} \\
&\quad - \frac{4392913}{298598400}  x ^{12}+ \frac{100260686730911}{45864714240000}  x ^{13}+ \frac{183830050986989969}{115579079884800000}  x ^{14}- \frac{224180083051247518037}{97086427103232000000}  x ^{15} \\
V'_3 &=  \frac{1}{16}  x ^6+ \frac{95}{576}  x ^7 + \frac{99}{512}  x ^8+ \frac{29105}{663552}  x ^9 - \frac{10508077}{79626240}  x ^{10}+ \frac{732947101}{9555148800}  x ^{11}+ \frac{941467680131}{1146617856000}  x ^{12} \\
&\quad + \frac{193042731654521}{137594142720000}  x ^{13} + \frac{2992354581488789}{4280706662400000}  x ^{14}- \frac{109470098444640062071}{97086427103232000000}  x ^{15} \\
V_4 &=  \frac{1}{8}  x ^4+ \frac{17}{64}  x ^5+ \frac{77}{768}  x ^6- \frac{3571}{9216}  x ^7 - \frac{59257}{110592}  x ^8+ \frac{1309517}{2654208}  x ^9+ \frac{31968767}{15925248}  x ^{10} + \frac{50142651521}{38220595200}  x ^{11}\\
&\quad- \frac{6944493534619}{2293235712000}  x ^{12}- \frac{135312316251467}{22015062835200}  x ^{13}+ \frac{87342511054149749}{115579079884800000} + x ^{14}+ \frac{6144395627935647622027}{388345708412928000000}  x ^{15} \\
V_5 &=  \frac{1}{32}  x ^6+ \frac{95}{1152}  x ^7 \frac{2677}{27648}  x ^8+ \frac{269}{12288}  x ^9- \frac{5314067}{79626240}  x ^{10}+ \frac{87702253}{2388787200}  x ^{11} \\
&\quad + \frac{467408584057}{1146617856000}  x ^{12}+ \frac{3555696264811}{5096079360000}  x ^{13}+ \frac{4984498011323963}{14447384985600000}  x ^{14}- \frac{52725527155315535453}{97086427103232000000}  x ^{15} \\
V_6 &=  \frac{5}{1152}  x ^8+ \frac{5005}{221184}  x ^9+ \frac{386029}{6635520}  x ^{10}+ \frac{7416311}{79626240}  x ^{11}+ \frac{19425481571}{191102976000}  x ^{12} \\ 
&\quad + \frac{1134494882761}{11466178560000}  x ^{13}+ \frac{263630347224169}{1541054398464000}  x ^{14}+ \frac{4296879240693151027}{10787380789248000000}  x ^{15} \\
V_7 &=  \frac{1}{64}  x ^6+ \frac{5}{72}  x ^7 \frac{3805}{27648}  x ^8+ \frac{59561}{442368}  x ^9+ \frac{601397}{159252480}  x ^{10}- \frac{2335042153}{19110297600}  x ^{11} + \frac{99836042591}{2293235712000}  x ^{12} \\
&\quad + \frac{51824530139999}{91729428480000}  x ^{13} + \frac{207944406615859127}{231158159769600000}  x ^{14} + \frac{10898694975478960837}{38834570841292800000}  x ^{15}\\
 V'_7 &=  \frac{1}{3072}  x ^{10}+ \frac{55667}{22118400}  x ^{11}+ \frac{7785821}{884736000}  x ^{12}+ \frac{1258120201}{63700992000}  x ^{13}+ \frac{1763035718197}{53508833280000}  x ^{14} \\
&\quad + \frac{10597817495588783}{224737099776000000}  x ^{15} \\
\end{align*}
\begin{align*}
V_8 &=  \frac{5}{2304}  x ^8+ \frac{1679}{147456}  x ^9+ \frac{1545691}{53084160}  x ^{10}+ \frac{295607977}{6370099200}  x ^{11}+ \frac{12773678617}{254803968000}  x ^{12} \\
&\quad + \frac{1475777364467}{30576476160000}  x ^{13}+ \frac{6509870778092779}{77052719923200000}  x ^{14}+ \frac{1846845836550193249}{9246326390784000000}  x ^{15} \\
t_2 &= -\frac{1}{96}x^6-\frac{11}{576}x^7-\frac{83}{4608}x^8-\frac{2447}{663552}x^9-\frac{1433}{13271040}x^{10}-\frac{303150173}{9555148800}x^{11}-\frac{6754153889}{76441190400}x^{12} \\ 
&\quad -\frac{14926695218519}{137594142720000}x^{13}-\frac{7849573000127}{158544691200000}x^{14}+\frac{1797434359986621511}{97086427103232000000}x^{15} \\
&\quad -\frac{232952050093988453401}{3398024948613120000000}x^{16}-\frac{24509424377792876369060333}{68504182964040499200000000}x^{17} \\
\mu_0 &= 1 -Jp^2-\frac{1}{2}x^3-\frac{1}{8}x^4+\frac{5}{32}x^5+\frac{3}{128}x^6-\frac{1699}{4608}x^7-\frac{35107}{55296}x^8-\frac{259061}{663552}x^9+\frac{974687}{6635520}x^{10} +\frac{1151870527}{4777574400}x^{11} \\ 
&\quad -\frac{23323161421}{38220595200}x^{12}-\frac{40392330558191}{22932357120000}x^{13}-\frac{102289196477738603}{57789539942400000}x^{14}-\frac{8209068928447247311}{48543213551616000000}x^{15}\\
&\quad +\frac{45063299341442551818889}{40776299383357440000000}x^{16} - \frac{26736958692466776560379961}{34252091482020249600000000}x^{17}
\end{align*}
\twocolumngrid


\begin{thebibliography}{10}
\bibitem{footnote}
Let us mention that the results by CORE \cite{capponi08} show a $1/8$ plateau in the phase diagram but
 the considered system sizes used in the exact diagonalizations forbid the presence of a $2/15$ plateau. 
As a consequence, one cannot judge whether a $1/8$ plateau is present in the phase diagram in the thermodynamic limit.
\end{thebibliography}
\end{document}